\documentclass{report}
\usepackage[dvips]{graphics}
\usepackage{ncsu-thesis}
\usepackage{mods}
\usepackage{natbib}
\newcommand\lesssim{\mathrel{\hbox{\rlap{\hbox{\lower4pt\hbox{$\sim$}}}\hbox{$<$}}}}
\newcommand\gtrsim{\mathrel{\hbox{\rlap{\hbox{\lower4pt\hbox{$\sim$}}}\hbox{$>$}}}}
\newcommand{\dopicture}[2]{\begin{center}\scalebox{#1}{\includegraphics{#2}}\end{center}}
\newcommand{\EffP}{\epsilon(>p)}
\newcommand{\EffOneFour}{\epsilon_{1.4}(>m_p c)}

\newcommand{\EffTen}{\epsilon_{10}(>10 m_p c)}
\newcommand{\rel}{relativistic}
\newcommand{\nonrel}{nonrelativistic}
\newcommand{\ultrarel}{ultrarelativistic}

\newcommand{\gamsk}{\gamma_{\mathrm{sk}}}
\newcommand{\mc}{Monte Carlo}

\newcommand{\ave}[1]{<\!\!{#1}\!\!>}
\newcommand\Alf{Alfv\'en}
\newcommand{\gamrel}{\gamma_{\mathrm{rel}}}
\newcommand{\betarel}{\beta_{\mathrm{rel}}}

\newcommand{\Nmin}{N_{\mathrm{min}}}
\newcommand{\fracinj}{f_{\mathrm{inj}}}
\newcommand{\GamNL}{\Gamma_{\mathrm{NL}}}
\newcommand{\GamTP}{\Gamma_{\mathrm{UM}}}
\newcommand{\gaminj}{\gamma_{\mathrm{inj}}}
\newcommand{\gamux}{\gamma_u(x)}
\newcommand{\delmax}{\delta \theta_{\mathrm{max}}}
\newcommand{\UDUave}{\ave{p_f/p_i}_{\mathrm{u-d-u}}}
\newcommand{\DUDave}{\ave{p_f/p_i}_{\mathrm{d-u-d}}}
\newcommand{\ppf}{p_{\mathrm{pf}}}

\newcommand{\Pret}{{\cal P_{\mathrm{ret}}}}
\newcommand{\rMC}{r_{\mathrm{MC}}}

\newcommand\rgz{r_{g,0}}
\newcommand\too{\, \rightarrow \, }
\newcommand{\thetaSK}{\theta_{\mathrm{sk}}}
\newcommand{\TBnZ}{\Theta_\mathrm{B0}}
\newcommand{\TbnZ}{\Theta_\mathrm{B0}}
\newcommand{\TuTwo}{\Theta_\mathrm{u2}}
\newcommand{\TBnTwo}{\Theta_\mathrm{B2}}
\newcommand{\TbnTwo}{\Theta_\mathrm{B2}}
\newcommand\vinj{v_\mathrm{inj}}
\newcommand\Einj{E_\mathrm{inj}}
\newcommand{\transrel}{trans-relativistic}
\newcommand{\rIso}{r_{\mathrm{iso}}}
\newcommand{\Ppar}{P_{\|}}
\newcommand{\Pperp}{P_{\perp}}

\newcommand{\Tmntot}{T^{\mu\nu}_{\mathrm{total}}}
\newcommand{\Tmnfluid}{T^{\mu\nu}_{\mathrm{fluid}}}
\newcommand{\TmnEM}{T^{\mu\nu}_{\mathrm{EM}}}
\newcommand{\FluxFluidEn}{F^{\mathrm{fluid}}_{en}}
\newcommand{\FluxFluidPx}{F^{\mathrm{fluid}}_{px}}
\newcommand{\FluxFluidPz}{F^{\mathrm{fluid}}_{pz}}
\newcommand{\FluxEmEn}{F^{\mathrm{EM}}_{en}}
\newcommand{\FluxEmPx}{F^{\mathrm{EM}}_{px}}
\newcommand{\FluxEmPz}{F^{\mathrm{EM}}_{pz}}

\newcommand\TBn{\Theta_{\mathrm{Bn}}}

\newcommand{\xx}[1]{\!\times\!10^{#1}}

\newcommand\itt{ }
\newcommand\bff{ }

\newcommand\TP{test-particle}

\newcommand\kmps{km s$^{-1}$}

\newcommand\pmax{p_\mathrm{max}}
\newcommand\pcc{cm$^{-3}$}
\newcommand\muG{$\mu$G}

\newcommand\egc{e.g.,}
\newcommand\etal{et al.}

\newcommand\alf{Alfv\'en}
\newcount\Inum
\newcount\IInum
\newcount\IIInum
\newcount\IVnum
\def\I{\global\multiply\IInum by 0 \global\multiply\IIInum by 0
            \global\multiply\IVnum by 0 \global\advance \Inum by 1
            {\the\Inum. }}
\def\II{\global\multiply\IIInum by 0\global\multiply\IVnum by 0
       \global\advance \IInum by 1 {\the\Inum.\the\IInum. }}
\def\III{\global\multiply\IVnum by 0\global\advance \IIInum by 1
            {\the\Inum.\the\IInum.\the\IIInum. }}
\def\IV{\global\advance \IVnum by 1
            {\the\IVnum. }}
\def\bfrac#1#2{\hbox{${{\displaystyle#1 \vphantom{(} }\over{
   \displaystyle #2 \vphantom{(} }}$}}                
\newcount\listnorom
\newcommand\listromanDE{\global\advance \listnorom by 1 
{\lowercase\expandafter{\ (\romannumeral\listnorom)}\ }}

\ssp
\begin{document}
\chapter*{\begin{center}Abstract\end{center}}
\pagestyle{empty}
\large
Double, Glen Paul. Investigations of Lepton and Baryon Acceleration in 
       Relativistic Astrophysical Shocks. (Under the direction of
       Donald C. Ellison.)\\

%
Gamma-ray bursts, occurring randomly each day anywhere in the universe,
may be the brightest objects in the sky during their short life.
Particle acceleration in mildly relativistic shocks, internal to the
main blastwave, may explain the early intensity peaks in gamma-ray bursts 
and the afterglow
may be explained by energetic particles accelerated by the main 
ultrarelativistic blastwave shock as it slows to the mildly relativistic 
range.
To help explain the phenomena, a nonlinear relativistic
Monte Carlo model was developed and used to study lepton and baryon 
acceleration by mildly relativistic modified shocks with the 
magnetic field parallel to the shock normal.
The study showed that for equal densities of
leptons and baryons, lepton acceleration is highly sensitive 
to the shock velocity profile. With the shock fully
modified by energetic baryons,
the injection efficiency of leptons, relative to baryons, increases with
Lorentz factor, and injection efficiency
will reach a maximum well below that of baryons
at the same momentum. Given the assumptions in this model, 
if the particles are energized by shock
acceleration, leptons will always carry far less energy than baryons 
when the lepton and baryon densities are of the same order. 
It was determined
that the lepton to baryon number density ratio must be approximately
$3\times 10^5$ for both species to equally share the kinetic  
energy of the shock. This energy equipartition
density ratio is independent of shock speed
over the range of mildly relativistic Lorentz factors used in the study,
but the result may extend to ultrarelativistic speeds.
The study was a special case of a
larger effort that will include relativistic oblique modified shocks
and computer generated gamma-ray spectra when the model is completed. 
%
%
New magnetohydrodynamic conservation laws and relativistic jump conditions
were developed for the model, along with a new equation of state 
and a new method for estimating the adiabatic index
in the mildly relativistic range. The present state of the model shows
smooth transitions of shock parameters from
nonrelativistic to highly relativistic unmodified shocks while allowing 
oblique magnetic fields and a pressure tensor.   
\thispagestyle{empty}

\newpage
%

\title{\bf Investigations of Lepton and Baryon Acceleration in 
       Relativistic Astrophysical Shocks}
\author{Glen Paul Double}
\thesistype{dissertation}
\degree{Doctor of Philosphy}
\dept{Department of Physics}
\foursigstrue
\beforepreface
\large
\setlength{\baselineskip}{24pt}
\ssp
%
\chapter*{\begin{center}Acknowledgements\end{center}}

I would like to thank my advisor, Don Ellison, for his guidance, 
and many helpful critiques and suggestions
throughout this course of study. I am
grateful to him for the use of his Monte Carlo program 
and for his patient explanations
of its sophisticated features, without which I could not have done
this work in any reasonable period of time.  
I would like to thank my advisory committee, including Mohamed Bourham,
Dean Lee, John Blondin, and Don Ellison who served as chairman.
I would also like to thank Steve Reynolds,
Frank Jones and Matthew Baring for many helpful suggestions. I
especially appreciate my wife's patient endurance and support during the
long hours spent in completing this work.


\tableofcontents
\listoffigures
\addchapter*{List of Symbols}
\begin{tabbing}
SYMBOL\qquad\= \quad\= DEFINITION\\[0.8ex]
$\alpha$ \> - \> anisotropy parameter for tensor pressure\\
$\beta$ \> - \> dimensionless speed, normalized by c\\
$c$ \> - \> speed of light (299792458 meters/second)\\
$\gamma$ \> - \> Lorentz factor $(1-\beta^2)^{-\frac{1}{2}}$\\
$\gamma_0$ \> - \> Lorentz factor of shock speed or far upstream flow speed;
also called  $\gamsk$\\
$\Gamma$ \> - \> adiabatic index\\
GRB \> - \> Gamma-ray burst\\
$\epsilon_0$ \> - \> permittivity of free space\\
$\epsilon_{\mathrm{inj}}$ \> - \> injection efficiency\\
$\eta$ \> - \> gyrofactor that scales the gyroradius $r_g$\\
$e$ \> - \> total energy density\\
$eV$ \> - \> electron volt\\
$f_s$ \> - \> normalized subshock size\\
ISM \> - \> Interstellar medium; the space between the stars\\
$k$ \> - \> Boltmann's constant\\
$\kappa$ \> - \> Diffusion coefficient\\
$\lambda$ \> - \> Mean free path\\
$\lambda_D$ \> - \> Debye length\\
$L$ \> - \> Diffusion Length\\ 
$\mu_0$ \> - \> permeability of free space\\
$n$ \> - \> particle number density\\
$p$ \> - \> particle momentum\\
$P_{\parallel}$ \> - \> fluid pressure along the magnetic field vector\\
$P_{\perp}$ \> - \> fluid pressure perpendicular to the magnetic field vector\\
$P_R$ \> - \> Probability of return function\\
$q$ \> - \> unit electric charge\\
$\rho_q$ \> - \> electric charge density\\ 
$\rho$ \> - \> rest mass density\\
$r$ \> - \> shock compression ratio ($u_0 / u_2$)\\
$r_g$ \> - \> gyroradius for a charged particle orbiting in a magnetic field\\
$r_s$ \> - \> subshock compression ratio ($u_1 / u_2$)\\
$\sigma$ \> - \> spectral index; the magnitude of a power law slope\\
$\Theta_{Bn}$ \> - \> angle between the magnetic field and the shock normal\\
$\mathcal{T}$ \> - \> temperature associated with kinetic energy of the particle ensemble\\
$u_0$ \> - \> shock speed or far upstream flow speed seen from the shock frame\\
$u_1$ \> - \> flow speed at the shock (subshock), 
as measured in the shock frame\\
$u_2$ \> - \> flow speed far downstream seen in the shock frame\\
$v$ \> - \> particle speed\\
$w$ \> - \> enthalpy ($e + P$) density or specific enthalpy\\
$Z$ \> - \> total electric charge number of an ion\\
\end{tabbing} 
\afterpreface
\Large
\ssp
%
\chapter{Introduction}

Shock waves occur in fluids whenever a disturbance propagates
faster than the characteristic speed of sound in the fluid. 
The same phenomena occur in space where 
the fluid is a plasma consisting of 
positively and negatively charged particles permeated by magnetic fields
and negligible electric fields.
 
In space where the particle density is low, typically a few
particles per cubic centimeter, shock waves are collisionless rather than
acoustic-ion shocks because the effective diffusion length due to charged 
particles interacting with the magnetic field is generally much shorter
than the mean free path between particles. 
An acoustic-ion shock wave propagates by Coulomb forces, 
for example shock waves in the Earth's atmosphere,
but collisionless shocks propagate through disturbances in the 
magnetic field of the space plasma. 

Shock waves are found throughout the universe.
Examples of nonrelativistic shocks range from transient
disturbances on the surface of the sun \citep{hollweg}, a standing
bowshock from the solar wind blowing across the
Earth's magnetic field \citep{Ellison85}, to
energetic shocks associated with supernova explosions \citep{DC98}
and other extreme types of exploding objects.  

Relativistic shocks originate from much more energetic events, 
including the most powerful supernova explosions 
\citep[hypernovae;][]{Rees00}, 
jets in extragalactic radio sources, and the mysterious objects 
that produce gamma-ray bursts. The study of astrophysical shock waves 
is important because the shock waves can accelerate charged particles,
by first-order
Fermi acceleration, to very high energies and possibly
explain the existence of the highest energy cosmic rays. The study of
relativistic shocks may bring some
understanding to the spectra seen in gamma-ray bursts.  

As an introduction, an overview will be presented of 
gamma-ray bursts, diffusive shock acceleration 
and the mathematical and computational tools 
that were developed by this author and others to study 
particle acceleration by relativistic shocks. 
The introduction is concluded with
the objectives and the overall organization of this dissertation. 
%

\section{Gamma-ray Bursts}
Gamma-ray bursts were discovered accidentally in the 1960's
by satellites intended for monitoring nuclear testing. Since then there has
been an abundance of research, both observational and theoretical, in an
attempt to better understand the nature and origin of the bursts. Much has
been published and an excellent review, among a number of others, 
was published by \citet{Rees00}. 

Gamma-ray bursts are extremely energetic blasts ($\sim10^{53}$ ergs if the
blast is isotropic; $\sim10^{51} $ ergs if beamed)
of primarily gamma and X-ray 
radiation that are often the brightest objects in the sky. They occur daily
and seem to come from every direction, from our galaxy to near
the edge of the observable universe. Gamma-ray bursts 
remain a great mystery, and only recently were optical
counterparts to the gamma-ray bursts observed. Virtually nothing is 
known about the engine that drives the burst. An understanding of gamma-ray
bursts will lead to a better understanding of the form and evolution of the
universe as a whole.

 
\subsection{Observational Overview}
\begin{figure}[!hbtp]
\dopicture{.65}{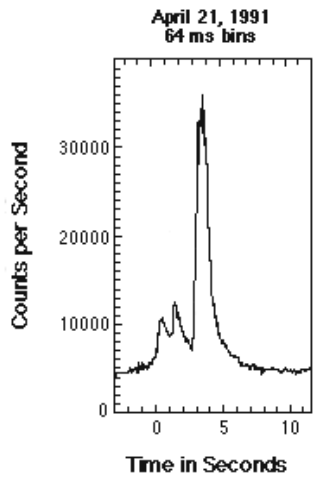}
\caption{Typical time plot of a gamma ray burst. 
Longer bursts tend to
have a more complex structure.
\label{fig:grb1}
}
\end{figure}

\begin{figure}[!hbtp]
\dopicture{.65}{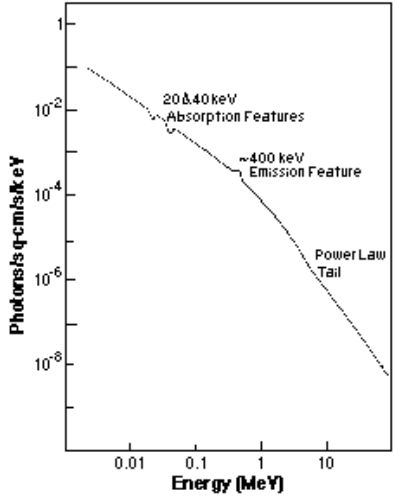}
\caption{Typical energy spectral plot of a gamma-ray burst.
\label{fig:spectra1}
}
\end{figure}

Gamma-ray bursts typically have a duration of seconds to minutes, as
shown in Figure \ref{fig:grb1}, 
but some have afterglows lasting weeks
or months. The bursts are isotropic on the sky and, for those which
have been associated with host galaxies, show large red shifts which
imply cosmological distances, and therefore extremely large energy
output to produce the brightness we see. 
Most of the power of the burst is non-thermal radiation
emitted in the 100 - 1000 keV range,
usually with a number of short duration intensity spikes which imply a
compact size with a radius not more than a few thousand kilometers. 
A``typical'' energy spectrum is shown in Figure \ref{fig:spectra1}, i.e.,
typical in the sense that The burst normally has a brief period of intense
multiple energy peaks, but every one is different and it makes it
difficult to summarize their basic features. 
The radiation energy is usually above 50 keV, sometimes in the MeV range,
but many gamma-ray
bursts have also exhibited an optical afterglow that follows the initial
burst, some with energies dipping into the infrared and microwave regions. 
The expansion rates of gamma-ray
bursts are explained by 
\ultrarel\ flows with Lorentz factors greater than 100,
possibly up to 1000 or more. 


\subsection{Proposed Models}	
\begin{figure}[!hbtp]
\dopicture{.44}{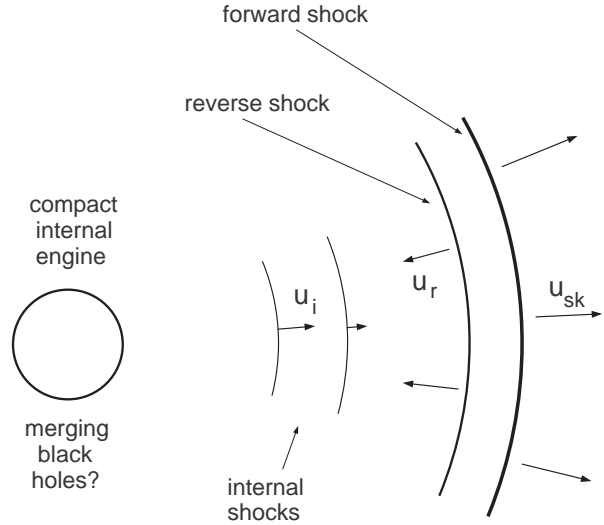}
\caption{The Fireball Model. A highly relativistic (quite likely beamed) 
forward shock moves
outward into the ISM and a reverse shock moves back
into the outward moving lepton wind. Various relativistic internal shocks
may interact with each other or with the reverse shock to produce intense
radiation spikes that make up the initial gamma-ray burst.
\label{fig:blast}
}
\end{figure}

The actual inner engine that produces the gamma-ray burst is unknown. 
Researchers speculate that compact objects (black holes, neutron stars,
etc.) somehow merge and produce the energy to power a gamma-ray burst.
The most widely accepted model appears to be the 
Fireball model \citep{Piran99} shown in Figure \ref{fig:blast}.
By some unknown means, a gamma-ray and/or neutrino
radiation-dominated highly relativistic 
blastwave gives rise to an external shock and an 
electron-positron pair-dominated  
\ultrarel\ wind in a region of low baryon density.   
A reverse shock, traveling backward into the lepton wind, and possibly
internal shocks interacting with each other or the reverse shock
at slower but still relativistic speeds, may be responsible for the
series of intensity peaks seen in the initial gamma-ray burst. The forward
shock slows quickly (on the order of minutes)
to mildly relativistic speeds as it picks up 
ambient mass in the interstellar medium.
Kinetic energy in the outflow  
is randomized by magnetic fields, probably collisionless shocks,
and is converted back into radiation seen as the afterglow. The 
computational methods used here allow the study of lepton and baryon
dynamics (i.e., Fermi acceleration and the ensuing nonthermal distributions
of energetic particles) in the internal shock regions
and the external afterglow regions where
the shocks speeds have Lorentz factors of 20 or less.
%

\section{First order Fermi acceleration}
Acceleration of a charged particle as a result of diffusive scattering in
converging flows was first suggested by \citet{Fermi49}. 
There were a number of important
papers that developed the concept of diffusive acceleration, or the
steady diffusive acceleration of test particles as they cross the shock
boundary numerous times and scatter
through stochastic processes. Due to the random
nature of the scattering process, information is destroyed (i.e., entropy is
increased) and the resulting energetic particle 
distribution is relatively independent of the details of the 
particle energy distribution prior to acceleration \citep{Drury83}. 
\citet{Krym77}, \cite{ALS77}, and
\citet{BO78} used a macroscopic approach to develop an analytical theory
of particle acceleration, assuming isotropy in the particle momentum
distribution in the fluid frame 
(because the particle velocity is assumed to be much higher
than the fluid velocity), and assuming collisionless, elastic scattering,
and assuming the scattering centers are frozen into the background plasma.
An accelerated particle momentum spectrum
can be generated by using the transport equation in the local fluid frame,
and matching boundary conditions for the steady-state solution 
on each side of the infinite plane shock. This approach is straightforward,
but it does not provide an understanding of the physical processes that
are responsible for the particle acceleration. 
 A microscopic derivation of diffusive particle acceleration was presented by
\citet{Bell78a}, \citet{Peacock81}, and also \citet{Michl81} which provides
much more insight into the physics of the particle acceleration process.
A number of papers, \citet{Drury83}, \citet{BE87}, and \citet{JE91} among 
others,
review both approaches in detail, covering the
effects of oblique magnetic fields, modification of the the shock
velocity profile by the backpressure of accelerated particles, and the
connection to cosmic rays. 
The papers cited above deal with nonrelativistic shocks, but they provide
a good introduction to shock acceleration of particles.

\begin{figure}
\dopicture{.7}{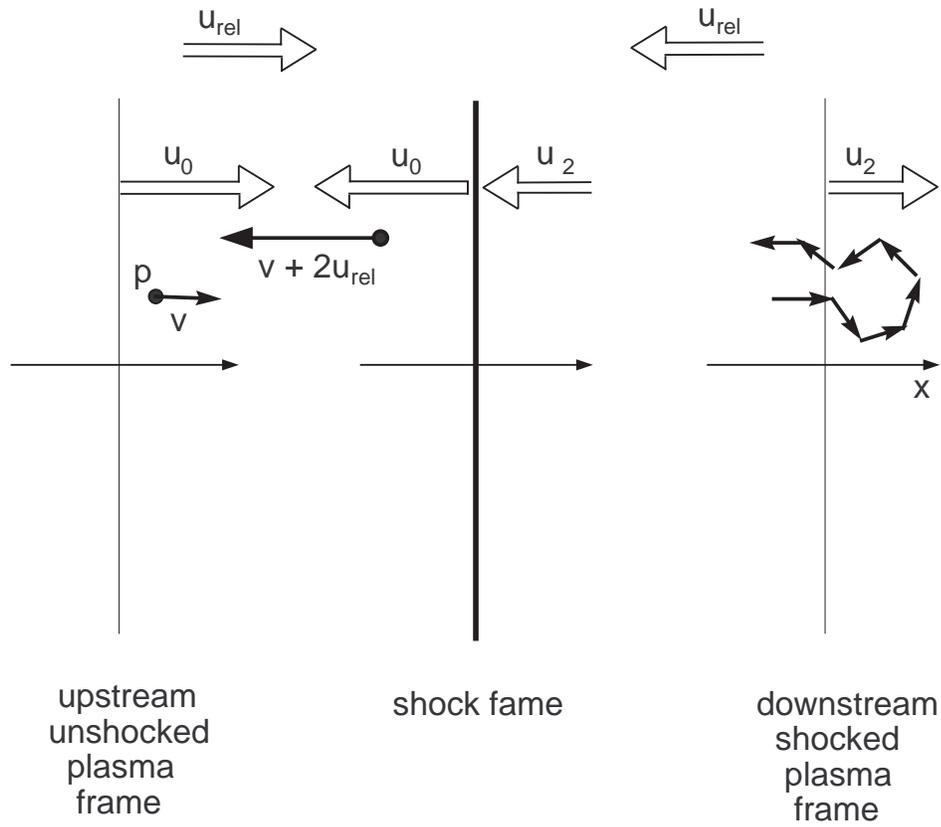}
\caption{Basic conceptional diagram of a plane shock moving through space
with a particle scattering elastically in the downstream frame and gaining
energy.
For nonrelativistic shocks, the upstream and downstream frames converge
with relative velocity $u_{\mathrm{rel}} = u_0 - u_2$.
\label{fig:basic}
}
\end{figure}

The most basic concept of nonrelativistic
first order Fermi acceleration is shown below.
A strong shock wave provides
the mechanism by which particles are scattered in pitch angle 
by magnetic wave turbulence on each side of the shock, as shown in
Figure \ref{fig:basic}.
Three frames of reference are shown. The upstream plasma frame sees the
shock frame moving toward it with velocity $-u_0$ and the downstream
frame moving toward it with relative velocity 
$u_{\mathrm{rel}} = u_2 -u_0$.
The shock frame sees the upstream frame as a flow moving toward it with
a velocity $+u_0$ and the downstream moving away with a velocity $+u_2$
The downstream frame sees the upstream frame moving toward it with
relative velocity 
$u_{\mathrm{rel}} = u_0 - u_2$ (i.e., the upstream and downstream
frames are converging at a speed $u_{\mathrm{rel}}$) 
and the downstream frame sees the shock moving away and to the left
with a velocity $-u_2$. A charged particle, for example a proton, 
initially at rest in the upstream frame crosses the shock 
from upstream to downstream because the shock is moving toward it. 
In the downstream frame, the particle has velocity 
$+u_{\mathrm{rel}}$. Suppose the
particle elastically scatters off of magnetic turbulence in the downstream
frame as shown in Figure \ref{fig:basic} and ends up moving to the left
with velocity $-u_{\mathrm{rel}}$. 
As the particle moves back across the shock, the
upstream frame sees the particle, which was originally at rest, now with
a velocity of $-2u_{\mathrm{rel}}$ 
(because the particle had a final velocity of
$-u_{\mathrm{rel}}$ 
in the downstream frame and the downstream frame is moving
toward the upstream frame with a velocity of $-u_{\mathrm{rel}}$. 
Hence, the particle
gained in kinetic energy by making a complete passage across the shock
and back into the upstream frame.
The same argument applies to a particle originally downstream;
the particle will also gain the same amount of kinetic energy. By
making numerous passages back and forth across the shock, a particle
can gain a large amount of kinetic energy, easily becoming relativistic.
Obviously, most particles are at some angle with respect to the shock
normal, therefore only the $x$ component 
of the particle's velocity contributes
to the particle's gain in energy. Since the process is completely random,
a distribution of particle energies will result that is independent of
the original state of the upstream conditions (\citep{Drury83}. 
In fact, the distribution
will result in a power law whose slope depends only on the compression
ratio, $r = u_0/u_2$, between the unshocked, upstream plasma
and the shocked downstream plasma, the details of which are discussed at
length in the papers cited above.

The main points of the derivation of the power law for accelerated particles 
are described in a review by \citet{Drury83}.
Suppose a particle has momentum $\vec{p}$, velocity $\vec{v}$ and 
pitch $\mu = \cos(\phi)$ between the momentum vector and the $x$ axis; 
then the average flux weighted change in a particle's
momentum with respect to the local plasma frame when it crosses the shock is
\begin{equation} 
\label{bell01}
\langle \triangle p\rangle = p\int\limits_{0}^{1}[\mu(u_0 - u_2)/v]2\mu d\mu
                           = \frac{2}{3}p(u_0 - u_2)/v
\end{equation}
Therefore, $p_1 \approx p_0[1 + \langle \triangle p\rangle]$, and after $N$
shock crossings (or $N/2$ returns from downstream), the average momentum 
will be
\begin{equation}
\label{bell05}
\langle p_N\rangle = \prod^N_{i = 1}[1 + \frac{2}{3}(u_0 - u_2)/v_i]p_0
\approx \prod^{N/2}_{i = 1}[1 + \frac{4}{3}(u_0 - u_2)/v_i]p_0
\end{equation}
Then
\begin{equation}
\label{bell10}
\ln\left[ \frac{p_N}{p_0}\right] = \frac{4}{3}(u_0 - u_2)\sum^{N/2}_{i=1}
\frac{1}{v_i}
\end{equation}
From the {\it{downstream}} frame, the flux of particles passing through
the shock from downstream to upstream is
\begin{equation} 
\label{bell15}
\left|n\int\limits_{-v}^{-u_2} (u_2 + v_x)dv_x\right| = 
\frac{n}{2}(u_2 - v)^2
\end{equation}
and the flux of particles passing through the shock from upstream to
downstream is 
\begin{equation}
\label{bell20}
n\int\limits_{-u_2}^{v} (u_2 + v_x)dv_x = \frac{n}{2}(u_2 + v)^2
\end{equation}
The probability of return from downstream to upstream is the
ratio of the two fluxes, or
\begin{equation}
\label{bell25}
P_R = \left(\frac{u_2 - v}{u_2 + v}\right)^2
\end{equation}
The probability that a particle has returned to the shock at least $N/2$
times is
\begin{equation}
\label{bell30}
P_R(N) = \left[\prod^{N/2}_{i=1}\left
(\frac{u_2 - v_i}{u_2 + v_i}\right)\right]^2
\end{equation}
Next, equate summations from the probability equation above and
the previous momentum equation:
\begin{equation}
\label{bell35}
\ln\left[P_R(N)\right] = -\frac{3u_2}{u_0 - u_2}
\ln\left(\frac{p}{p_0}\right)
\end{equation}
This gives the probability that a particle will reach at least 
momentum $p$:
\begin{equation}
\label{bell40}
P_R(p) = \left(\frac{p}{p_0}\right)^{-3u2/(u_0 - u_2)}
\end{equation}
The number density of particles accelerated to momentum p is the product
of the initial number density and the probability function:
\begin{equation}
\label{bell45}
n(p) = P_R(N)n(p_0) = 
n_0\frac{u_0}{u_2}\left(\frac{p}{p_0}\right)^{-3u_2/(u_0 - u_2)}
\end{equation}
Finally, the distribution function is 
\label{bell47}
\begin{equation}
\label{bell50}
f(p) = \frac{1}{4\pi p^2}\frac{\partial n}{\partial p} =
\frac{n_o}{4\pi}\left(\frac{3r}{r - 1}\right)
\left(\frac{p}{p_0}\right)^{-3r/(r-1)}
\end{equation}
Therefore, with the assumptions of test particles (which leave the plane
shock structure unmodified), particle velocities far greater than the 
nonrelativistic shock velocity and an 
isotropic distribution of particle momenta, the resulting
momentum distribution spectrum depends only on the compression ratio $r$,
i.e.,
\begin{equation}
\label{eq:fp}
f(p) \propto p^{-\sigma}\qquad\mathrm{where}\qquad\sigma = \bfrac{3r}{r-1} 
\end{equation}

Although the derivation of the functional relationship for the
momentum distribution with its accompanying assumptions were focused 
on nonrelativistic shocks, the same concepts of
particle acceleration can be applied to relativistic shocks, provided that
the new properties which apply to relativistic shocks are taken into
account. For example, particle velocities are close to those of the
shock velocity (i.e., the speed of light), and this leads to different
characteristics and criteria for the particles interacting with the shock.
The probability of return equation (\ref{bell25}) is true in general,
even at relativistic speeds, provided the downstream
momentum distribution is isotropic, and it should always be found to be
isotropic at least one or two mean free paths downstream from the shock. 
However, the momentum distribution at the shock is highly anisotropic.
Particle acceleration by relativistic shocks
will be addressed in a later chapter.
%

\section{Conservation laws and jump conditions}

Particles must obey conservation of momentum, energy and particle
continuity across the shock,
as well as obey Maxwell's equations and the thermodynamic equation of state.
The nonrelativistic jump conditions are well known;
for example, in \citet{EBJ96}, the jump conditions for nonrelativistic 
shocks with magnetic fields at oblique angles are given in full, and
implicitly incorporate the adiabatic equation of state.
The jump conditions
relate the upstream unshocked plasma to the downstream shocked plasma.
Hence, for particles scattering back and forth across the shock, the
jump conditions must be satisfied at each crossing and provide a 
self-consistent solution. 
This topic will be generalized for relativistic shocks and will be
thoroughly discussed in the next two chapters.
%

\section{The Monte Carlo approach}

Particle acceleration by shocks is inherently complicated and leads to
nonlinear differential equations. For example, the compression ratio
depends on the adiabatic index, which is affected by the same shock speed
that determines the compression ratio \citep{ER91}.
Except for special cases, the macroscopic equations that describe the
momentum distributions and jump conditions cannot be solved analytically.

Suppose particle acceleration is viewed microscopically and the individual
particles are allowed to scatter kinematically where momentum and energy
can be specifically tracked as the particle interacts with the shock
environment. This process lends itself nicely to computer techniques,
specifically to a Monte Carlo technique developed by \citet{E81} over
the last twenty years and is now a sophisticated tool for analyzing
both \nonrel\ and \rel\ shocks.

\begin{figure}[!hbtp]             
\dopicture{.7}{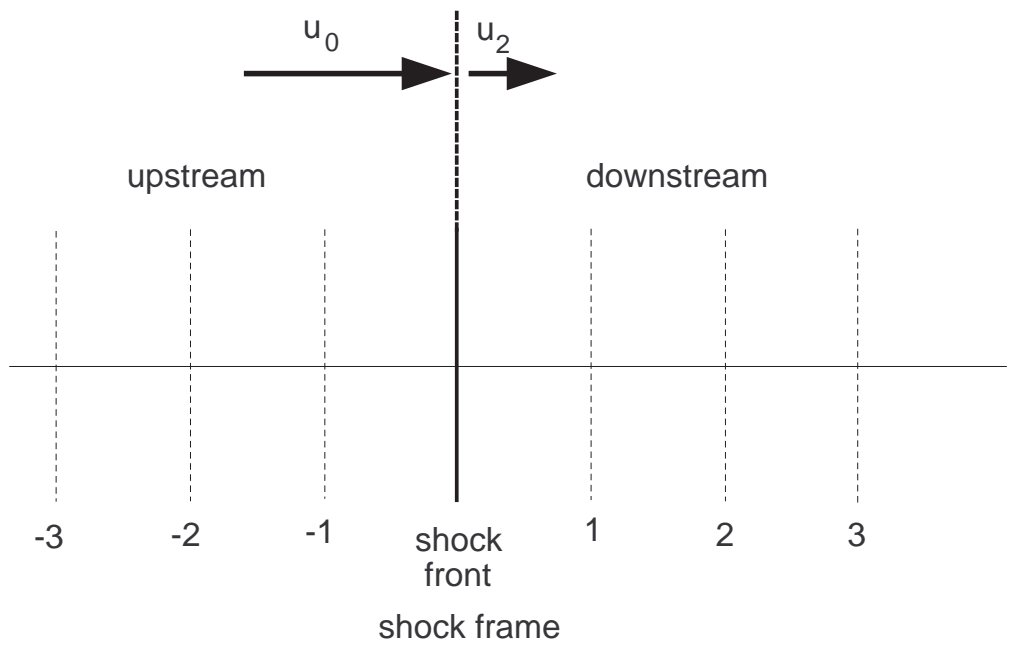}         
\caption{The shock environment showing the plasma regions divided up
into grid zones for flux calculations by the Monte Carlo model.
\label{fig:grid}
}
\end{figure}

The technique is briefly described here and will be discussed in more detail
in a later chapter. The shock region is divided into a number of grid
zones, many more than the few
shown in Figure \ref{fig:grid}. The model allows a number of 
particles to be injected into the shock environment far upstream,
which simulates 
unshocked particles in the interstellar medium being subjected to a
shock front. As the particles pass through the grid zones, the momentum
and energy of the particles are tabulated as corresponding ``fluxes''.
In the case of so-called test particles, which do not modify the shock
velocity profile, acceleration is turned off in the model, a 
compression ratio $r = u_0 / u_2$ is chosen and the fluxes are
tabulated at all the grid zones, both upstream and downstream. After
the particles have completely passed through the shock and if the momentum
and energy fluxes vary (i.e., are not conserved) across the
shock, then the process is repeated with a different compression ratio
until the fluxes appear to show continuity and remain at the same value
upstream to downstream.

The Monte Carlo model can also modify the shock velocity profile by 
allowing the backpressure of accelerated particles to affect the 
shock in a self-consistent manner. Again, after particles are injected,
the Monte Carlo model measures the momentum and energy flux at each
grid zone, compares the fluxes to those far upstream and estimates a
new shock velocity at each grid zone that may better conserve momentum
and energy flux. After a series of iterations, a final shock velocity
profile will result that self-consistently conserves momentum and
energy flux in the presence of accelerated particles, provided the
correct compression ratio was chosen. If the fluxes do not balance across
the shock, a new compression ratio is chosen and the process is repeated.
Another feature of the model allows the possibility of accelerated
particles to escape from the shock, carrying away momentum and energy
flux, as discussed in \citet{BE99}. These ideas will be discussed more
fully in later chapters.
%

\section{Overview and Objectives}

There are two separate areas of research activity for this dissertation,
and they are related in their common objective to understand and explain
the observed radiation in gamma-ray bursts.
 
The first research activity involves the ongoing development 
of a relativistic nonlinear Monte Carlo model, 
based on an earlier nonrelativistic version
\citep{EBJ96}, that will simulate the acceleration of 
charged particles by relativistic modified shocks 
with oblique magnetic fields when it is completed. It will be shown that
in its current state of development, the model very satisfactorily 
simulates the collisionless scattering and acceleration of 
charged particles in relativistic modified shocks with the
magnetic field parallel to the normal of the plane of the shock.
The development of the model, its operation, and the parameter
characteristics for unmodified shocks when shock speed, magnetic field
angle and pressure anisotropy are varied, are discussed in 
Chapters $2$ through $5$.
%
Chapter $2$ explains how and why the stress-energy tensor is used to handle
momentum and energy fluxes. The fluid and electromagnetic stress-energy
tensors are defined and a new equation of state is developed.
The new relativistic magnetohydrodynamic flux
relations for momentum and energy are derived in Chapter $3$,
along with the new jump conditions across the shock and a new method of
estimating the adiabatic index at all shock speeds. 
Also included in Chapter $3$ is a discussion of 
some of the compression ratio and magnetic field characteristics 
for unmodified shocks when the shock speed,
magnetic field angle and pressure anisotropy are varied. 
Chapter $4$ explains the new
relativistic momentum transformations between reference frames
and how modification of the shock velocity profile by energetic particles
occurs. Chapter $5$ describes the Monte Carlo technique for simulating the
kinematic scattering and acceleration of particles across a shock front
moving at relativistic speeds.
An application to parallel, modified shocks is described and the resulting
characteristics of the particle distributions, from nonrelativistic to highly
relativistic speeds, are discussed. 

The second area of research activity utilizes the relativistic shock
model to study lepton and baryon acceleration by relativistic modified
shocks with the magnetic field parallel to the shock normal. Chapter $6$
provides details of this study where the sensitivity
of lepton injection and acceleration to subshock size and shock speed is 
explored to find the maximum possible energy efficiency of leptons. 
New results are presented, given the assumptions in this model and if 
particles are accelerated by diffusive shock acceleration, that suggest 
leptons can never carry enough energy 
to explain the observed gamma-ray burst spectra when the
lepton to baryon number density is of the same order. Also,
the lepton to baryon particle number density ratio is investigated to find
the conditions under which leptons can carry the energy 
needed to produce the radiation observed in gamma-ray bursts.
A new result is shown where the equipartition of energy between leptons and
baryons is achieved in a relativistic shock when the lepton to baryon
number density ratio is approximately $3 \times 10^5$, independent of shock
speed, at least in the range of shock Lorentz factors of $2$ to $12$ used
in this study.
The results are discussed in the Conclusions in Chapter $7$, along with
a summary of all the accomplishments in the two research areas that
include the development of the relativistic Monte Carlo model and the
lepton-baryon acceleration study.  

%

\chapter{The Stress-energy Tensor}

The stress-energy tensor describes the condition of a medium at any 
point in spacetime. One advantage to using a tensor formulation is that
it is a very condensed and convenient way to express physical laws. A
second advantage, quoting \citet{T34}, ``..the
expression of a physical law by a tensor equation has exactly the same
form in {\it all} coordinate systems..'', due to the general 
transformation rules for tensors; i.e., 
\begin{equation}
\label{eq:Ttrans1}
T^{\mu\nu ...}_{\rho\sigma ...} = 0
\end{equation}
will be transformed into an equation of the same form
\begin{equation}
\label{eq:Ttrans2}
T^{'\mu\nu ...}_{\rho\sigma ...} = 0
\end{equation}
when the spacetime coordinates are transformed from 
$\{x^1, x^2, x^3, x^4\}$ to 
$\{x^{'1},x^{'2},x^{'3},x^{'4}\}$.

The stress-energy tensor is composed of two primary parts; the fluid
tensor and the electromagnetic tensor. Each tensor will be discussed
in the sections below.
Before discussing these tensors and the corresponding equation of state, 
some of their components require elaboration.
%

\section{The fluid tensor and the equation of state}
Aside from the first ($T^{00}$) component of the fluid
tensor which is the total energy density in the proper frame, 
the other {\it stress} components of the fluid tensor 
consist of momenta divided by a unit area. If the normal of
the unit area is in the direction of the momenta, the resulting component
is called {\it pressure}. If the normal of the unit area is perpendicular 
to the momenta, the resulting component is a {\it shear} stress.
  
Pressure is a macroscopic description of the particle momentum
in a given region of space. If the pressure is a scalar quantity, it
describes the average momentum or kinetic energy of an ideal
fluid in the absence of electromagnetic fields. If the fluid is an ensemble
of charged point particles (i.e., a plasma) in the presence of 
magnetic fields,  
the pressure may have different values in different directions. In this
case, pressure is described by a tensor. In either case, an adiabatic
compression or expansion of the region affects the pressure through an
adiabatic index. A plasma is dominated by magnetic field effects because
the particles move in such a way as to make electric fields negligible. 
Since a particle's energy does not ordinarily change by magnetic field 
deflections (assuming no radiation here),
the adiabatic approximation is, in general, a good assumption.

 
\subsection{The pressure tensor}
Consider a plasma with a magnetic field at some angle $\Theta_B$ with 
respect to the $x$ axis in the $xz$ plane.
The pressure tensor 
will be constrained to the gyrotropic case; i.e., the stress
can have one value parallel to
the magnetic field vector , but can have a different value perpendicular 
to the magnetic field vector (with rotational symmetry 
about the magnetic field vector)
as mentioned in equations (\ref{eq:P_para}) and (\ref{eq:P_perp}). Thus,
we have the space tensor
\begin{equation}
\label{ar:magPT}
\left\{\mathcal{P}^{ij}\right\}_m = 
\left( \begin{array}{ccc}
P_{\|} & 0 & 0\\
0 & P_{\perp} & 0\\
0 & 0 & P_{\perp}
\end{array} \right) 
\end{equation}
where subscript $m$ refers to the magnetic axis coordinate system.
Therefore, the pressure-stress tensor is diagonal
using the magnetic axis, but
rotation from the magnetic field coordinates to the {\it{xyz}} coordinates
about the $y$ axis
will produce a non-diagonal pressure-stress tensor as 
described by \citet{EBJ96}. Their 3-dimensional pressure-stress
tensor in the {\it{xyz}} plasma frame has components
\footnote{Note the small correction from the published reference.}

\begin{equation}
\label{ar:ROT13}
\left\{\mathcal{P}^{ij}\right\}_p = 
\left( \begin{array}{ccc}
 P_{\|}\cos^2{\Theta_{B}} + P_{\perp}\sin^2{\Theta_{B}} & 0 &
 (P_{\perp}-P_{\|})\sin{\Theta_{B}}\cos{\Theta_{B}}\\
 0 & P_{\perp} & 0 \\
 (P_{\perp}-P_{\|})\sin{\Theta_{B}}\cos{\Theta_{B}} & 0 & 
 P_{\perp}\cos^2{\Theta_{B}}+P_{\|}\sin^2{\Theta_{B}}
\end{array} \right) 
\end{equation}
corresponding to 
\begin{equation}
\label{ar:ROT113}
\left\{\mathcal{P}^{ij}\right\}_p = 
\left( \begin{array}{ccc}
 P_{xx} & 0 & P_{xz}\\
 0 & P_{yy} & 0 \\
 P_{zx} & 0 & P_{zz}
\end{array} \right) 
\end{equation}
where
\begin{equation}
\label{P21}
P_{\parallel} = P_{xx} - P_{xz}\bfrac{\sin{\Theta_B}}{\cos{\Theta_B}}
\end{equation}
and 
\begin{equation}
\label{P22}
P_{\perp} = P_{xx} + P_{xz}\bfrac{\cos{\Theta_B}}{\sin{\Theta_B}}
\end{equation}
The third equation completes the set:
\begin{equation}
\label{P25}
P_{zz} = P_{xx} + P_{xz}
\left(\bfrac{\cos^2{\Theta_B} -\sin^2{\Theta_B}}
{\sin{\Theta_B}\cos{\Theta_B}}\right) = P_{xx} + 2P_{xz}\cot{(2\Theta_B)}
\end{equation}


\subsection{The fluid stress-energy tensor}
The fluid stress-energy tensor in the proper frame is
defined with the following corresponding components:
\begin{equation}
\label{ar:FSET}
T^{\mu\nu}_{fluid} = 
\left( \begin{array}{cccc}
e & 0 & 0 & 0 \\[0.1cm]
0 & P_{xx} & 0 & P_{xz} \\[0.2cm]
0 & 0 & P_{yy} & 0 \\[0.2cm]
0 & P_{zx} & 0 & P_{zz}  
\end{array} \right) 
\end{equation}
using $P_{xx}$, $P_{zz}$ and $P_{xz} = P_{zx}$ from the previous section.

The $T^{00}$ component, $e$, is the total energy density in the proper 
frame or plasma frame. 
The other components
$P_{ij}$ are defined by \citet{T34} as the ``absolute stress'' components 
in the proper frame. $P_{ij}$ is the force parallel to the $i$-axis
exerted on a unit area normal to the $j$-axis. Hence, the diagonal components
can be considered a pressure, but the off-axis components are {\it{shear}}
stresses. The absolute stress components represent a different 
physical concept than the thermodynamic scalar pressure $P$.
Scalar pressure $P$ assumes a Maxwell-Boltzmann distribution and
is Lorentz invariant. The non-thermal components of the  
pressure-stress tensor $\mathcal{P}^{ij}$ embodied in the
fluid tensor above, in general,
transform significantly and pick up momentum flux components
in reference frames moving with respect to the proper frame.
It may be noted that at nonrelativistic speeds, 
the pressure-stress tensor is invariant under 
Galilean transformations because force and area are invariant.

A general adiabatic equation of state can be created from the conservation 
of energy density when oblique magnetic fields are present: 
\begin{equation}
\label{cons_nrgy}
e = \bfrac{Tr\{P_{ij}\}}{3(\Gamma - 1)} + \rho c^2
\end{equation}
where the trace of the pressure-stress tensor is 
$Tr\{\mathcal{P}^{ij}\} = P_{\|} + 2P_{\perp}$, $\Gamma$ is the adiabatic
index, and $\rho$ is the rest mass density. Using
equations (\ref{P21}) and (\ref{P22}) one can write
\begin{equation}
\label{nr_eos2}
\bfrac{Tr\{\mathcal{P}^{ij}\}}{3(\Gamma - 1)} = 
\bfrac{1}{3(\Gamma - 1)}(P_{\parallel} + 2P_{\perp}) =
\bfrac{1}{\Gamma - 1}\left[P_{xx} + \bfrac{P_{xz}}{3}
\left(2\cot{\Theta} - \tan{\Theta}\right)\right]
\ .
\end{equation}
In terms of magnetic field, where
$\cot\Theta_B = \bfrac{B_x}{B_z}$ and 
$\tan\Theta_B = \bfrac{B_z}{B_x}$, this becomes
\begin{equation}
\label{nr_eos3}
\bfrac{1}{3(\Gamma - 1)}(P_{\parallel} + 2P_{\perp}) =
\bfrac{1}{\Gamma - 1}\left[P_{xx} + \bfrac{P_{xz}}{3}
\left(2\bfrac{B_x}{B_z} - \bfrac{B_z}{B_x}\right)\right]
\ .
\end{equation}
Hence, the adiabatic equation of state, valid for both relativistic and
nonrelativistic shocks, is 
\begin{equation}
\label{obl_eos}
e = 
\bfrac{1}{\Gamma - 1}
\left[P_{xx} + 
\bfrac{P_{xz}}{3}
\left(2\bfrac{B_x}{B_z} - \bfrac{B_z}{B_x}\right)\right]
+ \rho c^2
\ .
\end{equation}


\subsection{Scalar pressure}
Scalar pressure is isotropic and Lorentz invariant. It is based on
either the ideal gas law $P = nkT$ where $k$ is Boltzmann's constant, 
or it is calculated by averaging the squares of the individual particle
momenta. When scalar pressure is
presented as a three dimensional tensor, it is diagonal with
equal components in all reference frames. The simplified equation of state, 
discussed in Appendix B, is stated again here:
\begin{equation}
\label{adiabat_eos}
e = \bfrac{P}{\Gamma - 1} + \rho c^2
\ .
\end{equation}
The fluid tensor in the plasma frame becomes
\begin{equation}
\label{simple_FT}
T^{\mu\nu}_{fluid} = 
\left(\begin{array}{cccc}
e & 0 & 0 & 0 \\
0 & P & 0 & 0 \\
0 & 0 & P & 0 \\
0 & 0 & 0 & P
\end{array}\right).
\end{equation}


\section{The Electromagnetic Stress-energy Tensor}
The electromagnetic field tensor composed by \citet{LL62} or the 
equivalent field-strength tensor of
\citet{Jksn75}, each effectively using the Lorentz gauge, 
can be stated as
\begin{equation}
\label{ar:EMF}
F^{\mu\nu} = 
\left(\begin{array}{cccc}
0 & -E_x & -E_y & -E_z \\
E_x & 0 & -B_z & B_y \\
E_y & B_z & 0 & -B_x \\
E_z & -B_y & B_x & 0
\end{array}\right)
\end{equation}
and it's dual,
\begin{equation}
\label{ar:EMFD}
\mathcal{F}^{\mu\nu} = 
\left(\begin{array}{cccc}
0 & -B_x & -B_y & -B_z \\
B_x & 0 & E_z & -E_y \\
B_y & -E_z & 0 & E_x \\
B_z & E_y & -E_x & 0
\end{array}\right)
\end{equation}
and from these field-strength tensors, 
the covariant form of the inhomogeneous
Maxwell equations may be written as
\begin{equation}
\label{eq:Max12}
F_{,\mu}^{\mu\nu} = \bfrac{4\pi}{c}J^{\nu}
\end{equation}
and the homogeneous equations may be written as
\begin{equation}
\label{eq:Max34}
\mathcal{F}_{,\mu}^{\mu\nu} = 0
\end{equation}
accompanied by the continuity equation
\begin{equation}
\label{eq:Cont}
J_{,\mu}^{\nu} = 0
\end{equation}
where $J^{\nu} = (c\rho_q,\vec{J})$ is the current four-vector, 
$\vec{J}$ is the current density three-space vector,
and $\rho_q$ is the electric charge density.

However, what is actually required is the electromagnetic energy
momentum tensor or electromagnetic stress-energy
tensor. This tensor can be constructed from the field tensors above as,
for example, \cite{AC88} did, but
referring to \citet{T34}, the electromagnetic stress-energy tensor may be
written directly as
\begin{equation}
\label{ar:EMSET}
T^{\mu\nu}_{\mathrm{EM}} = 
\bfrac{1}{4\pi}\left( \begin{array}{cccc}
\bfrac{E^2+B^2}{2} & \left[\vec{E}\times\vec{B}\right]_x & 
\left[\vec{E}\times\vec{B}\right]_y & 
\left[\vec{E}\times\vec{B}\right]_z \\[0.2cm]
\left[\vec{E}\times\vec{B}\right]_x & Q_{xx} & Q_{xy} & Q_{xz} \\[0.3cm]
\left[\vec{E}\times\vec{B}\right]_y & Q_{yx} & Q_{yy} & Q_{yz} \\[0.3cm]
\left[\vec{E}\times\vec{B}\right]_z & Q_{zx} & Q_{zy} & Q_{zz}  
\end{array} \right) 
\end{equation}

\noindent
where the {\it{Q}}'s are the Maxwell stresses defined as:
\begin{equation}
\label{the_Qs}
Q_{ii} = \bfrac{E^2 + B^2}{2} - E^2_i - B^2_i\quad \mathrm{and}\quad 
Q_{ij} = -(E_iE_j + B_iB_j)
\end{equation}
The overall electric field in a plasma is negligible and the plasma is
dominated by magnetic fields. In addition, the $xyz$ coordinate system is
oriented such that the magnetic field lies in the $xz$ plane; 
hence, the electromagnetic stress-energy
tensor can be simplified to

\begin{equation}
\label{ar:BSET}
T^{\mu\nu}_{\mathrm{EM}} = 
\bfrac{1}{4\pi}\left( \begin{array}{cccc}
\bfrac{B^2}{2} & 0 & 0 & 0 \\ 
0 & \bfrac{B^2_z - B^2_x}{2} & 0 & -B_xB_z \\
0 & 0 & \bfrac{B^2}{2} & 0 \\
0 & -B_zB_x & 0 & \bfrac{B^2_x - B^2_z}{2} 
\end{array} \right) 
\ .
\end{equation}
%

\section{Summary}
The concept of the stress-energy tensor was presented, and why and how it
is used to handle momentum and energy fluxes, both fluid and 
electromagnetic, in the Monte Carlo relativistic shock model. 
The stress-energy tensor was shown to consist of two parts. The first 
part contains total fluid energy and a gyrotropic pressure tensor. 
The second part contains magnetic fields,
with the assumption that electric
fields, over the large scale, are negligible in the space plasma.
From these tensors a new equation of state was established that is valid
for both nonrelativistic and relativistic shocks. The equation of state
includes a gyrotropic pressure tensor and oblique magnetic fields.
%

\chapter{Relativistic Magnetohydrodynamic Jump Conditions}

Relativistic shock jump conditions have been presented in a
variety of ways over the years.  The standard technique for deriving
the equations is to set the divergence of the stress-energy tensor
equal to zero on a thin volume enclosing the shock plane and use
Gauss's theorem to generate the jump conditions across the shock.  For
example, \citet{Taub48} developed the relativistic form of the
Rankine-Hugoniot relations, using the stress-energy tensor with
velocity expressed in terms of the Maxwell-Boltzmann distribution
function for a simple gas.  \citet{dHT50} presented a relativistic MHD
treatment of shocks in various orientations and a treatment of oblique
shocks for the nonrelativistic case, eliminating the electric field by
transforming to a frame where the flow velocity is parallel to the
magnetic field vector (now called the de Hoffmann-Teller frame).
\citet{Peacock81}, following \citet{LL59}, presented jump conditions
without electromagnetic fields, and \citet{BM76}, also using the
approach of \citet{LL59} and \citet{Taub48}, developed a concise set
of jump conditions for a simple gas using scalar pressure.
\citet{WZM87} provided a review of relativistic MHD shocks in
ideal, perfectly conducting plasmas, and in particular the treatment by
\citet{Lich67}, which used this approach to develop the relativistic
analog of Cabannes' shock polar \citep{Cab70} whose origins also lie
in \citet{LL59}.  
\citet{KW88} developed hydrodynamic equations using a
pressure tensor, and  
\citet{AC88} developed relativistic shock equations
for MHD jets using scalar pressure and magnetic fields with components
$B_z$ and $B_{\phi}$ (a parallel field with a twist).  \citet{BH91}
derived MHD jump conditions using the stress-energy tensor with
isotropic pressure and the Maxwell field tensor. By using a Lorentz
transformation to the de Hoffman-Teller frame, they restricted shock
speeds, $u_0$, to $u_0/c < \cos{\TBn}$, where $\TBn$ is the
angle between the shock normal
and the upstream magnetic field;
hence, this approach may only be used for mildly relativistic
applications.  
All of these approaches assumed that particles encountered by the
shock did not affect the shock velocity profile, i.e., shocked
particles were treated as test particles.

Here, previous work is extended 
by deriving a set of fully relativistic MHD jump
conditions with gyrotropic pressure and oblique magnetic fields.
The results are not restricted to the de Hoffmann-Teller frame and
apply for arbitrary shock speeds and arbitrary shock obliquities.
Solutions to the equations 
determine the downstream state of the gas
in terms of the upstream state for the special case of isotropic pressure
and, by parameterizing the pressure parallel and perpendicular to the
magnetic field, for gyrotropic pressure.
%
In this chapter analytic methods are used for determining
the fluid and electromagnetic characteristics of the shock and do not
include particle acceleration. Later work will combine these results
with Monte Carlo techniques \cite[\egc][]{EBJ96,ED2002} that will allow the
modeling of diffusive particle acceleration, including the
modification of the shock structure resulting from the back-reaction of
energetic particles on the upstream flow.
%

\section{Derivation of MHD jump conditions}
\subsection{Steady-state, plane shock}
\begin{figure}[!hbtp]              
\dopicture{.41}{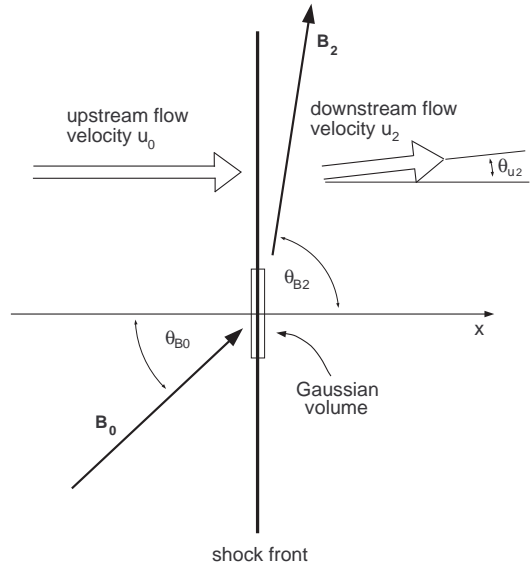} 
\caption{Schematic diagram of a plane shock, in the rest frame of the
shock, showing oblique
magnetic fields and a Gaussian volume over which the divergence of the
stress-energy tensor is integrated. In all of the examples the
upstream flow is parallel to the shock normal.
\label{fig:sk_box}
}
\end{figure}

Utilizing a Cartesian coordinate system with the $+x$-axis pointing to
the right, an infinite, steady-state, plane shock is shown in 
Figure \ref{fig:sk_box}
traveling to the left at a speed $u_0$ with it's velocity vector
parallel to the normal of the plane of the shock. In the rest frame of the
shock, the upstream fluid appears to be flowing to the right with a
speed of $u_0$. 

The upstream unshocked fluid
consists of a tenuous, plasma of baryons and leptons in
thermal equilibrium with 
$\mathcal{T}_{p0} = 
\mathcal{T}_{e0}$, where $\mathcal{T}_{p0}$ ($\mathcal{T}_{e0}$)
is the baryon (lepton) temperature.    
A uniform magnetic
field, $B$, makes an angle $\Theta_\mathrm{B0}$ 
with respect to the $x$-axis as seen from the upstream plasma frame.  
The field is kept weak enough to
insure high \alf\ Mach numbers and thus to insure that the magnetic
turbulence responsible for scattering the particles is frozen into the
plasma.  The $xyz$ coordinate system is oriented such that there are
only two components of magnetic field, $B_{x0}$ and $B_{z0}$, in the
upstream frame.  The field will remain co-planar in the downstream
frame and the downstream flow speed will be confined to the $x$-$z$
plane as well \cite[e.g.,][]{JE87}.

In the shock frame, the upstream flow is in the $+x$-direction and is
described by the normalized velocity four-vector $\beta^{\nu}_0 =
\gamma_0(1,\beta_0^x,0,0)$.  The downstream (i.e., shocked) flow
velocity four-vector is $\beta^{\nu}_2 = \gamma_2(1, \beta_2^x, 0,
\beta_2^z)$, where the subscript 0 (2) refers, here and elsewhere, to
upstream (downstream) quantities, $\beta = u/c$, and $\gamma = (1 -
\beta^2)^{-1/2}$ is the corresponding Lorentz factor associated with
the magnitude of the flow velocity, $u$.

The set of equations connecting the upstream and downstream regions of
a shock consist of the continuity of particle number flux (for
conserved particles), momentum and energy flux conservation, plus
electromagnetic boundary conditions at the shock interface, and the
equation of state.  The various parameters that define the state of
the plasma, such as pressure
and magnetic field, 
are
determined in the plasma frame and must be Lorentz transformed to the
shock frame where the jump conditions apply.  
In general, the six jump conditions plus the equation of state 
cannot be solved analytically because the adiabatic index (i.e., the
ratio of specific heats in the nonrelativistic  and ultrarelativistic
limits) is a
function of the downstream
plasma parameters, creating an inherently nonlinear problem
\citep[e.g.,][]{ER91}.  Even with the assumption of 
gyrotropic pressure, there are more unknowns than there are equations;
however, with additional assumptions, approximate 
analytic solutions may be obtained.


\subsection{Transformation Properties of the Stress-Energy Tensor}

Following \cite{T34}, equations (\ref{ar:FSET}) and (\ref{ar:BSET})
are combined to create the total stress-energy tensor,
$\Tmntot$, as the sum of fluid and electromagnetic parts, i.e.,
\begin{equation}
\label{SET3}
\Tmntot = \Tmnfluid + \TmnEM
\ .
\end{equation}
Then, by integrating over a thin volume containing the shock plane as
shown in Figure \ref{fig:sk_box}
and,
with Gauss' theorem, obtain the energy and momentum flux conditions
across the plane of the shock by using $T^{\mu\nu}n_{\nu} = 0$.
Thus, it can be seen that
$T^{0\nu}n_{\nu}$ yields the conservation of energy flux,
$T^{1\nu}n_{\nu}$ yields the $x$-contribution to momentum flux
conservation in the $x$-direction, $T^{3\nu}n_{\nu}$ yields the
$z$-contribution to momentum flux conservation in the $x$-direction,
and $n_{\nu} = (0,1,0,0)$ is the unit four-vector along the $x$-axis
in the reference frame of the shock. 
The Einstein summation convention is used here and throughout this thesis;
Repeated Greek indices are summed over four-space(time)  
and repeated English
indices are summed over three-space.  
 
The components of the fluid and electromagnetic tensors are defined in
the local plasma frame and are Lorentz transformed to the shock frame
where the flux conservation conditions apply, i.e.,
\begin{equation}
\label{ST4}
\left\{\Tmntot\right\}_s = 
\Lambda^{\mu}_{\alpha}
\left[\left\{T_{\mathrm{fluid}}^{\alpha\beta}\right\}_p
               +  \left\{T^{\alpha\beta}_{\mathrm{EM}}\right\}_p \right]  
\Lambda^{\nu}_{\beta}
\ ,
\end{equation}
where the subscript $p$ ($s$) refers to the plasma (shock) frame. 
Since the flow speeds in the model may have two space components, in
the $x$ and $z$-directions, the Lorentz transformation is
%
\begin{equation}
\label{LT0}
\Lambda^{\mu}_{\alpha} = 
\left( \begin{array}{cccc} 
\gamma & \gamma\beta_x & 0 & \gamma\beta_z\\[0.3cm]
\gamma\beta_x & 1 + \bfrac{\gamma-1}{\beta^2}\beta_x^2 & 
0 & \bfrac{\gamma-1}{\beta^2}\beta_x\beta_z\\[0.4cm]
0 & 0 & 1 & 0 \\[0.4cm]
\gamma\beta_z & \bfrac{\gamma-1}{\beta^2}\beta_z\beta_x & 
0 & 1 + \bfrac{\gamma-1}{\beta^2}\beta_z^2
\end{array}
\right) 
\end{equation}
where 
$\gamma = (1 - \beta^2)^{-\frac{1}{2}}$,  
$\beta^2 = \beta^i\beta_i$ 
is the square of the normalized flow speed
as seen from the shock frame, and $i$ is summed over the space
components.


\subsection{Flux Conservation Relations}
As discussed above, the energy and momentum conservation relations in
the shock frame can be derived by applying $T^{\mu\nu}n_{\nu} = 0$ to
equation (\ref{ST4}), individually on the Lorentz transformed fluid
and electromagnetic tensors.

The conservation of energy flux derives from
\begin{equation}
\label{TE1}
\left\{T^{0\nu}_{\mathrm{total}} \right\}_s n_{\nu} = 
\left\{T^{0\nu}_{\mathrm{fluid}} \right\}_s n_{\nu} +  
\left\{T^{0\nu}_{\mathrm{EM}} \right\}_s n_{\nu} = 0
\ .
\end{equation}
The fluid contribution to energy (scalar) flux conservation is 
\begin{eqnarray}
\label{PE42}
\FluxFluidEn =
\left\{T^{0\nu}_{\mathrm{fluid}} \right\}_s n_{\nu} = 
\gamma^2 \beta_x (e + P_{xx}) -
\gamma (\gamma-1) \bfrac{\beta_x \beta_z^2}{\beta^2}(P_{xx}-P_{zz}) + 
\hspace{2.5cm}\nonumber \\
\gamma \left[(2\gamma - 1)\beta_x^2 + \beta_z^2\right]
\bfrac{\beta_z}{\beta^2}P_{xz} 
\ , \hspace{1cm}
\end{eqnarray}
while the electromagnetic contribution is 
\begin{eqnarray}
\label{CE33}
\FluxEmEn = 
\left\{T^{0\nu}_{\mathrm{EM}}\right\}_s n_{\nu} = 
\bfrac{\gamma}{4\pi\beta^2}
\left[(\gamma -1)\beta_x\beta_z^2B_x^2 + 
(\gamma\beta_x^2 + \beta_z^2)\beta_xB_z^2 - \hspace{2.8cm}
\right . \nonumber \\
\left .
\{(2\gamma - 1)\beta_x^2 + \beta_z^2\}\beta_zB_xB_z\right]
\ . \hspace{0.5cm}
\end{eqnarray}

The conservation of momentum flux derives from
\begin{equation}
\label{TM1}
\left\{T^{i\nu}_{\mathrm{total}}\right\}_s n_{\nu} = 
\left\{T^{i\nu}_{\mathrm{fluid}}\right\}_s n_{\nu} +  
\left\{T^{i\nu}_{\mathrm{EM}}\right\}_sn_{\nu} = 0
\ .
\end{equation}
The $x$-component of the fluid tensor 
contributing to momentum flux is
\begin{eqnarray}
\label{PM44}
\FluxFluidPx =
\left\{T^{1\nu}_{\mathrm{fluid}}\right\}_s n_{\nu} = 
\gamma^2\beta_x^2(e + P_{xx}) + P_{xx} -
(\gamma-1)^2\bfrac{\beta_x^2\beta_z^2}{\beta^4}(P_{xx}- P_{zz}) +
\hspace{1cm} \nonumber \\
2(\gamma-1)(\gamma\beta_x^2 + \beta_z^2)
\bfrac{\beta_x\beta_z}{\beta^4}P_{xz} 
\ , \hspace{1cm}
\end{eqnarray}
and the $z$-component is
\begin{eqnarray}
\label{PM46}
\FluxFluidPz =
\left\{T^{3\nu}_{\mathrm{fluid}}\right\}_s n_{\nu} = 
\gamma^2\beta_x\beta_z(e + P_{xx}) - 
(\gamma-1)(\beta_x^2 + \gamma\beta_z^2)
\bfrac{\beta_x\beta_z}{\beta^4}(P_{xx} - P_{zz}) +
\qquad \qquad  \nonumber \\
\left[\gamma + 2(\gamma-1)^2\bfrac{\beta_x^2\beta_z^2}{\beta^4}\right]P_{xz} 
\ . \hspace{1cm}
\end{eqnarray}
The $x$-component of the electromagnetic tensor contributing to 
momentum flux is
\begin{eqnarray}
\label{CM32}
\FluxEmPx =
\left\{T^{1\nu}_{\mathrm{EM}}\right\}_s n_{\nu} = 
\bfrac{\gamma^2}{8\pi}\beta_x^2B^2 + 
\bfrac{1}{8\pi\beta^4}\left[(\gamma\beta_x^2 + \beta_z^2)^2 - (\gamma - 1)^2
\beta_x^2\beta_z^2\right](B_z^2 - B_x^2) - \hspace{1cm} 
\nonumber \\             
\bfrac{1}{2\pi}(\gamma-1)(\gamma\beta_x^2 + \beta_z^2)
\bfrac{\beta_x\beta_z}{\beta^4}B_xB_z 
\ ,  \hspace{0.8cm}
\end{eqnarray}
and the $z$-component is
\begin{eqnarray}
\label{CM34}
\FluxEmPz =
\left\{T^{3\nu}_{\mathrm{EM}}\right\}_sn_{\nu} = 
\bfrac{\gamma^2}{8\pi}\beta_x\beta_zB^2 +
\bfrac{1}{8\pi}(\gamma-1)^2(\beta_x^2 - \beta_z^2)
\bfrac{\beta_x\beta_z}{\beta^4}(B_z^2 - B_x^2) - \hspace{2cm}
\nonumber \\
\bfrac{1}{2\pi}(\gamma - 1)^2\bfrac{\beta_x^2\beta_z^2}{\beta^4}B_xB_z -
\bfrac{\gamma}{4\pi}B_xB_z
\ . \hspace{1cm}
\end{eqnarray}
In all cases where the \Alf\ Mach number is large,
the downstream flow velocity deviates only slightly from the shock
normal direction so $\beta_z \ll \beta_x$ as shown in the bottom frame of
Figure \ref{fig:jump_iso}. This allows a first-order
approximation in $\beta_z$ and the above equations become:
\begin{equation}
\label{PE43} 
\FluxFluidEn
\approx \gamma^2\beta_x(e + P_{xx}) + \gamma(2\gamma-1)\beta_zP_{xz}
\ ,
\end{equation}
\begin{equation}
\label{CE34}
\FluxEmEn
\approx 
\bfrac{\gamma^2}{4\pi}\beta_xB_z^2 - 
\bfrac{\gamma}{4\pi}(2\gamma-1)\beta_zB_xB_z
\ ,
\end{equation}
\begin{equation}
\label{PM45a}
\FluxFluidPx
\approx \gamma^2\beta_x^2(e + P_{xx}) + P_{xx} + 
2\gamma(\gamma-1)\bfrac{\beta_z}{\beta_x}P_{xz}
\ ,
\end{equation}
\begin{equation}
\label{CM32s}
\FluxEmPx
\approx \bfrac{\gamma^2}{8\pi}\beta_x^2B^2 +
\bfrac{\gamma^2}{8\pi}(B_z^2 - B_x^2) - 
\bfrac{\gamma(\gamma-1)}{2\pi}\bfrac{\beta_z}{\beta_x}B_xB_z
\ ,
\end{equation}
\begin{equation}
\label{PM47}
\FluxFluidPz
\approx 
\gamma^2\beta_x\beta_z(e + P_{xx}) + \gamma P_{xz} - 
(\gamma-1)\bfrac{\beta_z}{\beta_x}(P_{xx} - P_{zz}) 
\ ,
\end{equation}
and
\begin{equation}
\label{CM35}
\FluxEmPz
\approx 
\bfrac{\gamma^2}{8\pi}\beta_x\beta_zB^2 +
\bfrac{1}{8\pi}(\gamma-1)^2\bfrac{\beta_z}{\beta_x}(B_z^2 - B_x^2) - 
\bfrac{\gamma}{4\pi}B_xB_z
\ .
\end{equation}
As shown in the bottom frame of Figure \ref{fig:jump_aniso}, 
the approximation $\beta_z \ll \beta_x$ becomes
progressively better as the shock Lorentz factor increases.

When  the shock speed is ultrarelativistic, or when 
the magnetic field is parallel to the shock normal,
$\beta_z = 0$ and the above equations can be simplified further.

\noindent
For the ultrarelativistic case:
\begin{equation}
\label{PE43simp1} 
\FluxFluidEn = \gamma^2\beta_x(e + P_{xx})
\ ,
\end{equation}
\begin{equation}
\label{CE34simp1}
\FluxEmEn = \bfrac{\gamma^2}{4\pi}\beta_xB_z^2  
\ ,
\end{equation}
\begin{equation}
\label{PM44asimp}
\FluxFluidPx = \gamma^2\beta_x^2(e + P_{xx})  
\ ,
\end{equation}
\begin{equation}
\label{CM32ssimp1}
\FluxEmPx = \bfrac{\gamma^2}{4\pi}B_z^2
\end{equation}
\begin{equation}
\label{PM47simp}
\FluxFluidPz = \gamma P_{xz}
\ ,
\end{equation}
and
\begin{equation}
\label{CM35simp1}
\FluxEmPz = -\bfrac{\gamma}{4\pi}B_xB_z
\ .
\end{equation}
For parallel fields:
\begin{equation}
\label{PE43simp2} 
\FluxFluidEn = \gamma^2\beta_x(e + P_{xx})
\ ,
\end{equation}
\begin{equation}
\label{CE34simp2}
\FluxEmEn = 0
\ ,
\end{equation}
\begin{equation}
\label{PM45asimp}
\FluxFluidPx = \gamma^2\beta_x^2(e + P_{xx}) + P_{xx}
\ ,
\end{equation}
\begin{equation}
\label{CM32ssimp2}
\FluxEmPx = 0
\ ,
\end{equation}
\begin{equation}
\label{PM48simp}
\FluxFluidPz = 0
\ ,
\end{equation}
and
\begin{equation}
\label{CM35simp2}
\FluxEmPz = 0
\end{equation}\ .
%


\subsection{Jump conditions}
The jump conditions consist of the energy and momentum flux conservation
relations, the particle flux continuity, and the boundary conditions
on the magnetic field.
The conservation of particle number  
flux\footnote{Assuming  no pair creation nor annihilation, which is 
reasonable for a steady-state model.} 
is
\begin{equation}
\label{numdens}
\bigg[\gamma n\beta_x\bigg]^2_0 = 0
\ ,
\end{equation}
where the brackets are used as an abbreviation for
\begin{equation}
\gamma_2 n_2 \beta_{x2} -
\gamma_0 n_0 \beta_{x0} = 0
\ .
\end{equation}
This jump condition, as well as the ones that follow, are written in
the shock frame and, as always, the subscript 0 (2) refers to upstream
(downstream) quantities.  The remaining jump conditions are:
\begin{equation}
\label{jc_nrg}
\bigg[\FluxFluidEn  + \FluxEmEn \bigg]^2_0 = 0
\ , 
\end{equation}
\begin{equation}
\label{jcx_mom}
\bigg[ \FluxFluidPx + \FluxEmPx \bigg]^2_0 = 0
\ ,
\end{equation}
and
\begin{equation}
\label{jcz_mom}
\bigg[\FluxFluidPz + \FluxEmPz \bigg]^2_0 = 0
\ .
\end{equation}
Adding the boundary conditions on the magnetic field,
\begin{equation}
\label{s4Bnorm}
\bigg[B_x\bigg]^2_0 = 0
\end{equation}
and
\begin{equation}
\label{s4Etan2}
\bigg[\gamma(\beta_zB_x - \beta_xB_z)\bigg]^2_0 = 0
\ ,
\end{equation}
completes the set of six jump conditions.
%

\section{Solving the jump condition equations}

At this point there are eight unknown downstream quantities
($\beta_{x2}$, $\beta_{z2}$, $B_{x2}$, $B_{z2}$, $P_{xx2}$, $P_{xz2}$,
$e_2$, and $n_2$) and only six equations.\footnote{$P_{zz}$ can be
eliminated with equation~(\ref{P25}).} 
Assuming isotropic pressure replaces $P_{xx}$ and $P_{zz}$ with $P$,
sets $P_{xz} = 0$, and yields
\begin{equation}
\label{ST1}
T^{\mu\nu}_{\mathrm{total}} = (e + P)u^{\mu}u^{\nu} + Pg^{\mu\nu} + 
T^{\mu\nu}_{\mathrm{EM}}
\ ,
\end{equation}
where $g^{\mu\nu}$ is the Minkowski metric \citep[e.g.,][]{RL79}
\begin{equation}
\label{metric}
g^{\mu\nu} = \left( \begin{array}{cccc}
-1 & 0 & 0 & 0 \\ 0 & 1 & 0 & 0 \\ 0 & 0 & 1 & 0 \\ 0 & 0 & 0 & 1
\end{array} \right) 
\ .
\end{equation}
Unfortunately, this only removes one unknown and the equations are
still under constrained.  To proceed further, the equation
of state is utilized to find an approximate
expression for $\Gamma$.

Oblique shock jump conditions cannot, in general, be solved
analytically even for isotropic pressure because the downstream
adiabatic index, $\Gamma_2$, depends on the total downstream energy
density and the components of the pressure tensor (or scalar
pressure), which are not known before the solution is obtained. The
problem is inherently nonlinear except in the nonrelativistic and
ultrarelativistic limits where $\Gamma_2 = 5/3$ and $4/3$,
respectively.  Furthermore, the gyrotropic pressure components are
determined by the physics of the model and do not easily lend
themselves to analytic interpretation, although \citet{KW88} provided
equations based on a power law distribution in momentum for the
pressure tensor components in the special case of a parallel
relativistic shock with test particle diffusive shock acceleration.

An excellent approximation can be obtained if $v_{th} \ll u_0$, where
$v_{th}$ is the average thermal speed of the unshocked plasma. In this
case, after scattering in the downstream frame, all particles have 
\begin{equation}
\label{eq:gamrel}
\gamrel = (1 - \betarel^2)^{-1/2}
\ ,
\end{equation}
where
\begin{equation}
\betarel = \frac{v_{\mathrm{rel}}}{c} =
\frac{\beta_0 - \beta_2}{1-\beta_0\beta_2}
\end{equation}
is the relative $\beta$ between the converging plasma frames.
Using simple kinetic theory, it can be shown (see Appendix B) 
that the pressure 
\begin{equation}
\label{eq:KE_P}
P = \frac{n}{3}<\vec{p}\cdot\vec{v}>
\end{equation}
where $n$ is the particle number density, and $\vec{p}$ and $\vec{v}$
are the particle momentum and velocity, respectively.  Then, with the
approximation for particle velocity and the isotropic pressure 
version of the equation of
state (equation \ref{adiabat_eos}),
\begin{equation}
\Gamma_2 = 
\frac{P}{e - \rho c^2} + 1 =
\frac{(1/3) p_{\mathrm{rel}} v_{\mathrm{rel}}}{(\gamrel - 1)m c^2} + 1 =
\frac{\gamrel \betarel^2}{3 (\gamrel - 1)} + 1
\ ,
\end{equation}
or,
\begin{equation}
\label{SpecHeat}
\Gamma_2 = \frac{4 \gamrel + 1}{3 \gamrel}
\ .
\end{equation}
This approximation allows a direct numerical solution for isotropic
pressure, arbitrary obliquity, and arbitrary flow speed.  
It might be noted that equation (\ref{SpecHeat}) provides an upper limit
to the adiabatic index because any particles accelerated by the shock 
would tend to raise the average Lorentz factor of the particles and cause
the adiabatic index to decrease slightly.

\begin{figure}[!hbtp]              
\dopicture{.7}{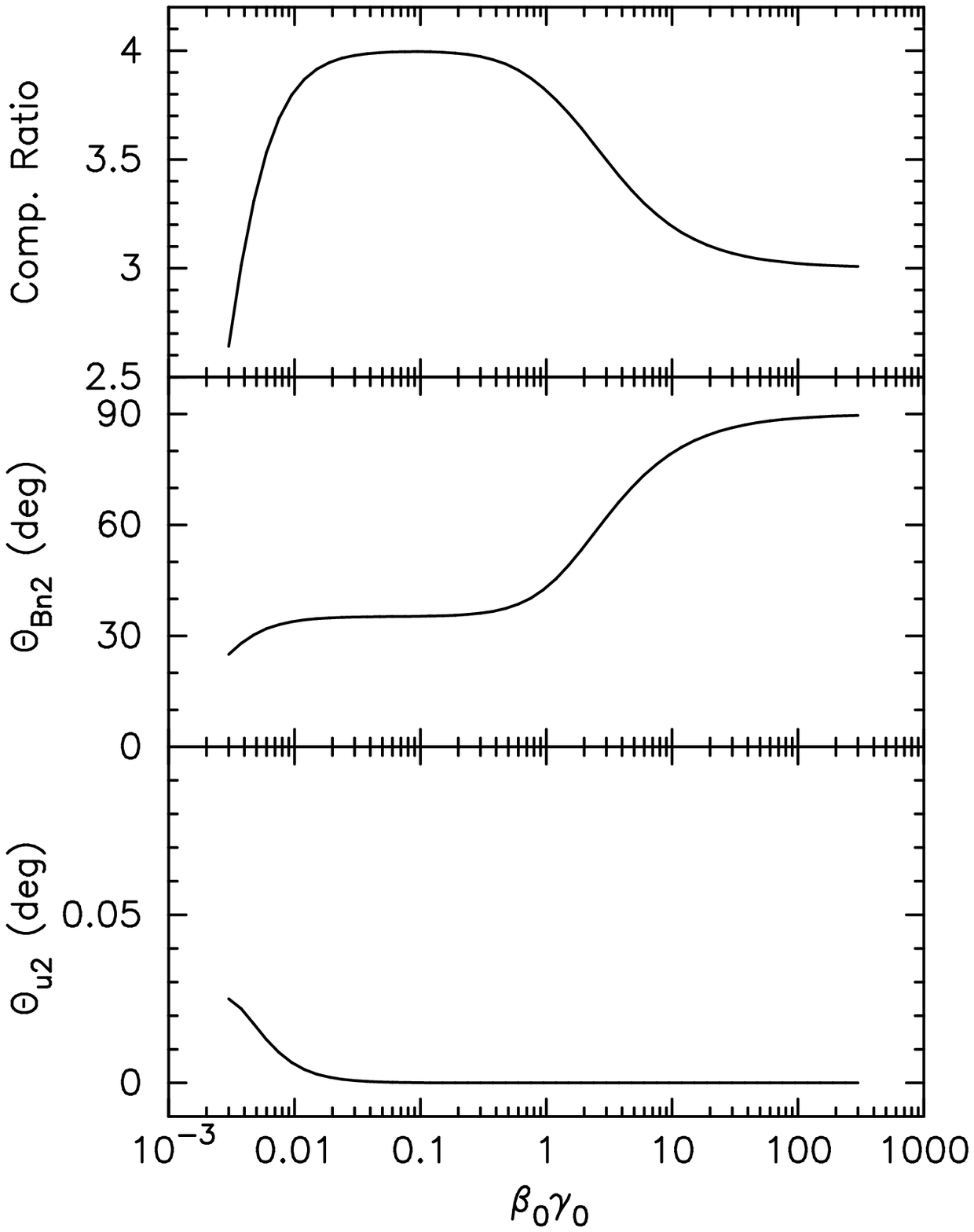} 
\caption{Compression ratio, $r$, $\TbnTwo$, $\TuTwo$, and
$\Gamma_2$ versus $\beta_0 \gamma_0$.  All examples have $n_0=0.1$
\pcc, 
$\mathcal{T}_0=10^6$ K (with equal electron and proton temperatures), and
$B_0=10$ \muG. The solid curves have $\TbnZ=0.1^\circ$, the dashed
curves have $\TbnZ=10^\circ$, and the dotted curves have
$\TbnZ=60^\circ$. In the bottom panel, the three cases are virtually
identical as they are in the top panel for 
 $\beta_0 \gamma_0 \gtrsim
0.02$. 
The solid dots in the bottom panel are values calculated using
the \mc\ simulation described in \citet{ED2002}.
\label{fig:jump_iso}
}
\end{figure}
In
Figure~\ref{fig:jump_iso}, the results are shown as a function of
$\beta_0 \gamma_0$ for a particular set of ambient parameters as
listed in the figure caption. 
The approximate equations~(\ref{PE43})-(\ref{CM35}) have been used,
although the exact equations can be solved if necessary. 

There are a number of important characteristics of the
solution. 
First, the solution goes smoothly from fully \nonrel\ to
\ultrarel\ shock speeds and obtains the canonical values for the
compression ratio $r \equiv \beta_{x0}/\beta_{x2} = 4$ for high Mach
number, \nonrel\ shocks, and $r=3$ for \ultrarel\ flows.  
Three cases were shown with different upstream magnetic field
obliquities: $\TbnZ= 0.1^\circ$ (solid curves), $\TbnZ= 10^\circ$
(dashed curves), and $\TbnZ= 60^\circ$ (dotted curves).  
In all cases, the downstream magnetic field angle shifts toward
$\TbnTwo=90^\circ$ as the shock Lorentz factor increases, indicating
the importance of treating oblique fields in highly \rel\ shock
acceleration.  
The value of $r$ is weakly dependent on $\TbnZ$ at \nonrel\ speeds,
and essentially independent of $\TbnZ$ (or other ambient parameters)
at \rel\ speeds.
The compression ratio is within 10\% of 3 for $\gamma_0 \gtrsim 2.1$ and
within 1\% of 3 for $\gamma_0 \gtrsim 7$.
For the ambient conditions used in this example, the angle the
downstream flow makes with the shock normal, $\TuTwo$, is small at all
$\beta_0\gamma_0$ (note the logarithmic scale for $\TuTwo$),
consistent with the assumption that $\beta_z \ll \beta_x$.

\begin{figure}[!hbtp]              
\dopicture{.65}{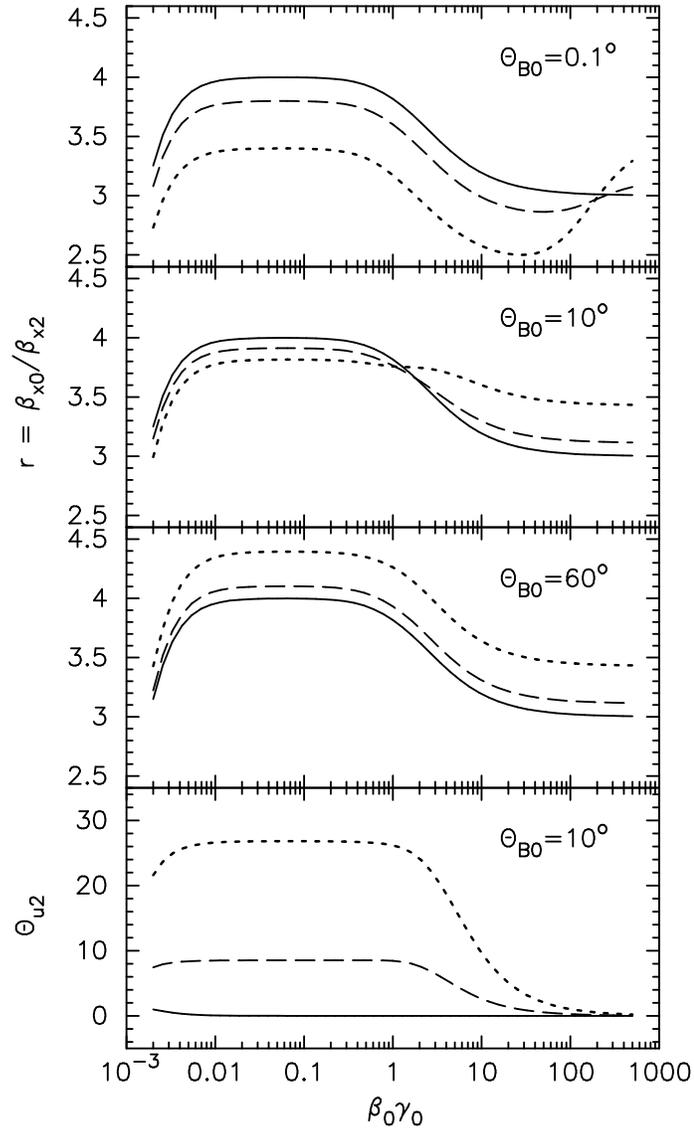} 
\caption{The top three panels show the compression ratio, $r$,
versus $\beta_0 \gamma_0$ for shocks with varying anisotropy,
$\alpha$, and obliquity, $\TBnZ$.  The bottom panel shows the angle
the downstream flow makes with the shock normal, $\TuTwo$, versus
$\beta_0 \gamma_0$. As in Figure~\ref{fig:jump_iso}, $n_0=0.1$ \pcc,
$\mathcal{T}_0=10^6$ K (with equal electron and proton temperatures),
and $B_0=10$ \muG.  The solid curves have $\alpha=1$, the dashed
curves have $\alpha=0.7$, and the dotted curves have $\alpha=1.3$.
\label{fig:jump_aniso}
}
\end{figure}

In the bottom panel of Figure~\ref{fig:jump_iso}
$\Gamma_2$ determined from equation~(\ref{SpecHeat}) is compared with a
calculation using the \mc\ simulation described in detail in
\citet{ED2002} (solid points). The \mc\ value for $\Gamma_2$ is
determined, without Fermi acceleration, directly from the shocked
particle distribution using the $r$ shown in the top panel. It is
essentially identical to that given by equation~(\ref{SpecHeat}) at
all Lorentz factors confirming the utility of this approximation.
As noted above, if Fermi acceleration does occur, $\Gamma_2$ will
approach $4/3$ at lower $\beta_0 \gamma_0$ due to the contribution of
energetic particles.

%
For gyrotropic pressure, an additional assumption is needed to obtain
an analytical solution and this assumption is taken to be
\begin{equation}
\label{PressConst}
\Pperp = \alpha \Ppar \ ,
\end{equation}
where $\alpha$ is an arbitrary parameter and $\alpha = 1$ gives
isotropic pressure.  Equation~(\ref{PressConst}) allows the illustration 
of the effects of anisotropic pressure but is not suggested as
a meaningful physical model.
 
Using equations (\ref{P21}), (\ref{P22} and (\ref{PressConst}) yields
\begin{equation}
P_{xx} =
\Ppar (\cos^2{\TBn} + \alpha \sin^2{\TBn})
\ ,
\end{equation}
\begin{equation}
P_{xz} =
\Ppar (\alpha - 1) \sin{\TBn} \cos{\TBn}
\ ,
\end{equation}
and the result is a closed set of equations for the jump conditions for
shocks with gyrotropic pressure, arbitrary obliquity, and arbitrary
flow speed. Results for various $\TbnZ$'s and $\alpha$'s are shown in
Figure~\ref{fig:jump_aniso} (the other ambient parameters are the same
as used for Figure~\ref{fig:jump_iso}). The solid curves (which are
identical to the curves in the top panel of Figure~\ref{fig:jump_iso})
have $\alpha=1$, the dashed curves have $\alpha=0.7$, and the dotted
curves have $\alpha=1.3$.
In all cases, the pressure in the unshocked, upstream plasma is taken to be
isotropic and $\alpha \ne 1$ is only applied downstream.

The effects of anisotropic pressure on the compression ratio come
about mainly through changes to $P_{xx}$ and this depends strongly on
the downstream obliquity, $\TBnTwo$. In the top panel of
Figure~\ref{fig:jump_aniso}, $\TBnZ=0.1^\circ$ and $\TBnTwo \sim
0.1^\circ$ for \nonrel\ and mildly \rel\ shock speeds. Therefore, at
these speeds $P_{xx} \sim \Ppar$ and since $\Ppar = \Pperp/\alpha$,
the fraction of downstream pressure in $P_{xx}$ is inversely
proportional to $\alpha$.  As shown in the Figure, $r$ is less than
the isotropic value, $\rIso$, for $\alpha= 0.7$ and greater than
$\rIso$ for $\alpha=1.3$.
As the shock becomes fully \rel, however, $\TBnTwo$ approaches
$90^\circ$ for any $\TBnZ>0$ (see the middle panel of
Figure~\ref{fig:jump_iso}) and $P_{xx} \sim \Pperp$.
When this is the case, the fraction of pressure in $P_{xx}$ will be
proportional to $\alpha$ and $r < \rIso$ for $\alpha >1$ and $r >
\rIso$ for $\alpha <1$.

The transition where  $r$ crosses $\rIso$ occurs at slower
shock speeds as $\TBnZ$ increases (middle panel of
Figure~\ref{fig:jump_aniso}) until 
$\TBnZ$ is large enough (bottom panel of Figure~\ref{fig:jump_aniso})
so no transition occurs.
The change in $\TBnTwo$ is relatively small for the examples shown in
Figure~\ref{fig:jump_aniso} but $\TuTwo$ changes significantly, as
shown in the bottom panel for $\TBnTwo = 10^\circ$. Despite the larger
$\TuTwo$ for $\alpha=0.7$, the approximation $\beta_z \ll \beta_x$
should still be valid.

%

\section{Correspondence with Nonrelativistic Jump Conditions}
The relativistic jump conditions [i.e., equations (\ref{numdens}) - 
(\ref{s4Etan2})] must correspond to the nonrelativistic jump conditions
when the shock speed drops into the nonrelativistic range. Here, the
pertinent relativistic equations are (\ref{PE42}) - (\ref{CM34}) and
not their approximations because at nonrelativistic shock speeds, the
$z$ component of the downstream flow is, in general, not negligible.
Therefore, to reduce the relativistic jump conditions
to their nonrelativistic counterparts, start by rewriting  equations 
(\ref{PE42}) - (\ref{CM34}) to second order in $\beta$; i.e.,
$\gamma \approx 1 + \bfrac{\beta^2}{2}$. Next, combine the resulting equations
with the equation of state (\ref{obl_eos}) and eliminate the energy density
$e$, then carefully watch the
terms as $\beta$ is reduced with proper algebra to the nonrelativistic range.
The resulting equations will match those published in \citet{EBJ96}, i.e.
\begin{equation}
\label{EBJ_px}
\left[\rho u_x^2 + P_{xx} + \bfrac{B_z^2}{8\pi}\right]^2_0 = 0
\end{equation}
\begin{equation}
\label{EBJ_pz}
\left[\rho u_xu_z + P_{xz} - \bfrac{B_xB_z}{4\pi}\right]^2_0 = 0
\end{equation}
and\footnote{Note a small typographical error in the energy flux equation
in the referenced paper is corrected here.}   
\begin{eqnarray}
\label{EBJ_nrg}
\left[\bfrac{\Gamma}{\Gamma - 1}P_{xx}u_x + 
P_{xz}\left\{u_z + \bfrac{u_x}{3
(\Gamma - 1)}\left(\bfrac{2B_x}{B_z} - \bfrac{B_z}{B_x}\right)\right\}+
\hspace{3cm}
\right.\nonumber\\
\left.
\bfrac{1}{2}\rho u_x^3 + \bfrac{1}{2}\rho u_xu_z^2 +
\bfrac{u_xB_z^2}{4\pi} - \bfrac{u_zB_xB_z}{4\pi} + Q_{esc}\right]^2_0 = 0
\end{eqnarray}
where $Q_{esc}$ is inserted to account for any energy flux lost due to
escaping particles at a free escape boundary (FEB). The remaining jump 
conditions are straightforward.


\section{Summary}
The new relativistic magnetohydrodynamic flux relations for 
momentum and energy were derived, along with the new shock jump 
conditions. 
Setting the divergence of the stress-energy tensor
equal to zero leads to the momentum and energy conservation laws across
the shock. The equations were written in full and also to first
order in small $\beta_z$ and $\beta_z = 0$ because $\beta_z$ is usually
small for relativistic shocks and $\beta_z = 0$ for all parallel shocks.

The continuity equation and electromagnetic boundary conditions,
the new equation of state derived in the previous chapter, and a newly
dervied approximation for the adiabatic index over the 
trans-relativistic range were combined with the conservation laws 
across the shock to create a set of general relativistic
jump conditions relating the upstream and downstream regions across the
plane of the shock. The equations were solved analytically for the case
of isotropic pressure. For the case of gyrotropic pressure, an additional
anisotropy parameter $\alpha$ was introduced to allow a solution.

Some initial results were shown for unmodified shocks to demonstrate how
the compression ratio, adiabatic index, downstream magnetic field angle
and flow angle vary with shock speed and upstream magnetic field. Also,
the gyrotropic pressure tensor anisotropy was varied to show the effects
on the compression ratio and field obliquity.

Finally, a correspondence was established between the relativistic
and nonrelativistic jump conditions. It was shown that all of the 
shock parameters in the relativistic shock model
vary smoothly over the entire range of shock speeds from
nonrelativistic to highly relativistic. This is a new feature, 
especially for modified shocks.
%

\chapter{Particle Acceleration at Relativistic Shocks}

\section{Comparisons between relativistic and nonrelativistic shocks}
Relativistic shocks, where the flow speed Lorentz factor 
$\gamma_0 = [1 - (u_0/c)^2]^{-1/2}$ is significantly greater than 1,
are likely to be much less
common than nonrelativistic shocks, but may occur in extreme objects
such as pulsar winds, hot spots in radio galaxies, and gamma-ray
bursts (GRBs).  Largely motivated by the application to GRBs,
relativistic shocks have received much attention by a
number of researchers \citep[e.g.,][]
{BedOstrow96,KGGA2000,SD00, AGKG2001,TMM01}.
However, except for some preliminary work done over a decade ago
\citep[][]{SK87,EllisonJapan91,EllisonPoland91} and, aside from
\citet{ED2002},
current descriptions
of relativistic shocks undergoing first-order Fermi acceleration are
test particle analytical approximations that do not include the 
back-reaction of the accelerated particles on the shock structure 
\citep[e.g.,][]{Peacock81,HD88,KW88}; however, \citet{KS87b} and
\citet{BH91} used 
\mc\ techniques to calculate their test particle results. 
This may be a serious limitation of \rel\ shock theory in
applications, such as GRBs, where high particle acceleration
efficiencies are often assumed and test particles are, by their very
definition, negligible. 

In collisionless shocks, charged particles gain energy by scattering
back and forth between the converging upstream and downstream plasmas.
This basic physical process of diffusive or first-order Fermi
shock acceleration, is the same in \rel\ and \nonrel\ shocks.
Differences in the mathematical description and outcome of the process
occur, however, because energetic particle distributions are nearly
isotropic in the shock reference frame in \nonrel\ shocks (where $v
\gg u_0$; $v$ is the individual particle speed), but are highly
anisotropic in \rel\ shocks (since $v \sim u_0 \sim c$)
\citep[e.g.,][]{Peacock81,KS87b}.

The most important results from the theory of test-particle
acceleration in \ultrarel\ shocks are:
(i) regardless of the state of the unshocked plasma, particles can
pick up large amounts of energy 
$\Delta E \sim \gamma_0^2$ in their
first shock crossing cycle \citep{Vietri95}, but will
receive much smaller
energy boosts ($\ave{E_f/E_i} \sim 2$) for subsequent crossing cycles
\citep[e.g.,][]{GA99,AGKG2001}\footnote{$E_i$ ($E_f$) is the particle
energy at the start (end) of an upstream to downstream to upstream (or
a downstream to upstream to downstream) shock crossing cycle.};
(ii) the shock compression ratio, defined as $r \equiv u_0 / u_2$,
tends to $3$ as $u_0 \rightarrow c$ \citep[e.g.,][]{Peacock81,Kirk88},
where $u_2$ is the flow speed of the shocked plasma measured in the
shock frame;\footnote{Note that the density ratio across the 
relativistic shock
$\rho_2/\rho_0 \ne u_0/u_2$,
in contrast with
\nonrel\ shocks, because
the Lorentz factors associated with the relativistic flows modify the
particle flux jump condition. Here and elsewhere the subscript 0 (2)
is used to indicate far upstream (downstream) values.}
 and
(iii) a so-called `universal' spectral index, $\sigma \sim 4.2-4.3$
(in equation~\ref{eq:TPpowerlaw}) exists in the limits of $\gamma_0 \gg
1$ and $\delta\theta \ll 1$, where $\delta\theta$ is the change in direction
a particle's momentum vector makes at each pitch angle scattering
\citep[e.g.,][]{BedOstrow98,AGKG2001}.

\citet{ED2002} found that these results are modified in 
mildly \rel\ shocks, even in the test-particle approximation, 
and in fully \rel\ shocks (at
least for 
$\gamma_0 \lesssim 10$) when the back-reaction of the
accelerated particles is treated self-consistently, which causes the
shock to smooth and the compression ratio to change from test-particle
values.
In mildly \rel\ shocks, $f(p)$ remains a power law in the \TP\
approximation but both $r$ and $\sigma$ depend on the shock Lorentz
factor, $\gamma_0$.  When efficient particle acceleration occurs in
mildly \rel\ shocks 
(i.e., $\gamma_0 \lesssim 3$), large increases in $r$ can
result and a power law is no longer a good approximation to the
spectral shape.  In these cases, the compression ratio is determined by
balancing the momentum and energy fluxes across the shock with the
\mc\ simulation.
For larger Lorentz factors, accelerated particles smooth the
shock structure just as they do in slower shocks,
%
but $r$ approaches 3 as $\gamma_0$ increases.
In general, efficient particle acceleration results in spectra very
different from the so-called `universal' power law found in the
test-particle approximation unless $\gamma_0 \gtrsim 10$.
%

\section{Relativistic momentum transformations}

Elastic scattering and the subsequent changes in the 
direction\footnote{The magnitude of momentum remains constant in
an elastic collision.} 
of a particle's momentum 
takes place in the plasma frame, but all particle positions
and momenta, and the corresponding jump conditions are handled
in the shock frame; hence, relativistic
frame transformations must be made from the shock frame to the plasma frame
and from the plasma frame to the shock frame.

\begin{figure}[!hbtp]             
\dopicture{.41}{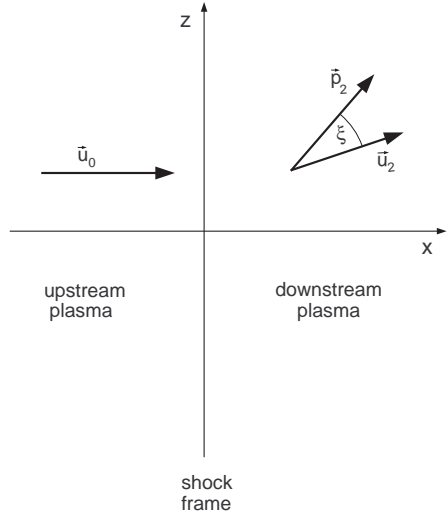}       
\caption{Schematic diagram showing the shock frame and a downstream
particle with momentum 
$\vec{p_2}$ in the downstream frame.
The downstream frame is moving with velocity 
$\vec{u_2}$ with respect to the shock frame.
\label{fig:u_p}
}
\end{figure}

Referring to Figure \ref{fig:u_p}, the downstream plasma frame is moving with
velocity $\vec{u}_2 = \hat{x}u_{x2} + \hat{z}u_{z2}$ with respect to the 
shock frame, and a downstream particle has 
momentum $\vec{p}_2 = \hat{x}p_{x2} + \hat{y}p_{y2} + \hat{z}p_{z2}$
in the downstream plasma frame, with the angle $\xi$ between these two
vectors, where
$\hat{x}$, $\hat{y}$ and $\hat{z}$ are unit vectors in the $x$, $y$ 
and $z$ directions. The particle momentum is defined as 
$\vec{p} = \gamma_vm\vec{v}$ where $v$ is the particle velocity,
$\gamma_v = \left[\left(\bfrac{p}{mc}\right)^2 + 1\right]^{\frac{1}{2}}$
is the corresponding Lorentz factor, 
$p^2 = p_{x}^2 + p_{y}^2 + p_{z}^2$, $m$ 
is the particle's rest mass, and the
Lorentz factor of the relative frame velocity, 
$\gamma_{u2} = \left(1 - \bfrac{u_2^2}{c^2}\right)^{-\frac{1}{2}}$.
 
Before the relativistic transformations can be performed, the
momentum components of the particle parallel and perpendicular to the
frame velocity vector $\vec{u}_2$ are required.
The parallel momentum component is simply 
the scalar product $\hat{u}\cdot\vec{p}$ where the frame velocity unit
vector is $\hat{u} = \hat{x}\bfrac{u_x}{u} + \hat{z}\bfrac{u_z}{u}$, and
it is understood that in these equations and the ones below where no
number subscripts are used, all quantities are downstream plasma
frame quantities.
The momentum vector parallel to $\vec{u}$ is
$\vec{p}_{\parallel} = \hat{u}(\hat{u}\cdot\vec{p})$ or
\begin{equation}
\label{eq:para_p}
\vec{p}_{\parallel} = 
\hat{x}\left[\left(\bfrac{u_{x}}{u}\right)^2 p_x + 
\left(\bfrac{u_x u_z}{u^2}\right)p_z\right] +
\hat{z}\left[\left(\bfrac{u_x u_z}{u^2}\right)p_x + 
\left(\bfrac{u_z}{u}\right)^2 p_{z}\right]\ .
\end{equation}
The component of particle momentum perpendicular to the frame velocty 
vector is $\vec{p}_{\perp} = \vec{p} - \vec{p}_{\parallel}$ or
\begin{eqnarray}
\label{eq:perp_p}
\vec{p}_{\perp} = \hat{x}\left[\left\{1 - 
\left(\bfrac{u_x}{u}\right)^2\right\}p_{x} -
\left(\bfrac{u_x u_z}{u^2}\right)p_{z}\right] +
\hat{y}p_y + \hspace{2.5cm} \nonumber\\
\hat{z}\left[\left\{1 - \left(\bfrac{u_{z}}{u}\right)^2\right\}p_{z} -
\left(\bfrac{u_x u_z}{u^2}\right)p_{x}\right]\ .
\end{eqnarray}
The relativistic momentum frame transformations from the downstream
frame to the shock frame are:
\begin{equation}
\label{eq:p_frm_par}
\vec{p}_{\parallel,{sk}} = 
\gamma_{u}\left[\vec{p}_{\parallel} + \bfrac{E\vec{u}}{c^2}\right] =
\gamma_{u}\left[\vec{p}_{\parallel} + \gamma_{v}m\vec{u}\right]
\end{equation}
and the invariant perpendicular component of momentum
\begin{equation}
\label{eq:p_frm_prp}
\vec{p}_{\perp,{sk}} = \vec{p}_{\perp}\ .
\end{equation}
This leads to 
\begin{equation}
\label{eq:p_sk}
\vec{p}_{sk} =  \vec{p}_{\parallel,{sk}} + \vec{p}_{\perp,sk}
\end{equation}
where
\begin{eqnarray}
\label{eq:p_par_sk}
\vec{p}_{\parallel,{sk}} =
\gamma_{u}\hat{x}\left[\left(\bfrac{u_{x}}{u}\right)^2p_{x} +
\left(\bfrac{u_{x}u_{z}}{u^2}\right)p_{z} + \gamma_vmu_x\right] + 
\hspace{5cm} 
\nonumber \\
\gamma_u\hat{z}\left[\left(\bfrac{u_{x}u_{z}}{u^2}\right)p_{x} + 
\left(\bfrac{u_{z}}{u}\right)^2p_{z} + \gamma_vmu_z\right]\ . 
\end{eqnarray}
The particle momentum, now seen from the shock frame, requires the
corresponding $x$, $y$ and $z$ components, where 
$p_{x,sk} = \hat{x}\cdot\vec{p}_{sk}$, 
$p_{y,sk} = \hat{y}\cdot\vec{p}_{sk}$, and
$p_{z,sk} = \hat{z}\cdot\vec{p}_{sk}$. The resulting components
are:
\begin{equation}
\label{eq:p_sk_x}
p_{x,{sk}} = \left[\left(\gamma_{u} - 
1\right)\left(\bfrac{u_{x}}{u}\right)^2 + 1\right]p_{x} +
\gamma_{u}\gamma_{v}mu_{x} +
\left(\gamma_{u} - 1\right)\left(\bfrac{u_{x}u_{z}}{u^2}\right)p_{z}
\end{equation}
and
\begin{equation}
\label{eq:p_sk_z}
p_{z,{sk}} = \left[\left(\gamma_{u}-
1\right)\left(\bfrac{u_{z}}{u}\right)^2 + 1\right]p_{z} +
\gamma_{u}\gamma_{v}mu_{z} +
\left(\gamma_{u} - 1\right)\left(\bfrac{u_{x}u_{z}}{u^2}\right)p_{x}\ .  
\end{equation}
The $y$ component of momentum
will remain invariant under the Lorentz transformations.

Finally, the momentum components are made dimensionless by dividing
through by $m_pu_0$, where $m_p$ is the proton rest mass and $u_0$ is
the shock flow speed:
\begin{equation}
\label{eq:pxsk}
p'_{x,sk} = \left[\left(\gamma_{u}-1\right)
\left(\bfrac{u_{x}}{u}\right)^2 + 1\right]
\left[\bfrac{p_{x}}{m_pu_0}\right] + 
\bfrac{\gamma_{u}\gamma_{v}mu_{x}}{m_pu_0} +
\left(\gamma_{u} - 1\right)
\left(\bfrac{u_{x}u_{z}}{u^2}\right)
\left[\bfrac{p_{z}}{m_pu_0}\right]
\end{equation}
\begin{equation}
\label{eq:pysk}
p'_{y,sk} = \bfrac{p_y}{m_pu_0}
\end{equation}
\begin{equation}
\label{eq:pzsk}
p'_{z,sk} = \left[\left(\gamma_{u}-1\right)
\left(\bfrac{u_{z}}{u}\right)^2 + 1\right]
\left[\bfrac{p_{z}}{m_pu_0}\right] +
\bfrac{\gamma_{u}\gamma_{v}mu_{z}}{m_pu_0} + 
\left(\gamma_{u} - 1\right)
\left(\bfrac{u_{x}u_{z}}{u^2}\right)
\left[\bfrac{p_{x}}{m_pu_0}\right]  
\end{equation}

The relativistic transformations from the shock frame
to the downstream plasma frame are
generated in the same way, using the transformation
\begin{equation}
\label{eq:sk_frm_par}
\vec{p}_{\parallel} = 
\gamma_{u}\left[\vec{p}_{\parallel_{sk}} - 
\gamma_{v,sk}m\vec{u}_{sk}\right]
\end{equation}
with the final normalized equations:
\begin{equation}
\label{eq:sk_to_px2}
p'_{x,pf} = \left[\left(\gamma_{u}-1\right)
\left(\bfrac{u_{x}}{u}\right)^2 + 1\right]
p_{x,sk} - 
\bfrac{\gamma_{u}\gamma_{v,sk}}mu_{x}{m_pu_0} +
\left(\gamma_{u} - 1\right)
\left(\bfrac{u_{x}u_{z}}{u^2}\right)p_{z,sk}
\end{equation}
\begin{equation}
\label{eq:pypf}
p'_{y,pf} = \bfrac{p_y}{m_pu_0}
\end{equation}
\begin{equation}
\label{eq:sk_to_pz2}
p'_{z,pf} = \left[\left(\gamma_{u}-1\right)
\left(\bfrac{u_{z}}{u}\right)^2 + 1\right]
p_{z,sk} -
\bfrac{\gamma_{u}\gamma_{v,sk}mu_{z}}{m_pu_0} + 
\left(\gamma_{u} - 1\right)
\left(\bfrac{u_{x}u_{z}}{u^2}\right)p_{x,sk} 
\end{equation}\\
where it is understood that 
$p'_{x,sk}$, $p'_{y,sk}$ and $p'_{z,sk}$ 
($p'_{x,pf}$, $p'_{y,pf}$ and $p'_{z,pf}$)
are momentum components
in the shock frame (plasma frame)
normalized by $m_pu_0$, $m$ is the mass of the particle
under consideration, 
and $\gamma_{v,sk}$ is
the Lorentz factor associated with the particle velocity as seen in the
shock frame.
Equations for frame transformations between the upstream plasma
frame and the shock frame are generated in the same manner.
%

\section{Method of calculating momentum and energy flux for parallel
relativistic shocks}
\cite{ER91} describe the Monte Carlo procedure for numerically
calculating the various fluxes in the shock frame, showing the flux sums
$A(x,r)$ corresponding to the number flux, $B(x,r)$ corresponding to the
momentum flux, and $C(x,r)$ corresponding to the energy flux, where x
refers to the x component of momentum measured in the shock frame,
and $r$ is the compression ratio $u_0/u_2$.
 
\cite{ER91} also describe the general process for flux conservation
across the shock, assuming no explicit particle escape. A 
compression ratio $r$ is chosen, the fluxes A,B, and C are calculated
first for an unmodified shock, then the backpressure is used to 
modify the shock velocity profile through several iterations until the
flux profiles converge to a stable solution. 
If the calculated fluxes are not constant across the shock, $r$ is varied
and the process is repeated until equations (\ref{mom_flx}, 
\ref{nrg_flx}, and \ref{pt_flx}) are satisfied.
  
Recalling the flux conservation relations for parallel shocks in Chapter 3,
specifically, equations (\ref{PE43simp2}) - (\ref{CM35simp2}), where
$P_{xx}$ is now called $P$ and $w = e + P$,
jump conditions are created across the
shock.  The possibility of escaping particles which carry away
particle, momentum, and energy fluxes as discussed in \cite{BE99}, 
is now included, to give the following equations:
\begin{equation}
\label{mom_flx}
\gamma_0^2w_0\frac{u_0^2}{c^2} + 
P_0 = \gamma_x^2w_x\frac{u_x^2}{c^2} + P_x + F_p
\end{equation} 
\begin{equation}
\label{nrg_flx}
\gamma_0^2w_0u_0 = \gamma_x^2w_xu_x + F_e
\end{equation} 
The third equation results from the requirement that the number of
particles is conserved across the shock [\cite{LL59}]:
\begin{equation}
\label{pt_flx}
\gamma_0n_0u_0 = \gamma_xn_xu_x + F_n.
\end{equation} 
The three equations above represent the  
momentum density flux, the energy density flux, and the particle
number density flux respectively, as viewed from
the shock frame, while the number density $n$ and the enthalpy $w = e + P$ 
are both measured in their respective local plasma frames \citep{ER91}. 
The terms $F_p$, $F_e$ and $F_n$ represent escaping momentum, energy and
particle fluxes, respectively, all referenced to the shock frame.
Note that the Lorentz factors are referenced to the 
flow velocity (as a result of the three space components of velocity). 
The subscript 0 refers to the far upstream frame and the subscript x
refers to any frame to the right of a grid plane, as shown in 
Figure \ref{fig:grid}. 
Recall that the grid zones come from the original shock frame divisions
along the $x$ axis within which the flow velocity is constant.
Also, if both sides of equation (\ref{pt_flx})
are multiplied by the particle rest mass
\footnote{One could say that $\gamma_vm_p$, where $\gamma_v$ is associated 
with the particle speed, but viewed from the shock frame $\gamma_v$ is the
same on both sides of the equation and drops out.},
the third equation could be written as 
$\gamma_0\rho_0u_0 = \gamma_x\rho_xu_x$.

In addition to the conservation relations, there is a relation that
combines the adiabatic equation of state and the conservation
of total energy, i.e. a parameterization of the ratio of specific heats 
appropriate to the region of interest,
\begin{equation}
\label{Gam}
\Gamma = 1 + P/(e + \rho c^2), 
\end{equation}
where $P$ is the kinetic pressure,
$e$ is the total internal energy density including the rest mass energy,  
$\rho c^2$ is the rest mass energy density, and
$\Gamma$ is the true ratio of specific heats under special conditions  
\citep{BM76}. These ideas are discussed in more detail in Appendix B.
Note that the
compression ratio $r = u_0/u_2$ depends on $\Gamma$, making the solution
to these equations nonlinear \citep{ER91}.
%

\section{Method for modifying the relativistic shock velocity profile}
This section describes a method for modifying the shock velocity
profile using the relativistic flux conservation relations, including the 
fluxes from escaping particles.
 
From equations (\ref{mom_flx} and \ref{nrg_flx}), one can write
\begin{equation}
\label{mom2_flx}
\gamma_x^2w_x\frac{u_x^2}{c^2} + P_x = B(x,r),
\end{equation}
and 
\begin{equation}
\label{nrg2_flx}
\gamma_x^2w_xu_x = C(x,r)
\end{equation} 
where $B(x,r)$ and $C(x,r)$ are calculated as described by \cite{ER91}.
Enthalpy $w$ can be eliminated by writing the last equation as
\begin{equation}
\label{enthalpy}
w_x = \frac{C(x,r)}{\gamma_x^2u_x}
\end{equation}
and substituting this relation into equation (\ref{mom2_flx}). 
Solving for pressure $P$ we have 
\begin{equation}
\label{prssr}
P_x = B(x,r) - \gamma_x^2\frac{u_x^2}{c^2}\frac{C(x,r)}{\gamma_x^2u_x}
= B(x,r) - C(x,r)\frac{u_x(old)}{c^2}
\end{equation}
where $u_x(old)$ is the existing velocity at grid zone position $x$.   

So far, using $u_x(old)$, we have an estimate of the pressure. Going
back to equation (\ref{mom_flx}), it can be noted that the left hand 
side of this equation (far upstream from the shock) 
is a constant momentum flux because there are no escaping particles, so
$K_p$ is calculated:
\begin{equation}
\label{Kmom1}
\gamma_0^2w_0\frac{u_0^2}{c^2} + P_0 = K_p. 
\end{equation}
Then one can rewrite equation (\ref{mom_flx}) as 
\begin{equation}
\label{Kmom2}
K_p = \gamma_x^2w_x\frac{u_x^2}{c^2} + P_x + F_p
\end{equation}
The left hand side of equation (\ref{nrg_flx}), far upstream where there
are no escaping particles, yields a constant energy flux, so
$K_e$ is calculated:
\begin{equation}
\label{Knrg1}
\gamma_0^2w_0u_0 = K_e. 
\end{equation}
Then one can rewrite equation (\ref{nrg_flx}) as
\begin{equation}
\label{Knrg2}
K_e = \gamma_x^2w_xu_x + F_e
\end{equation}
with 
\begin{equation}
\label{Kenthalpy}
w_x = \frac{K_e - F_e}{\gamma_x^2u_x}.
\end{equation}
$K_p$ and $K_e$ are numerically calculated in the same way
as are $B(x,r)$ and $C(x,r)$, but far upstream and far from all velocity
variations where there are no escaping particles.
$F_p$ and $F_e$, the momentum and energy fluxes of escaping
particles, are numerically calculated separately from $B(x,r)$ and $C(x,r)$,
but in the same way as the other values.

Substituting (\ref{Kenthalpy}) into (\ref{Kmom2}), 
we have
\begin{equation}
\label{KP}
K_p = (K_e - F_e)\frac{u_x}{c^2} + P_x + F_p\quad ,
\end{equation}
but $u_x$ is an estimate of the {\it new} flow velocity: 
\begin{equation}
\label{newu} 
u_x = c^2\frac{(K_p - F_p - P_x)}{K_e - F_e} = u_x(new)\quad . 
\end{equation}

This equation can now be combined 
with equation (\ref{prssr}) to give the new flow velocity
estimate in terms of the calculated flux values:

\begin{equation}
\label{KPunew}
u_x(new) = \frac{c^2}{(K_e - F_e)}\left[K_p - F_p - B(x,r) + 
C(x,r)\frac{u_x(old)}{c^2}\right].
\end{equation}

This last equation estimates the new velocity profile for the next
iteration for a given compression ratio. After a number of iterations,
the velocity and flux profiles should stop varying, i.e., $u_x(new)$
should become closer and closer to the previously calculated $u_x$.  
If, after the series of iterations, the flux profiles
are constant across the shock (i.e., from upstream to downstream), then
the correct compression ratio was used. If the flux profiles are not
constant, the compression ratio must be varied up or down. In general,
if the downstream side shows a jump upwards in momentum flux after the
iterations, it means the compression ratio was too high and needs to be 
lowered.
%

\section{Calculating $\Gamma$}
After the iterations are completed and, assuming conservation of the
particle, momentum and energy fluxes has been achieved, the parameter
$\Gamma$ can be calculated. If the pressure and energy density are 
known, equation (\ref{Gam}) can be used to calculate $\Gamma$ as shown
in section 2.3 of \cite{ER91}. For Lorentz factors of 10 or greater,
the $\Gamma$ derived in this manner
should agree very closely with the $\Gamma$ calculated from 
equation (\ref{SpecHeat}):
\begin{equation}
\label{eq:Gr}
\Gamma = 1 +\beta_2 = 1+\frac{1}{r}
\end{equation}
%

\section{Summary} 
Relativistic momentum transformation equations were developed that
relate the upstream and downstream plasma reference frames to the frame
at rest with the shock. The equations allow oblique flows and arbitrary
particle momentum directions. 

A method was developed for modifying the relativistic shock velocity
profile by using the relativistic flux conservation relations,  with
arbitrary magnetic field angle, to treat the momentum and energy flux 
at every grid zone, plus the fluxes from escaping energetic particles.
%

\chapter{The Monte Carlo Technique and Computer Simulations}

The description of particle diffusion and energy gain is far more
difficult when $\gamma_0 \gg 1$ because the diffusion approximation,
which requires nearly isotropic distribution functions, cannot be
made. Because of this, \mc\ simulations, where particle scattering and
transport are treated explicitly, and which, in effect, solve the
Boltzmann equation with collective scattering
\citep[e.g.,][]{EE84,KS87b,EJR90,ER91,Ostrow91,BedOstrow96}, offer
advantages over analytic methods.  This is true in the test-particle
approximation, where analytic results exist, but is even more
important for nonlinear \rel\ shocks. 

In \nonrel\ shocks, for $v \gg u_0$, a diffusion-convection equation
can be solved directly 
for infinite, plane shocks
\citep[e.g.,][]{ALS77,BO78}, yielding the
well-known result
\begin{equation}
\label{eq:TPpowerlaw}
f(p) \, d^3 p\propto
p^{-\sigma} \, d^3p
\quad \hbox{with} \quad
\sigma = 3 r/ (r-1)
\ ,
\end{equation}
where $r$ is the shock compression ratio, $p$ is the momentum, and
$f(p)\, d^3 p$ is the number density of particles in $d^3p$.
Equation~(\ref{eq:TPpowerlaw}) is a steady-state, test-particle result
with an undetermined normalization, but the spectral index, $\sigma$,
in this limit is independent of the shock speed, $u_0$, or any details
of the scattering process as long as there is enough scattering to
maintain isotropy in the local frame. To obtain an absolute injection
efficiency, or to self-consistently describe the nonlinear
back-reaction of accelerated particles on the shock structure (at least
when the seed particles for acceleration are not fully \rel\ to begin
with), techniques which do not require $v \gg u_0$ must be
used. Furthermore, 
for particles that do not obey $v \gg u_0$ 
additional assumptions must be made for how
these particles  
interact with the background
magnetic waves and/or turbulence, i.e., the so-called ``injection
problem'' must be considered \citep[see, for
example,][]{JE91,Malkov98}.
The \mc\ techniques described here make the simple assumption that
all particles, regardless of energy, interact in the same way, i.e.,
all particles scatter elastically and isotropically in the local
plasma frame with a mean free path proportional to their gyroradius.
These techniques and assumptions have been used to calculate nonlinear
effects in \nonrel\ collisionless shocks for a number of years with
good success comparing model results to spacecraft observations
\citep[e.g.,][]{EE84,EMP90,EJB99}.

Early work on relativistic shocks was mostly analytical in the test
particle approximation \citep[e.g.,][]{BM76,Peacock81,KS87a,HD88},
although the analytical work of \citet{SK87} explored modified shocks.
Test-particle Monte Carlo techniques for \rel\ shocks were developed
by \citet{KS87b} and \citet{EJR90} for parallel, steady-state shocks,
i.e., those where the shock normal is parallel to the upstream
magnetic field, and extended to include oblique magnetic fields by
\citet{Ostrow91}.  Some preliminary work on modified relativistic
shocks using Monte Carlo techniques was done by
\citet{EllisonJapan91,EllisonPoland91}.

Monte Carlo techniques are used to model nonlinear particle
acceleration in parallel collisionless shocks of various speeds,
including mildly relativistic ones. 
When the acceleration is efficient, the back-reaction of accelerated
particles modifies the shock structure and causes the compression
ratio, $r$, to increase above test-particle values.  Modified shocks
with small 
Lorentz factors 
can have compression ratios
considerably greater than $3$ and the momentum distribution of
energetic particles no longer follows a power law relation.  These
results may be important for the interpretation of gamma-ray bursts if
mildly \rel\ internal and/or afterglow shocks play an important role 
accelerating particles that produce the observed radiation.
For $\gamma_0 \gtrsim 10$, $r$ approaches $3$ and the so-called
`universal' test-particle result of $N(E) \propto E^{-2.3}$ is
obtained for sufficiently energetic particles. In all cases, the
absolute normalization of the particle distribution follows directly
from the model assumptions and is explicitly determined.
%

\section{\mc\ Model}

\begin{figure}[!hbtp]             
\dopicture{.45}{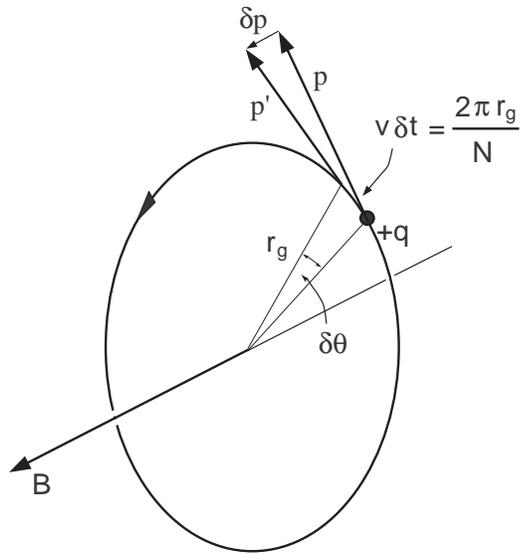}       
\caption{A charged particle in a helical orbit around a magnetic field vector 
$\vec{\mathrm{B}}$. The particle moves through an angle $\delta\theta$ 
in time $\delta t$. N is an input parameter chosen to contol the pitch
angle granularity. 
\label{fig:gyro_cyc01}
}
\end{figure}

The techniques used here are essentially identical to those described in
\citet{EBJ96} and \citet{EJB99}. The differences are that the code has
been made fully \rel\ and only results for parallel shocks with
pitch-angle diffusion are presented in this chapter.
The code is steady-state, includes a uniform magnetic
field, and moves particles in helical orbits, as shown in Figure
\ref{fig:gyro_cyc01}.  
The \alf\ Mach number 
is assumed to be large, i.e., any effects from \alf\ wave
heating in the upstream precursor are neglected. 
This also means the second-order acceleration of particles
scattering between oppositely propagating \alf\ waves is neglected. 
Such an effect in \rel\ plasmas with strong magnetic fields is
proposed for nonlinear
particle acceleration in GRBs by \citet{Pelletier99}
\citep[see also][]{PellMar98}.

The pitch angle diffusion is
performed as described in \citep{EJR90} and is shown in 
Figure \ref{fig:gyro_cyc01}. That is, after a small
increment of time, $\delta t$, a particles' momentum vector,
$\mathbf{p}$, undergoes a small change in direction, $\delta\theta$.  If
the particle originally had a pitch angle, $\theta$ (measured relative
to the shock normal direction), it will have a new pitch angle
$\theta'$ such that
\begin{equation}
\label{eq:cosprime}
\cos{\theta'} = \cos{\theta} \, \cos{\delta\theta} +
\sqrt{1 - \cos^2{\theta}} \sin{\delta\theta} \cos{\phi}
\ ,
\end{equation} 
where $\phi$ is the azimuth angle measured with respect to the
original momentum direction. All angles are measured in the local
plasma frame. 
\begin{figure}[!hbtp]             
\dopicture{.45}{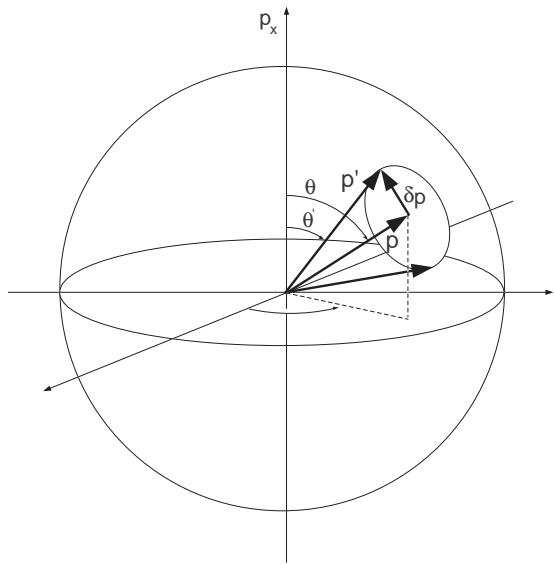}       
\caption{Conceptual diagram showing a particle with momentum $\vec{p}$\
elastically scattering on magnetic turbulence through an 
angle $\delta\theta$. 
\label{fig:scat01}
}
\end{figure}

If $\delta\theta$ is chosen randomly from a uniform
distribution between 0 and $\delmax$ and $\phi$ is chosen from a
uniform distribution between 0 and $2 \pi$, the tip of the momentum
vector, referring to Figure \ref{fig:scat01}, 
will perform a random walk on a sphere of radius $p$.  As shown
by \citet{EJR90}, the
angle $\delmax$ is determined by
\begin{equation}
\label{eq:thetamax}
\delmax =
\left ( 6 \, \delta t/t_c \right )^{1/2} =
\left ( 12 \pi/ N \right )^{1/2}
\ ,
\end{equation}
where $N = \tau_g / \delta t \gg 1$ is the number of gyro-segments,
$\delta t$, dividing a gyro-period $\tau_g = 2 \pi r_g/v$. The time
$t_c$ is a ``turn around'' time defined as $t_c = \lambda / v$, where
$\lambda$ is the particle mean free path.  The mean free path is taken
to be proportional to the gyroradius $r_g = pc/(QeB)$ ($e$ is the
electronic charge, $Q$ is the ionic charge number, and $B$ is the
local uniform magnetic field), i.e.,
%
$\lambda = \eta \, r_g$,
%
where $\eta$ determines the strength of scattering. In all of the
examples given here $\eta=1$ is set, or, in other words, the strong
scattering Bohm limit is assumed.

For a downstream particle to return upstream, it's velocity vector
must be directed within a cone with opening angle $\theta_2$ such that
$|v_2\cos{\theta_2}| > u_2$, where $v_2$ and $\theta_2$ are measured
in the downstream frame and $\theta_2=0^{\circ}$ is in the
$-x$-direction, i.e., along the shock normal direction.  
For 
fully relativistic
shocks with $v_2 \simeq c $ and $u_2
\simeq c/3$, $\cos{\theta_2} \gtrsim 1/3$ for a downstream particle to
cross the shock into the upstream region.
When the particle enters the upstream region it 
must satisfy essentially the same constraint, i.e.,
$|v_0 \cos{\theta_0}| > u_0$, 
where now $v_0$ and $\theta_0$ are measured in the upstream
frame.\footnote{In the \TP\ approximation, $u_0$ is just the shock
speed. In nonlinear shocks, the flow speed just upstream from the
subshock at $x=0$ will be less than the far upstream shock speed,
$u_0$, as measured in the shock reference frame.}
Since both the particle and shock
have high Lorentz factors, one can write
\begin{equation}
\label{eq:ret_pitch}
\cos{\theta_0} = u_0/v_0 = 
\frac{ \left ( 1 - 1/\gamma^2_0 \right )^{1/2}}
{\left (1 - 1/ \gamma^2_v \right )^{1/2}} \simeq
\frac{1 - 1/(2 \gamma^2_0)}{1- 1/(2\gamma^2_v)}
\ ,
\end{equation}
where $\gamma_v \equiv [1-(v_0/c)^2]^{-1/2}$ is the particle Lorentz
factor. Since $\cos{\theta_0}
\simeq 1 - \theta_0^2/2$ for small $\theta_0$, we have
\begin{equation}
\theta_0^2 \simeq \frac{1}{\gamma^2_0} - \frac{1}{\gamma^2_v}
\ .
\end{equation}
For ultrarelativistic particles with $\gamma_v \gg \gamma_0$, 
$\theta_0 \simeq 1/\gamma_0$ \citep[e.g.,][]{GA99},
but $\theta_0$ can be much smaller for mildly \rel\ particles.

\begin{figure}[!hbtp]              
\dopicture{0.8}{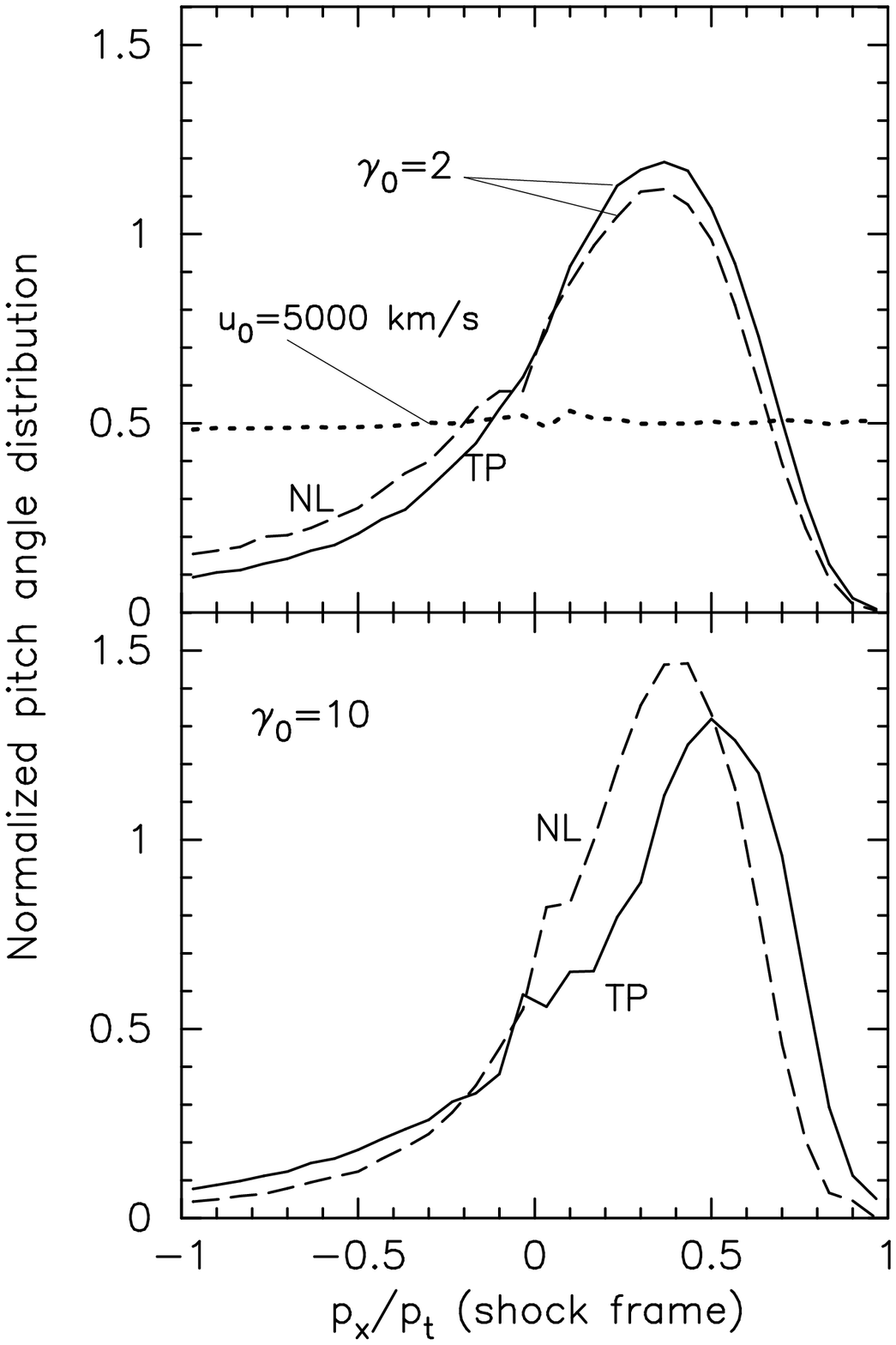} 
\caption{Distribution of the cosine of the pitch angle, i.e.,
$p_x/p_t$, for particles crossing $x=0$. The solid and dashed curves
in the top panel are for a shock with Lorentz factor $\gamma_0=2$. The
dotted curve in the top panel is for a \nonrel\ shock with speed
$u_0=5000$ \kmps. The bottom panel shows curves for a fully \rel\
shock with $\gamma_0=10$. In all cases, the curves are normalized so
that the areas under them is 1 and the $x$-component of momentum,
$p_x$, is positive when directed downstream. The nonlinear (NL) and
unmodified (UM) shock results are labeled.
\label{fig:angle}
}
\end{figure}

In order to re-cross into the downstream region, particles must
scatter out of the upstream cone defined by $\theta_0$ and
\citet{AGKG2001} show that most particles are only able to change the
angle they make with the upstream directed shock normal by
$|\delta\theta| \sim \theta_0$ before being sweep back downstream, making
the distribution of shock crossing particles highly anisotropic.
Therefore, if the shock Lorentz factor $\gamma_0 \gg 1$, a larger
fraction of particles re-cross the shock into the downstream region
with highly oblique angles (as measured in the shock
frame) compared to lower speed shocks (see Figure~\ref{fig:angle}
discussed below).
Particles crossing at such oblique angles receive smaller energy gains
than would be the case for an isotropic pitch angle distribution and
\citet{AGKG2001} go on to show that $\ave{E_f/E_i} \sim 2$ for a shock
crossing cycle (after the first one).

\begin{figure}[!hbtp]              
\dopicture{0.8}{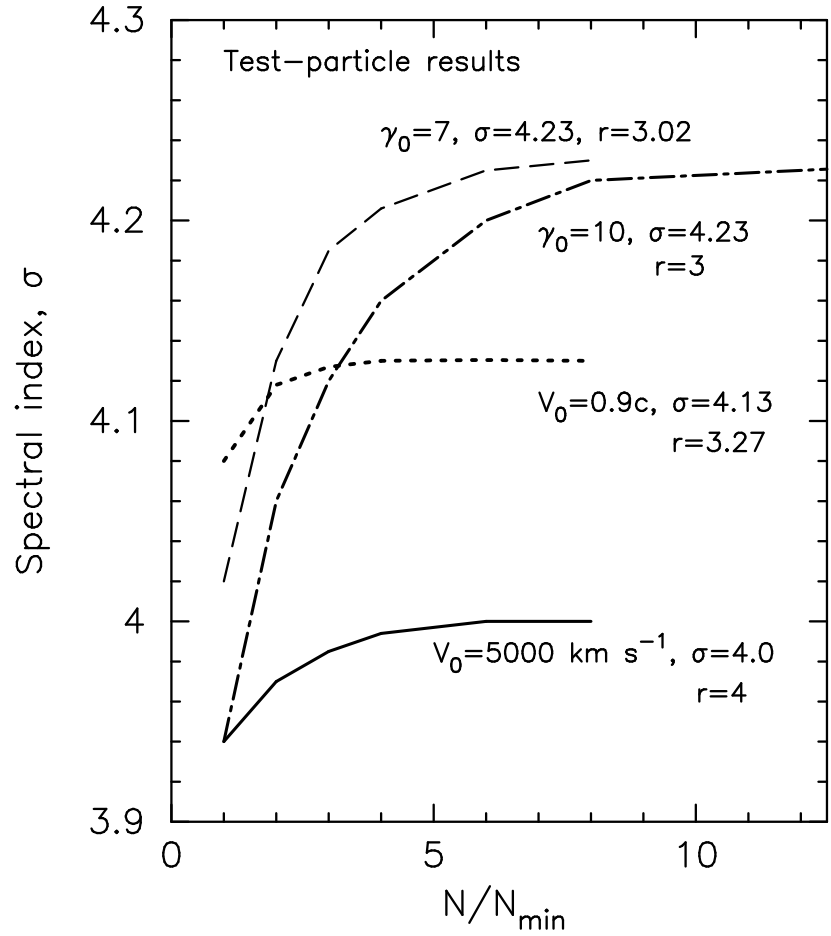} 
\caption{Power law spectral index, $\sigma$, versus number of
gyro-segments, $N$, for various test-particle shocks as labeled. In
all cases, as $N$ is increased the spectral index converges asymptotically
and obtain 
the known results of $\sigma \simeq 4$ for \nonrel\ shocks
with $r=4$, and $\sigma \simeq 4.23$ for fully \rel\ shocks with
$r=3$. The parameter $\Nmin$ is defined in equation~(\ref{eq:gyroeq}).
\label{fig:gyro}
}
\end{figure}

With these concepts in mind, the requirement must be 
$\delmax < \theta_0$, or
\begin{equation}
\label{eq:gyroeq}
N > \Nmin = 12\pi\gamma_0^2
\ .
\end{equation}
The result (as shown in Figure~\ref{fig:gyro}) is that the power law
spectral index, $\sigma$, asymptotically approaches a maximum value as
$N$ is increased.  If $N$ is less than the value required for
convergence to the asympotic value
(and the gyro-segments are too large), the distribution
will be flatter than produced with the convergent value of $N$ because
more particles are able to cross from upstream to downstream with
$\thetaSK \ll 90^\circ$ and receive unrealistically large energy
boosts.
This effect has long been known from the comparison of pitch-angle
diffusion to large-angle scattering in \rel\ shocks
\citep[e.g.,][]{KS87b,EJR90}.
For all of the examples reported on here, $N$ is chosen large enough
so it makes no difference if $\delta\theta$ is chosen uniformly between
$0^\circ$ and $\delmax$ or if $\cos{\delta\theta}$ is chosen uniformly
between $\cos{\delmax}$ and 1.
Figure~\ref{fig:gyro} shows how the results depend on $N$ for shock
speeds ranging from fully \nonrel\ to fully \rel.  In all cases, as
$N$ is increased the spectral index approaches a maximum and for
$\gamma_0 \gtrsim 7$ the well known result $\sigma =
4.2$--$4.3$ is obtained.
The fact that the computation time for the Monte Carlo simulation scales
as $N$ and $N \propto \gamma_0^2$ places limits on modeling \ultrarel\
shocks with this technique.

In Figure~\ref{fig:angle} pitch-angle distributions
(measured in the shock reference frame) of particles crossing the
shock are compared.
The curves are normalized such that the area under each curve equals
one and the $x$-component of particle momentum in the shock frame,
$p_x$, is positive when directed downstream to the right (see
Figure~\ref{fig:profG2} for the shock geometry). Here, $p_t$ is the
magnitude of the total particle momentum also measured in the shock
frame.
In the top panel, unmodified (UM) test particle and nonlinear (NL)
mildly \rel\ shock ($\gamma_0 = 2$) with a \nonrel\ one ($u_0=5000$
\kmps) are compared.
Particles crossing the $\gamma_0 = 2$
shock are highly anisotropic with $p_x/p_t$ strongly peaked near $\sim
0.35$. In the \nonrel\ shock, the particles are nearly isotropic
except for a slight flux-weighting effect. 
There is little difference
in the distributions between the UM and NL shocks.
In the bottom panel, the pitch-angle distributions for UM and NL shocks 
with $\gamma_0 = 10$ are shown. While the distributions are somewhat more
sharply peaked, they are quite similar to those for $\gamma_0 = 2$ and
show relatively small variations between the UM and NL shocks.

The main difference between the present code and an earlier code used
by \citet{EJR90} to model test-particle \rel\ shocks is that the
previous code used a guiding center approximation with an emphasis on
large-angle scattering rather than the more explicit orbit calculation
of pitch-angle diffusion used here.  Other than the far greater range
in $\gamma_0$ and the nonlinear results presented here, the work of
\citet{EJR90} is consistent with this work.

The particle transport is performed as follows.  Particles of some
momentum, $\ppf$, (measured in the local plasma frame) are injected
far upstream from the shock and pitch-angle diffuse and convect until
they cross a grid zone boundary, i.e., a plane with its normal parallel 
to the upstream flow speed
\citep[see][for a full discussion]{EBJ96}. For unmodified shocks, the
flow speed changes from the upstream to downstream value in a 
discontinuous step at the shock boundary,
but for nonlinear shocks the bulk flow speed changes in small steps, 
each separated by a grid zone boundary, from $u_0$ far
upstream to $u_2$ downstream.
In the unmodified case, the shock thickness is essentially zero (i.e.,
shorter than the distance a particle diffuses in $\delta t$) but in
the nonlinear, modified case, the shock precursor extends over the
entire region of varying bulk flow speeds and a small scale
``subshock'' (at position $x=0$ in this simulation) exists where most
of the entropy production occurs.
When a particle crosses a grid zone boundary, $\ppf$ is transformed to
the new local frame moving with a new speed relative to the subshock
and the particle continues to scatter and convect.  Each particle is
followed until it leaves the system in one of three ways. It can
convect far downstream and not return to the subshock, it can obtain a
momentum greater than some $\pmax$ and be removed, or it can diffuse
far enough upstream to cross an upstream free escape boundary (FEB)
and be removed. Both $\pmax$ and the position of the FEB are free
parameters in this model \citep[see][for a discussion of the
self-consistent determination of the maximum
particle energy in \nonrel\ supernova remnant shocks]{BerezV97}.
For the nonlinear calculations, the shock structure and compression 
ratio are iterated until the number, momentum, and energy fluxes are
conserved across the shock.  This procedure has been detailed many
times for \nonrel\ shocks \citep[see][and references therein]{EBJ96}
and the modifications required for \rel\ shocks were given in
\citet{ER91}.

To avoid excessive computation, a probability of return
calculation as described in detail in \citet{EJR90} is used.  That is,
the standard expression obtained by \citet{Peacock81} is used,
\begin{equation}
\label{eq:ProbRet}
P_R = \left ( \frac{v - u_2}{v + u_2} \right )^2
\ ,
\end{equation}
to determine the probability, $P_R$, that a particle, having crossed a
particular point in the uniform downstream flow, will return back
across that point.
Equation~(\ref{eq:ProbRet}) is fully relativistic and independent of
the diffusive properties of the particles as long as they are
isotropic in the $u_2$ frame. This isotropy is ensured by only
applying equation~(\ref{eq:ProbRet}) once a particle has diffused
several mean free paths downstream from the shock.
For \ultrarel\ shocks with $v\simeq u_0 \simeq c$ and $r \simeq 3$,
equation~(\ref{eq:ProbRet}) gives $P_R \simeq 0.25$.  

For clarity, it is noted that this is not the same probability, $\Pret$,
used by \citet{AGKG2001} to determine the \TP\ power law
index, i.e.,
\begin{equation}
s = 1 + \frac{\ln (1/\Pret)}{\ln \ave{E_f/E_i}}
\ ,
\end{equation}
where $N(E) \propto E^{-s}$ and, for fully \rel\ particles, $s =
\sigma - 2$. The quantity $\ave{E_f/E_i}$ is the average energy ratio
for a particle undergoing a shock crossing cycle.
Although not explicitly stated in \citet{AGKG2001}, it is clear from
the context that $\Pret$ is calculated just behind the shock where the
particle distribution is highly anisotropic.  In this case, particles
that have just crossed from upstream to downstream will be more likely
to recross back into the upstream region than indicated by
equation~(\ref{eq:ProbRet}) because their pitch angles are more likely
to be highly oblique relative to the shock normal than in the
isotropic distributions further downstream (compare the solid or
dashed curves to the dotted curve in the top panel of
Figure~\ref{fig:angle} discussed below). For $\gamma_0 = 10$,
\citet{AGKG2001} find $\Pret = 0.435 \pm 0.005$ and $\ave{E_f/E_i} =
1.97 \pm 0.01$ giving the standard result $s = 2.230 \pm 0.012$.  As
shown in Figure~\ref{fig:gyro}, the unmodified results are consistent
with this spectral index for $\gamma_0 \gtrsim 7$.

\begin{figure}[!hbtp]              
\dopicture{0.8}{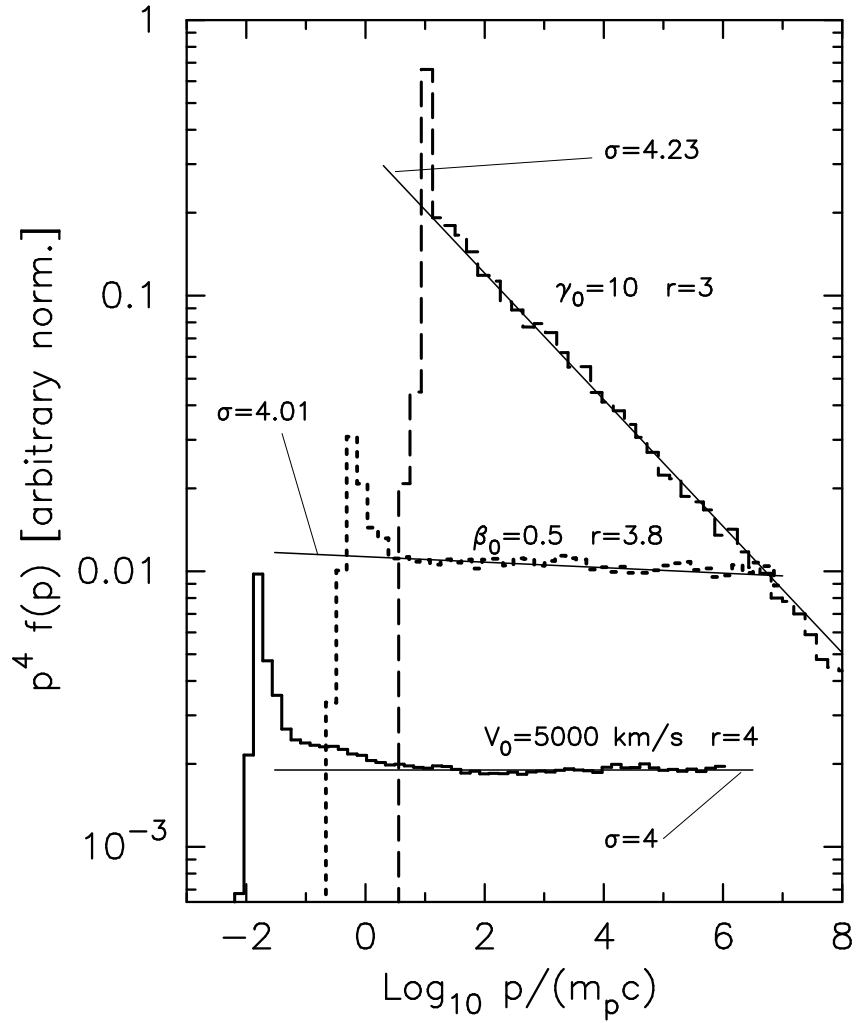} 
\caption{Particle spectra, $p^4 f(p)$, versus momentum for various
unmodified shocks with speeds as indicated. The test-particle
compression ratios, $r$, and spectral indices, $\sigma$, are
noted. The far upstream plasmas in the 5000 \kmps, and $\beta_0=0.5$
shocks are thermal at $10^6$ K. The $\gamma_0=10$ shock has a far
upstream plasma which is a $\delta$-function at 1 MeV.  All spectra
are calculated at the shock, in the shock frame, and the relative
normalization is arbitrary.
\label{fig:specTP}
}
\end{figure}
%

\section{Test-Particle Results}

In Figure~\ref{fig:specTP} particle distributions for unmodified shocks 
are shown with speeds ranging from fully \nonrel\ ($u_0 = 5000$ \kmps) 
to mildly \rel\ ($\beta_0 = u_0/c = 0.5$) to fully \rel\
($\gamma_0 = 10$). The \nonrel\ distribution matches the standard \TP\
Fermi result of $\sigma = 4$ for $r=4$ and the fully \rel\ result is
consistent with the well-known limit of $\sigma \too 4.2 - 4.3$ as
$\gamma_0 \too \infty$ for $r=3$. In the trans-relativistic regime, both
the compression ratio and the spectral index vary with $u_0$.
%
For the $u_0 = 0.5c$
distribution shown in Figure~\ref{fig:specTP}, the compression ratio 
was determined by balancing the mass, momentum, and energy fluxes
across the shock in the test-particle limit, i.e., by ignoring any
effects from the accelerated particles. This technique is described in
detail in \citet{ER91}. It was found that 
$r = 3.8 \pm 0.1$ and the spectral index is 
$\sigma \simeq 4.01$, slightly flatter than $3 \, r / (r-1)
\simeq 4.07$, the \nonrel\ result.

\begin{figure}[!hbtp]              
\dopicture{0.8}{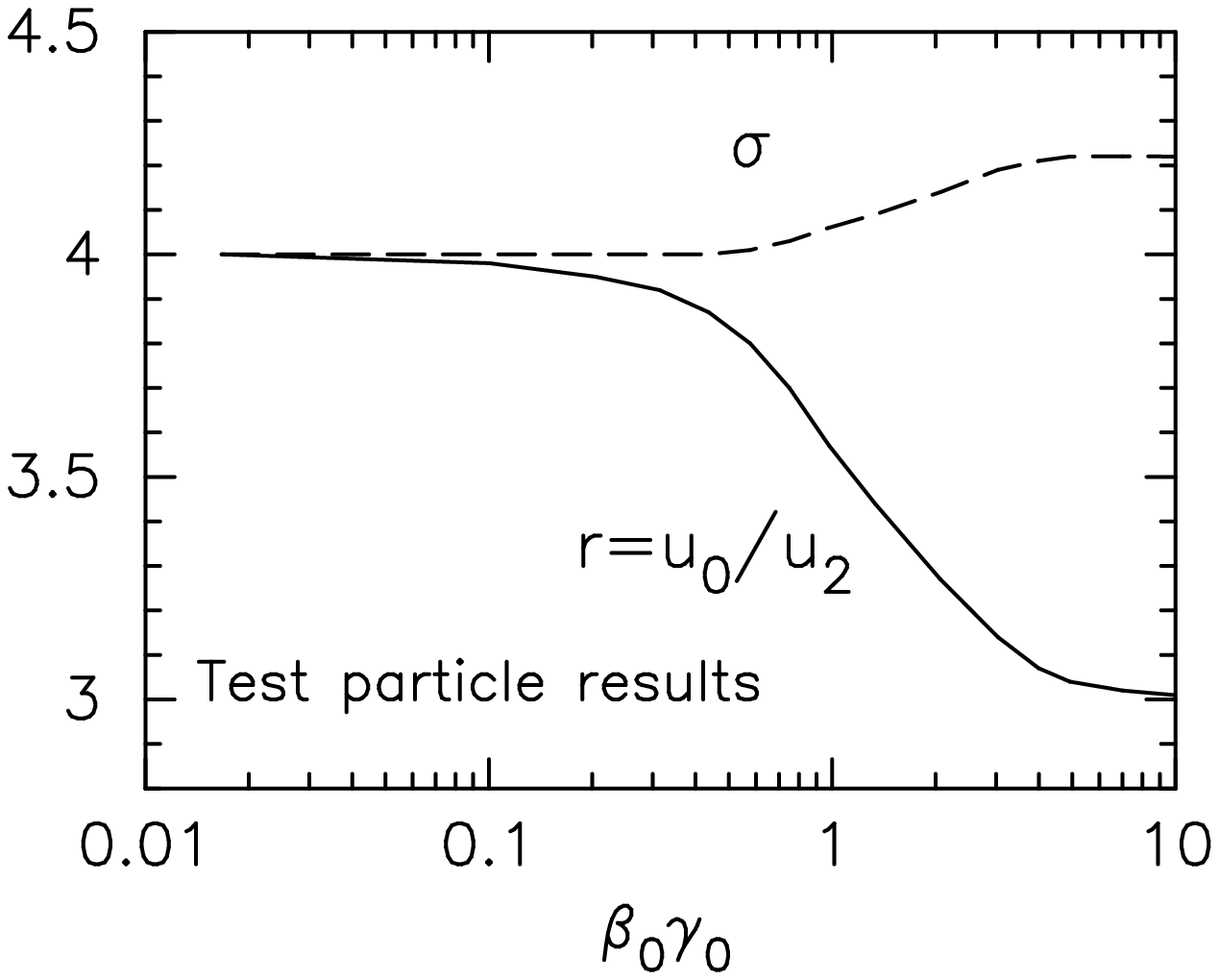}
\caption{The solid line is the compression ratio, $r$, and the
dashed line is the spectral index, $\sigma$, for unmodified (i.e.,
\TP) shocks. The solid dots show $r$ for shocks undergoing efficient
particle acceleration. In all cases, $r$ is determined for each
$\beta_0 \gamma_0$ by balancing the momentum and energy fluxes across
the shock. The maximum cutoff momentum, $\pmax$, is unimportant for
the \TP\ shocks, but is set to $10^4\, \beta_0 \gamma_0 \, m_p c$ for
the nonlinear shocks.
\label{fig:ratio}
}
\end{figure}

Figure~\ref{fig:ratio} shows the compression ratio as
a function of $\beta_0 \gamma_0$ (solid curve), still ignoring the
effects of accelerated particles.  
The compression ratio is determined self-consistently by balancing the
momentum and energy fluxes across the shock with no restriction on the
shocked or unshocked adiabatic index.
As expected, $r$ decreases smoothly
from $4$ for fully \nonrel\ shocks to $\sim 3$ for fully \rel\ shocks.
The power law index, $\sigma$, is also shown (dashed curve)
and this varies slowly from $\sigma =4$ to $\sigma \simeq 4.23$
between the two extremes.
For comparison, $r$ (solid dots) is shown for shocks undergoing
efficient particle acceleration. For these points, all shock
parameters 
except $\beta_0\gamma_0$ and $\pmax$ are kept constant. For the
nonlinear shocks, the maximum cutoff momentum is set to $\pmax= 10^4
\, \beta_0\gamma_0 \, m_p c$ in all cases.\footnote{The value of
$\pmax$ can have a large influence on the shock characteristics at low
$\beta_0 \gamma_0$ because $r$ is large enough (and $\sigma$ small
enough)
that particles escaping
at $\pmax$ are dynamically important. When $\beta_0 \gamma_0$ becomes
large enough so that $r$ drops below $\sim 4$, $\pmax$ is no longer an
important parameter for the shock structure.}
The nonlinear results are discussed in detail below, but here
only emphasize that $r$ in nonlinear shocks will be larger than the
\TP\ value for $\gamma_0 \lesssim 10$.
The test-particle results shown in Figure~\ref{fig:ratio} are in close
agreement with recent analytic results of 
\citet{KGGA2000} and \citet{Gallant2002}.
It seems that no analytic results exist for nonlinear, \rel\ shocks.

\begin{figure}[!hbtp]              
\dopicture{0.8}{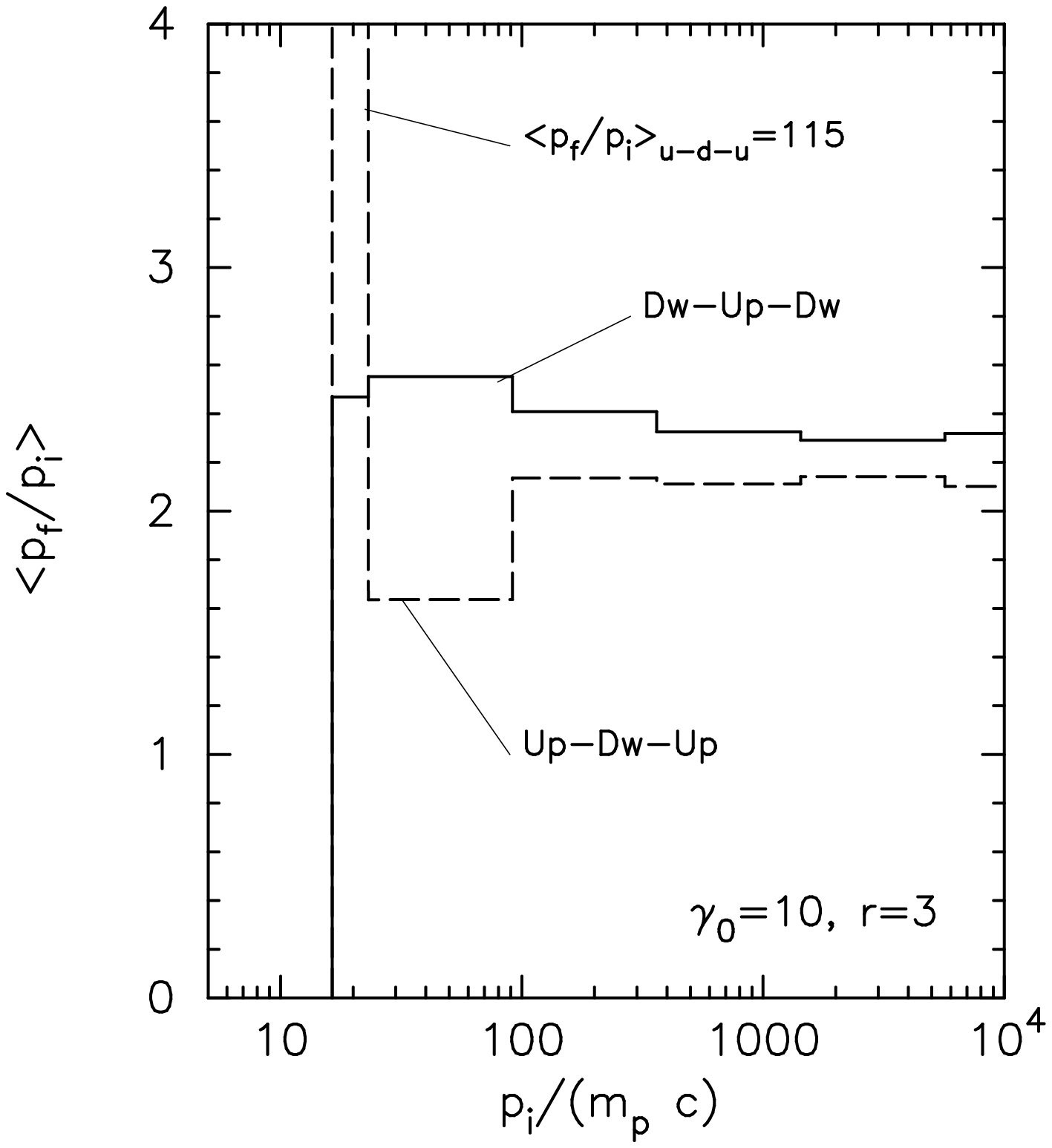}
\caption{Average ratios of final (f) to initial (i) momentum for
downstream to upstream to downstream and upstream to downstream to
upstream shock crossing cycles. The histograms are for a test-particle
shock with $\gamma_0=10$ and $r=3$. Note the large momentum gain
($\UDUave = 115$) in the first shock crossing cycle.
Note also that this particular value depends on the choice of the 
injection momentum.
\label{fig:pfpi}
}
\end{figure}

In Figure~\ref{fig:pfpi} the average ratios of momenta are shown
(measured in the local plasma frame) for particles executing upstream
to downstream to upstream cycles across the shock, \hbox{$\UDUave$},
and downstream to upstream to downstream cycles, \hbox{$\DUDave$}.
These results are for $\gamma_0 = 10$ ($r=3$) and show a slight momentum
dependence at low momenta but converge to $\ave{p_f/p_i} = 2.2 \pm
0.1$ at high momenta. This value is close to $\ave{E_f/E_i} = 1.97 \pm
0.01$ reported by \citet{AGKG2001} for $\gamma_0 = 10$.
As mentioned above, in the first upstream to downstream to upstream
cycle, particles achieve a large boost in momentum as indicated in the
figure.

The difference between this value of $\ave{p_f/p_i}$ at large $p_i$ and
that of \citet{AGKG2001} is greater than the uncertainties and
probably stems from the different assumptions made in the simulations.
As discussed above, $\ave{p_f/p_i}$ depends critically on the average
angle a particle makes when crossing the shock. While the large
majority of particles in these simulations gain energy by crossing from
downstream to upstream and then immediately (within a few
$\delta t$'s) re-crossing back into the downstream region at oblique
angles, a few manage to diffuse farther upstream (see
Figure~\ref{fig:traj09c} below). When these particles re-cross the
shock into the downstream region, they can do so at flatter angles and
receive larger energy gains. One might speculate that differences in how
these few particles are treated in the simulations might produce the
differences in $\ave{p_f/p_i}$.

\begin{figure}[!hbtp]              
\dopicture{0.8}{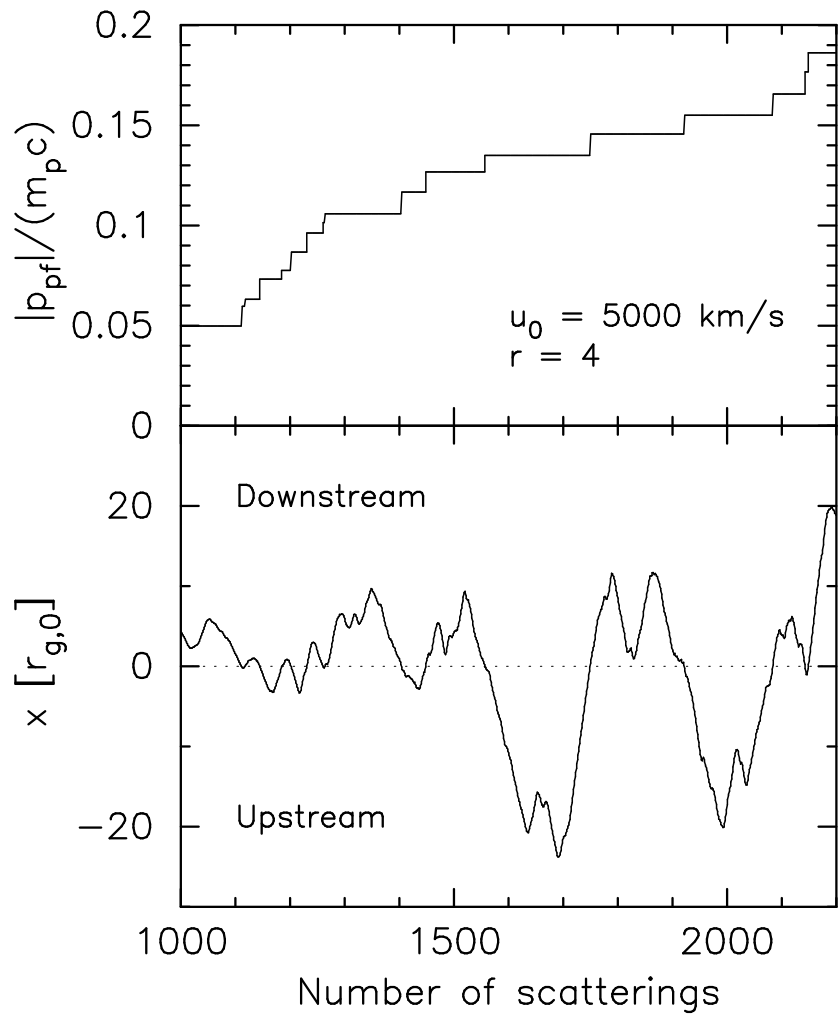}
\caption{Particle trajectory (lower panel) and momentum (upper
panel) in an unmodified shock of speed $u_0=5000$ \kmps. The momentum
is calculated in the local plasma frame, either upstream or downstream
from the shock.
\label{fig:traj5000}
}
\end{figure}

As an illustration of how particles interact with nonrelativistic and
relativistic unmodified shocks, trajectories for two
individual particles are shown in Figures~\ref{fig:traj5000} and
\ref{fig:traj09c}. The lower panel in each figure shows a trace of the
particle trajectory and the upper panels show the particle momentum,
always measured in the local plasma frame. For the \nonrel\ shock
(Figure~\ref{fig:traj5000}), the speed of the particle is far greater
than the shock speed and it diffuses easily on both sides of the
shock. When it crosses $x=0$, it does so nearly isotropically (except
flux weighting makes crossings with flat trajectories slightly more
likely) and essentially always gains momentum. The momentum gain in a
single shock crossing is small, but a particle can stay in the system
for many crossings.

When the shock speed, $u_0$, is close to $c$, the particle will be
convected downstream much more rapidly than in \nonrel\ shocks and few
particles will be able to cross the shock many times. However,
downstream particles that do manage to cross the shock into the
upstream region do so with much flatter trajectories, as discussed
above, and can receive large momentum boosts in a single shock
crossing due to the shock's Lorentz factor (note the logarithmic scale
in the top panel of Figure~\ref{fig:traj09c}). In a typical shock
crossing cycle, downstream particles gain momentum when they cross
into the upstream region (see positions labeled $a$ and $b$ in
Figure~\ref{fig:traj09c}), lose momentum when they cross back
downstream because they cross with oblique pitch angles,
but end up with a net momentum boost.
However, as shown by the position labeled $c$, it is possible for a
particle to diffuse farther upstream before being convected back to
the shock.  In this case, it can cross the shock with a flat
trajectory and gain momentum upon entering the downstream region.
If the acceleration is efficient, the few particles that diffuse far
upstream carry enough pressure to produce the shock smoothing 
discussed next.

\begin{figure}[!hbtp]              
\dopicture{0.8}{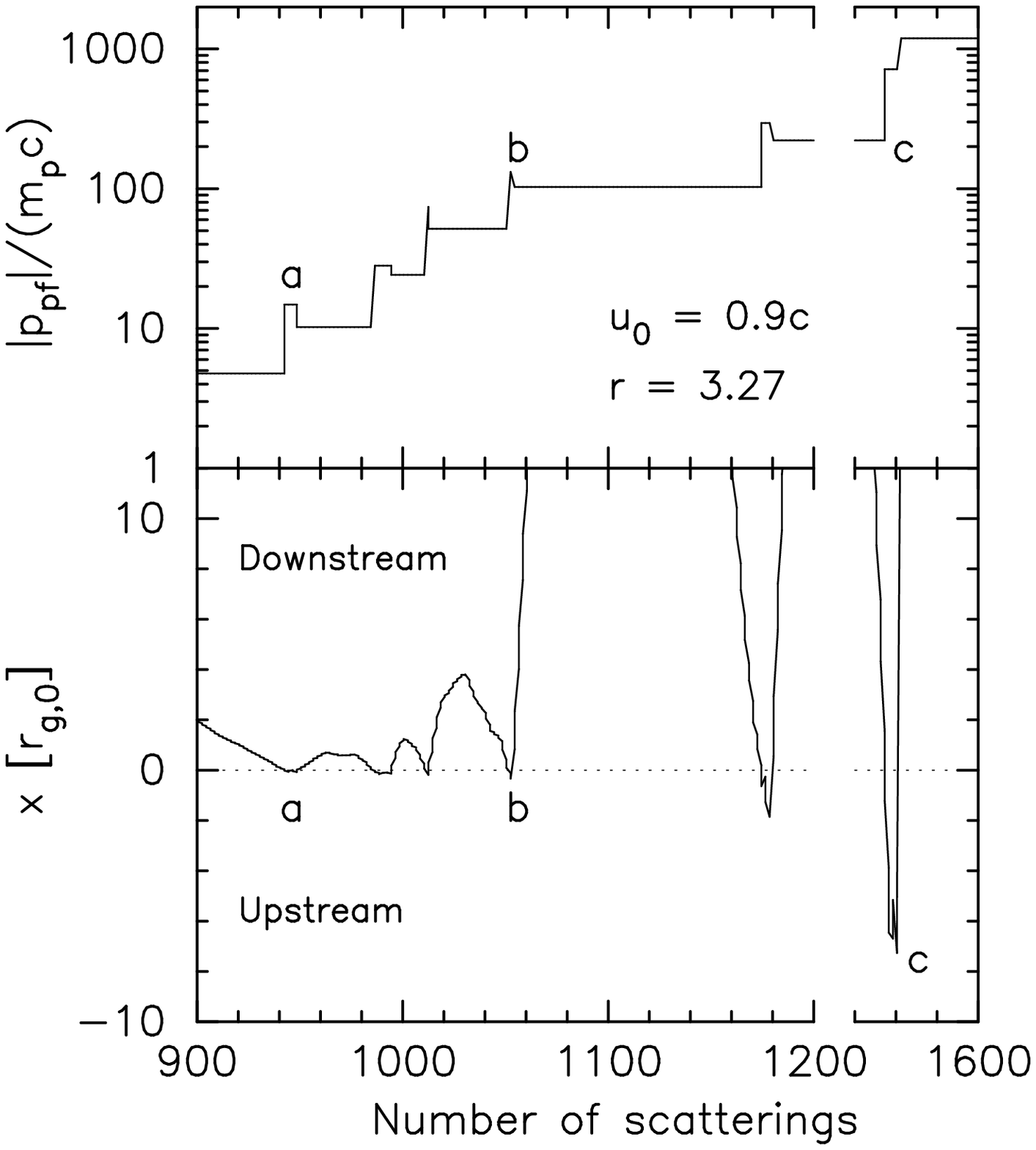}
\caption{Trajectory and momentum for a particle in an unmodified
shock with speed $u_0=0.9c$. Note that the horizontal axis is split at
1200 scatterings. Note also that the momentum scale is logarithmic
because the momentum excursions are very large.
\label{fig:traj09c}
}
\end{figure}
%

\section{Non-Linear Results}

As noted by \citet{AGKG2001}, the large energy boost particles receive
in their initial crossing of the shock provides a natural injection
process for further acceleration and suggests that \rel\ shocks may be
efficient accelerators.
However, just as with \nonrel\ shocks, efficient acceleration limits
the use of test-particle approximations and requires that the
nonlinear back-reaction of the accelerated particles be treated
self-consistently \citep[e.g.,][]{JE91}.
These nonlinear effects will result in a smoothing of the shock and a
change in the overall shock compression ratio, just as they do in
\nonrel\ shocks.
The differences between test-particle and nonlinear results, for 
the parameter ranges investigated here, are large enough to
produce spectra noticeably different from the often quoted $N(E)
\propto E^{-2.3}$ and to influence applications to GRB models where
high shock efficiencies are assumed.  This is illustrated with two
examples, one mildly \rel\ ($\gamma_0 = 1.4$) and one more fully \rel\
($\gamma_0 = 10$).  The far upstream conditions have relatively little
influence on the results as long as $\gamma_p \ll \gamma_0$, where
$\gamma_p$ is the plasma frame Lorentz factor of the far upstream,
injected particles. For concreteness, in both examples the far
upstream plasma is taken
to be a thermal distribution of leptons and baryons
at a temperature of $10^6$ K, but only baryon acceleration is considered;
the electrons only being included for charge neutrality.
 
\subsection{Mildly relativistic shock: $\gamma_0 = 1.4$}

\begin{figure}[!hbtp]              
\dopicture{0.7}{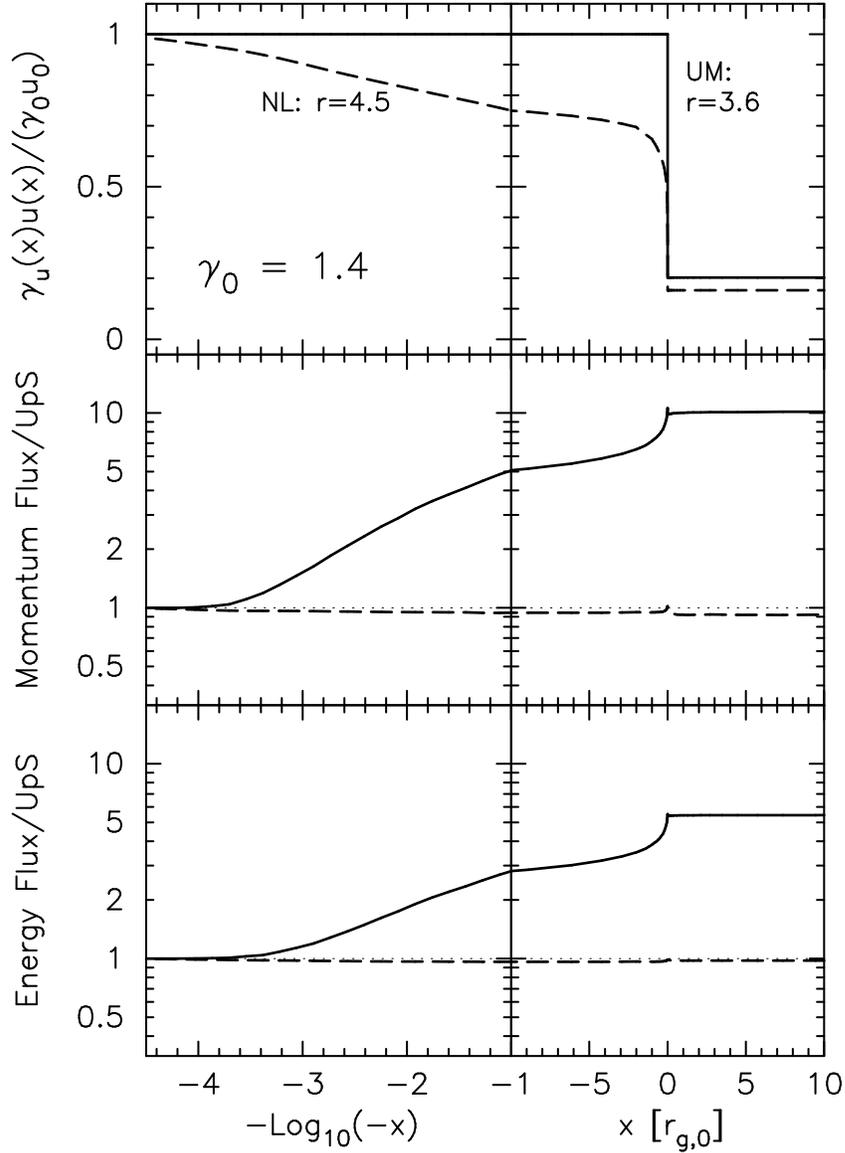}
\caption{Unmodified (UM) and nonlinear (NL) shock profiles, i.e.,
$\gamux u(x)$, and momentum and energy fluxes versus position,
$x$. All quantities are scaled to far upstream values and in all
panels the solid curves are results from unmodified shocks and the
dashed curves are nonlinear results. The NL momentum and energy fluxes
are 3 to 4\% below the far upstream values because particles escape at
a maximum momentum, 
$\pmax = 10^4 \beta_0 \gamma_0 m_p c \simeq
10^{13}$ eV/c. $N/\Nmin \sim 10$ have been used in both cases.
\label{fig:profG2}
}
\end{figure}

Figure~\ref{fig:profG2} shows unmodified and nonlinear shock
structures for $\gamma_0 = 1.4$. The top panel shows $\gamux \, u(x)$,
where $\gamux = \{1 - [u(x)/c]^2\}^{-1/2}$, the middle panel is
momentum flux, and the bottom panel is the energy flux, all scaled to
far upstream values.  All curves are plotted versus $x$, where $x=0$
is the position of the sharp subshock.  A logarithmic scale is used
for $x < -10 \, \rgz$ and a linear scale is used for $x > -10\,\rgz$,
where $\rgz \equiv m_p u_0/(eB)$.\footnote{In these parallel
shock simulations where \alf\ wave production is ignored, the magnetic
field, $B$, has no other effect than setting arbitrary length and time
scales.}
In each panel, the solid curve is from an unmodified shock with $r
\simeq 3.6$ (see Figure~\ref{fig:ratio}), while the dashed curve is
the momentum and energy flux conserving result.

For the pitch angle diffusion model, where particles interact
elastically and isotropically in the local frame according to
equation~(\ref{eq:cosprime}) and the discussion following it,
particles are accelerated efficiently enough at the unmodified shock
that the momentum and energy fluxes are not conserved and rise well
above the allowed far upstream values. In order to conserve these
fluxes, the shock structure must be smoothed and the overall
compression ratio increased above the test-particle value.  The
computational scheme calculates this compression ratio and flux
conserving profile and the result is the dashed curve in the top panel
with the corresponding momentum and energy fluxes in the middle and
bottom panels.

The source of the non-conservation of momentum and energy is the
efficient acceleration of particles by the sharp flow speed
discontinuity. While the actual injection and acceleration efficiency
depends on the particular pitch angle diffusion model, once the
scattering assumptions are made, the kinematics determine the
injection and acceleration of the particles without additional
parameters.\footnote{Other input parameters, such as the Mach number
and $\pmax$ or the position of the FEB, influence the acceleration
efficiency, but these are ``environmental'' parameters rather than
parameters needed to describe the plasma interactions.}
Of course it would have been possible to make assumptions which
resulted in an acceleration efficiency low enough that momentum and
energy are approximately conserved without significantly smoothing the
shock structure or changing the compression ratio from the
test-particle value. For example, one could have only allowed shocked
particles above some Lorentz factor $\gaminj$ to recross into the
upstream region or arbitrarily restricted the number of particles that
recrossed into the upstream to a small fraction, $\fracinj$, of all
downstream particles. By making $\fracinj$ low enough or $\gaminj$
high enough one could make the efficiency as low as desired.
However, there are at least three reasons for not making such an
assumption. The first is that restricting the acceleration efficiency
requires additional parameters (i.e., $\gaminj$ and/or $\fracinj$) to
those needed to describe pitch angle
diffusion.\footnote{Alternatively, a far more complex model of the
plasma interactions can be postulated than done here, inevitably
requiring additional parameters \citep[e.g.,][]{Malkov98}.}
The second is that models with {\it inefficient} acceleration will not
help explain GRBs (or other objects) that require high efficiencies.
If \rel\ shocks are inefficient accelerators they are not very
interesting. If they are efficient, they will have a qualitative
resemblance to the results shown here even if, as is likely, the
actual plasma processes are far more complex than the simple model we
use.
A third, perhaps less compelling, reason is that identical
scattering assumptions as used here have been used for some time in
\nonrel\ shocks and shown to match both spacecraft observations
\citep[e.g.,][]{EMP90} and hybrid plasma simulations
\citep[e.g.,][]{EGBS93} of collisionless shocks.

The increase in compression ratio from $r \simeq 3.6$ to $\simeq 4.5$
shown in Figure~\ref{fig:profG2} comes about, in part, because the
particles escaping at $\pmax$ carry away momentum and energy fluxes
which make the shocked plasma more compressible.
This effect is countered to some degree by the fact that the
self-consistent shock produces a downstream distribution with a
smaller fraction of \rel\ particles than the \TP\ shock so that the
downstream ratio of specific heats $\Gamma > 4/3$. This tends to
produce a smaller compression ratio.
The escaping fluxes show up as a lowering of the dashed curves below
the far upstream values, as shown in the bottom two panels of
Figure~\ref{fig:profG2}, and amount to about 3\% of the far upstream
values for both momentum and energy.  Including the escaping fluxes,
the momentum and energy fluxes are conserved to within a few percent
of the far upstream values.
While the changes seen here are similar to those seen and discussed
for many years in efficient, \nonrel\ shock acceleration
\citep[see][and references therein]{BE99}, it must be noted that there
are no known analytic expressions relating escaping fluxes and
$\Gamma$ to $r$ in this \transrel\ regime. One obvious difference is
that for \nonrel\ shocks, the escaping momentum flux is generally much
less than the escaping energy flux (when both are measured as
fractions of incoming flux) since $\rho_e v_e^3/(\rho_0 u_0^3) \gg
\rho_e v_e^2/(\rho_0 u_0^2)$ when $v_e \gg u_0$
\citep[see][]{Ellison85}. Here, $v_e$ is the velocity of the escaping
particle.  In \rel\ shocks, $v_e \sim u_0 \sim c$ so the escaping
fluxes are about equal as shown in Figure~\ref{fig:profG2}.

\begin{figure}[!hbtp]              
\dopicture{0.8}{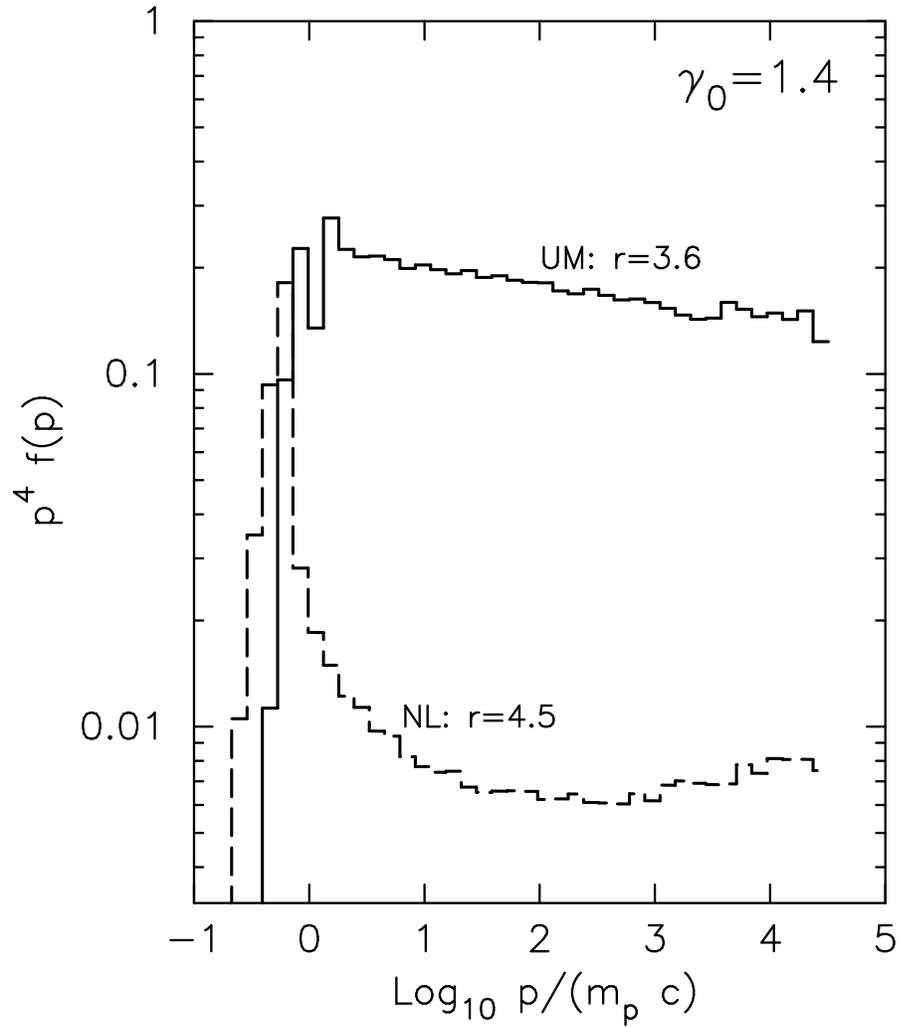}
\caption{Particle distributions, $p^4 f(p)$, for the shocks
shown in Figure~\ref{fig:profG2}. The nonlinear spectrum (dashed
curve) shows the distinctive concave shape seen in efficient
\nonrel\ shock acceleration, and has a greater fraction of low
momentum particles than the spectrum from the unmodified shock.
As in Figure~\ref{fig:specTP}, the spectra are calculated at the shock
in the shock frame and truncated with a $\pmax$. Unlike
Figure~\ref{fig:specTP}, the normalization here shows the actual
acceleration efficiency.
\label{fig:specG2}
}
\end{figure}

In Figure~\ref{fig:specG2} $p^4 f(p)$ is plotted for the $\gamma_0=1.4$
shock. The solid curve is from an unmodified (UM) shock and the dashed
curve is the nonlinear (NL) result.
The shock smoothing and increase in $r$ produce substantial
differences in the spectra even though both shocks have exactly the
same input conditions.
(i) The overall normalization of the NL
spectrum is less, reflecting the conservation of energy
flux. 
(ii) The NL result has the distinctive concave curvature seen in
\nonrel\ shocks stemming from the fact that higher momentum particles
have a longer upstream diffusion length and get accelerated more
efficiently than lower momentum particles in the smooth shock.
(iii) The slope at the highest momentum in the NL spectrum reflects the
overall compression ratio and is flatter than the TP spectrum because
$r$ is greater.
(iv) The ``thermal'' peak is shifted to lower momentum in the NL
result and contains a larger fraction of mildly \rel\ particles than
in the UM result, i.e., $\Gamma \simeq 1.41$ for the NL shock
compared to $\Gamma \simeq 1.36$ for the UM shock.


\subsection{Fully relativistic, nonlinear shock: $\gamma_0 = 10$}

\begin{figure}[!hbtp]              
\dopicture{0.8}{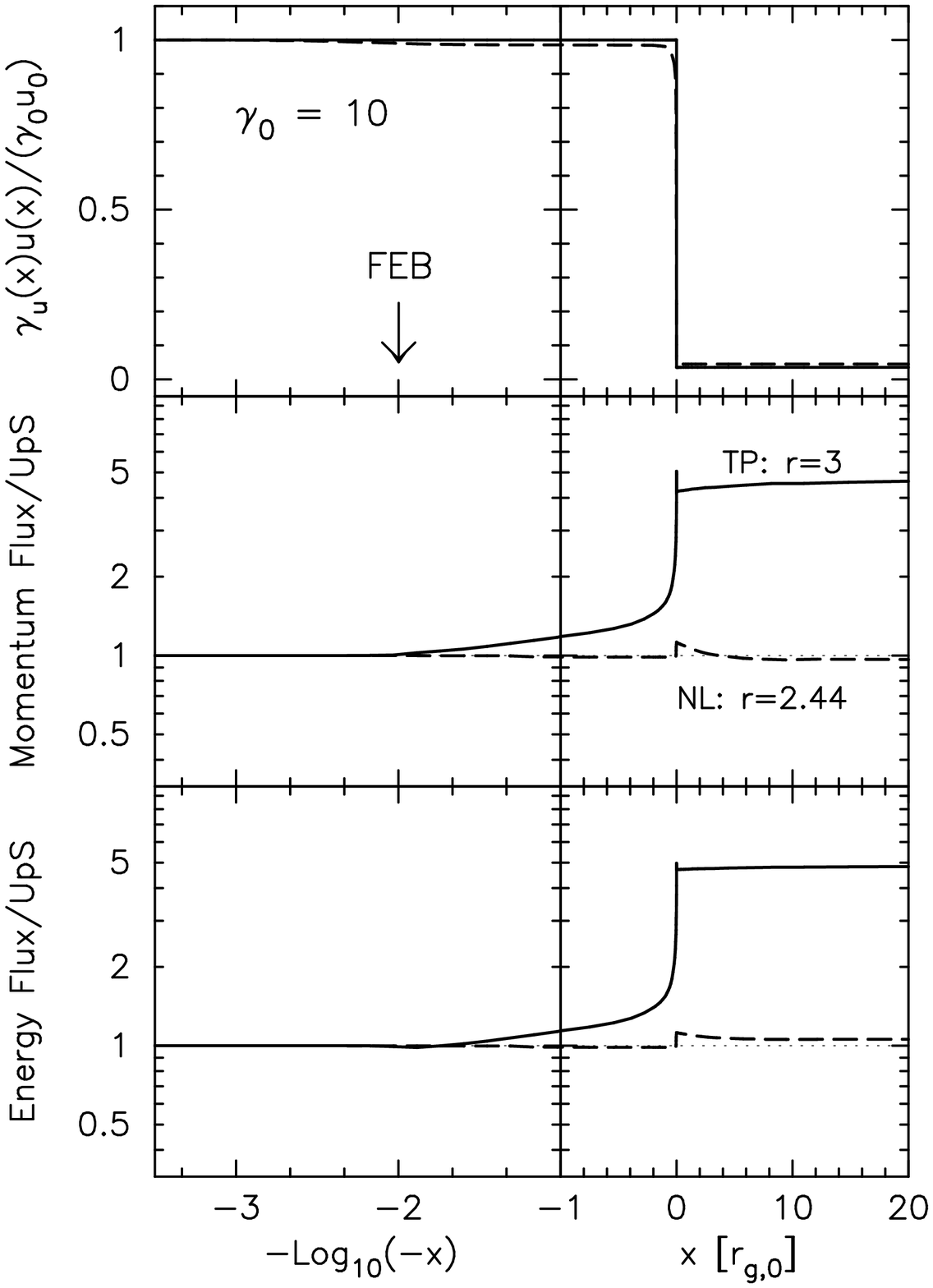} 
\caption{Unmodified (UM: solid curves) and nonlinear (NL: dashed
curves) shock profiles as in Figure~\ref{fig:profG2} for
$\gamma_0=10$. The acceleration is truncated by $\pmax = 10^4 \beta_0
\gamma_0 m_p c$ and $N/\Nmin \sim 10$ in both cases.
\label{fig:profG10}
}
\end{figure}

\begin{figure}[!hbtp]              
\dopicture{0.8}{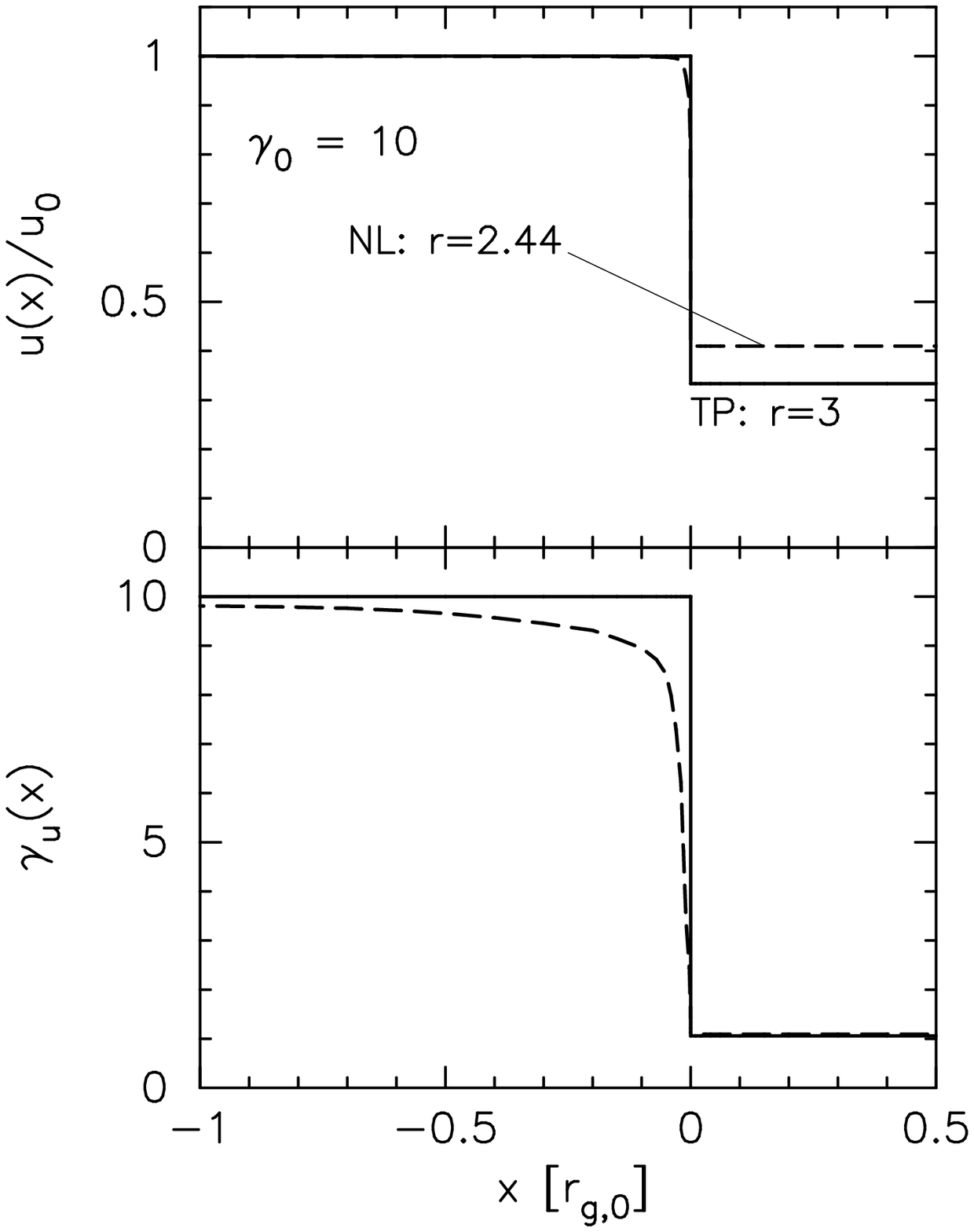}
\caption{The top panel is the flow speed at $x$ normalized to the
far upstream shock speed, $u_0$ versus $x$, for the $\gamma_0 = 10$
shocks shown in Figure~\ref{fig:profG10}. 
The bottom panel is the flow Lorentz factor,
$\gamma_u(x)$, versus $x$. In both panels, the solid curves are the
unmodified shock results with $r=3$ and the dashed curves are the
nonlinear results with $r\simeq 3.03 \pm 0.01$.
\label{fig:ugam}
}
\end{figure}

Figure~\ref{fig:profG10} shows results for $\gamma_0 = 10$.  In all
panels, the solid curves are for the unmodified shock, while the
dashed curves are for the flux conserving smoothed shock, both with
identical input parameters. As before, the acceleration is truncated
with a $\pmax = 10^4 \beta_0 \gamma_0 m_p c \simeq 9.3\xx{13}$ eV/c. In
Figure~\ref{fig:ugam}, $u(x)$ and $\gamux$ are plotted separately.
Even though the length-scale of the shock smoothing is only a few
$\rgz$ and
the change in $r=3.03\pm 0.01$ is small, they
bring the momentum and energy fluxes into balance from values a factor
of five too large in the unmodified shock. The less than 1\% difference
between $r=3.03\pm 0.01$ and the canonical value of 3 is significant; 
a self-consistent solution does not exist outside of this range.

As discussed above, both the shock structure and the overall
compression ratio may be modified when particle acceleration is
efficient. For mildly \rel\ shocks, such as the $\gamma_0 = 1.4$ example,
a unique solution can be determined directly from the conserved
momentum and energy fluxes. For larger $\gamma_0$'s however, the fluxes
are less sensitive to changes in the structure and compression ratio
and an additional constraint is needed to obtain a unique
solution. This constraint is provided by the \rel\ jump conditions 
for momentum and energy, i.e.,
\begin{equation}
\label{eq:mom_flx}
\gamma_{u0}^2 w_0 \frac{u_0^2}{c^2} + P_0 = 
\gamma_{u2}^2 w_2 \frac{u_2^2}{c^2} + P_2
\ ;
\end{equation}
\begin{equation}
\label{eq:en_flx}
\gamma_{u0}^2 w_0 u_0 = \gamma_{u2}^2 w_2 u_2
\ ,
\end{equation}
where $w = e + P$ is the enthalpy density, $e$ is the total energy
density, $P$ is the pressure, and it is assumed that escaping fluxes are
negligible.  The energy density and pressure are related
through a combination of the adiabatic equation of state and the
conservation of energy,
i.e., 
\begin{equation}
P = (\Gamma -1)(e - \rho c^2)
\ ,
\end{equation}
where $\rho c^2$ is the rest mass energy density and $\Gamma$ 
is, in special cases, 
the ratio of specific heats \citep[e.g.,][]{ER91}.
%
%
Dividing equation (\ref{eq:mom_flx}) by 
equation (\ref{eq:en_flx}) yields
\begin{equation}
\label{vel_1}
\frac{u_0}{c^2} + \frac{P_0}{\gamma_{u0}^2(e_0 + P_0)u_0} =
\frac{u_2}{c^2} + \frac{P_2}{\gamma_{u2}^2(e_2 + P_2)u_2}
\ ,
\end{equation}
or, in terms of beta's,
\begin{equation}
\label{betas1}
\beta_0 + \left(\frac{P_0}{e_0 + P_0}\right)
\frac{1 - \beta_0^2}{\beta_0} =
\beta_2 + \left(\frac{P_2}{e_2 + P_2}\right)
\frac{1 - \beta_2^2}{\beta_2}
\ .
\end{equation}

The second term on the left hand side is small compared to
$\beta_0$ for unshocked upstream particle temperatures less than
$10^9$K
and upstream particle densities less than $100$ cm$^{-3}$. Neglecting
this term, equation~(\ref{betas1}) becomes
\begin{equation}
\label{betas2}
\beta_0 = 
\beta_2 + \left(\frac{P_2}{e_2 + P_2}\right)
\frac{1 - \beta_2^2}{\beta_2}
\ ,
\end{equation}
or,
\begin{equation}
\label{betaquad}
e_2\beta_2^2 - \beta_0(e_2 + P_2)\beta_2 + P_2 = 0
\ ,
\end{equation}
which has the shock solution
\begin{equation}
\label{betasol}
\beta_2 = \beta_0/r = 
\frac{1}{2e_2}
\left [ \beta_0(e_2 + P_2) - 
\sqrt{\beta_0^2(e_2 + P_2) - 4e_2P_2} \right ]
\ .
\end{equation}
For $\gamma_0 \gtrsim 10$, $\beta_0 \simeq 1$ and $r \simeq
e_2/P_2$. 
Furthermore, if \ultrarel\ downstream particle speeds can be assumed
so that $e_2 \gg \rho_2c^2$, equation (\ref{betasol}) reduces to,
\begin{equation}
r \simeq \frac{1}{\Gamma_2 - 1}
\ .
\end{equation}
For $\Gamma_2=4/3$, $r \simeq 3$, the standard \ultrarel\ result.

\begin{figure}[!hbtp]              
\dopicture{0.8}{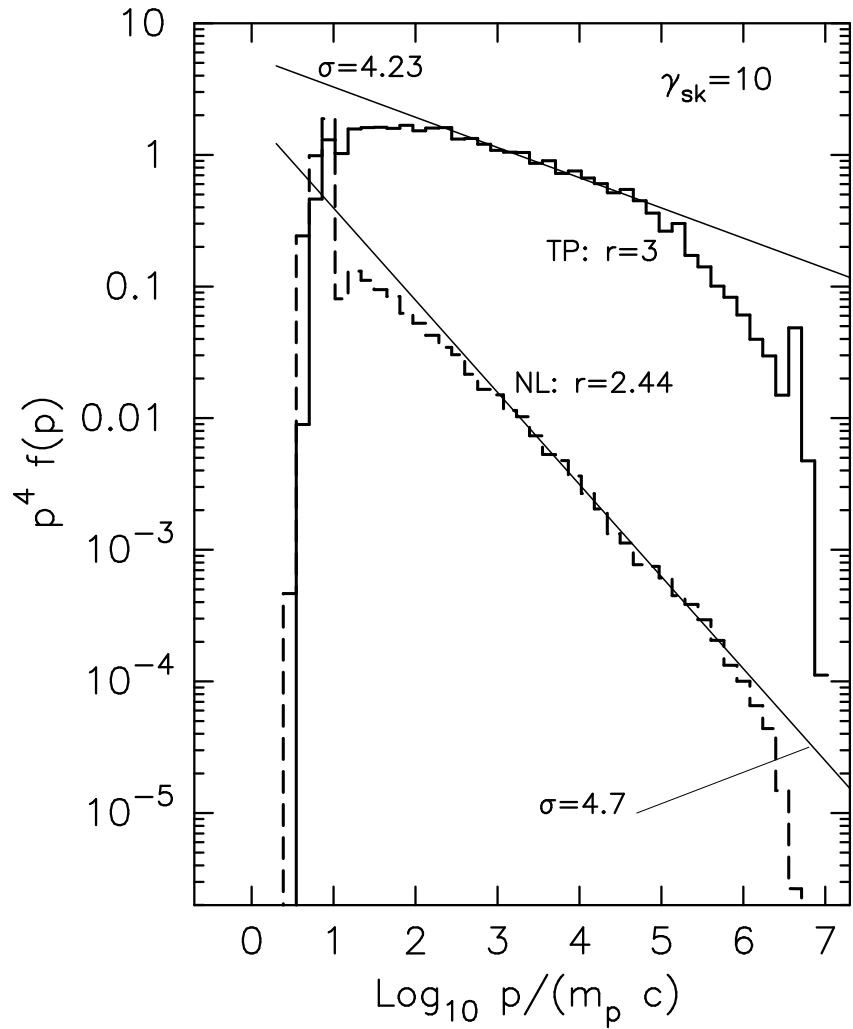}
\caption{Particle distributions, $p^4 f(p)$, for the shocks shown
in Figures~\ref{fig:profG10} and \ref{fig:ugam} with $\gamma_0 = 10$. The
spectra for the nonlinear (NL) and unmodified (UM) shocks are
labeled and both are calculated at $x=0$ in the shock frame.
The light-weight solid lines show spectral indexes, $\sigma=4.23$.
\label{fig:specG10}
}
\end{figure}

To obtain a unique shock solution for $\gamma_0 \gtrsim 3$, both the shock 
structure and the compression ratio, $\rMC$, are modified for the
Monte Carlo simulation, $P_2$ and $e_2$ are calculated from the
resultant downstream particle distributions, and $\rMC
\simeq r$ as determined from equation~(\ref{betasol}) are checked.
If $\rMC \ne r$, the shock structure and $\rMC$ are varied until 
a consistent solution is found.
For smaller $\gamma_0$'s, equation~(\ref{betasol}) doesn't apply because
escaping fluxes become significant. In this case, however, the changes
in the momentum and energy fluxes from changes in the shock structure
and $\rMC$ are large enough that a unique solution can be found easily,
as in the $\gamma_0 = 1.4$ example.

Despite the fact that $u(x)$ is modified on a fairly small
length scale, the resultant particle distribution function is changed
substantially, as indicated in Figure~\ref{fig:specG10}.
In this figure, the solid curve is from the unmodified shock and the
dashed curve is from the nonlinear shock, both having exactly the same
input conditions. The unmodified spectrum in Figure~\ref{fig:specG10} is
similar to that shown with a dashed curve in Figure~\ref{fig:specTP}
only now 
the far upstream plasma is taken to be a thermal gas at a temperature
of $10^6$ K rather than a delta function distribution of particles
with speeds, $\vinj = (2 \Einj/m_p)^{1/2}$, with $\Einj= 1$ MeV, as
was assumed for the example in Figure~\ref{fig:specTP}.
%

The shock smoothing has caused the low energy portion of the
distribution to steepen and the overall intensity to decrease,
reflecting the fact that the NL spectrum conserves momentum and energy
while the UM one doesn't.
The peaks in the two distributions at low momenta are also very
different, with the NL spectrum having a larger fraction of slower
particles than the UM one. These peaks result from the first shock
crossing where all particles receive a large energy gain.  In the UM
case, a far greater fraction of the accelerated downstream particles
are able to receive further energization by recrossing back into the
upstream region than in the NL shock.
The different speed distributions result in different $\Gamma$'s and
it was found, 
by directly calculating $\Gamma$ from the distributions, that
$\GamTP \simeq 1.34$ while $\GamNL \simeq 1.36$. 


\subsection{Acceleration efficiency}

\begin{figure}[!hbtp]              
\dopicture{0.8}{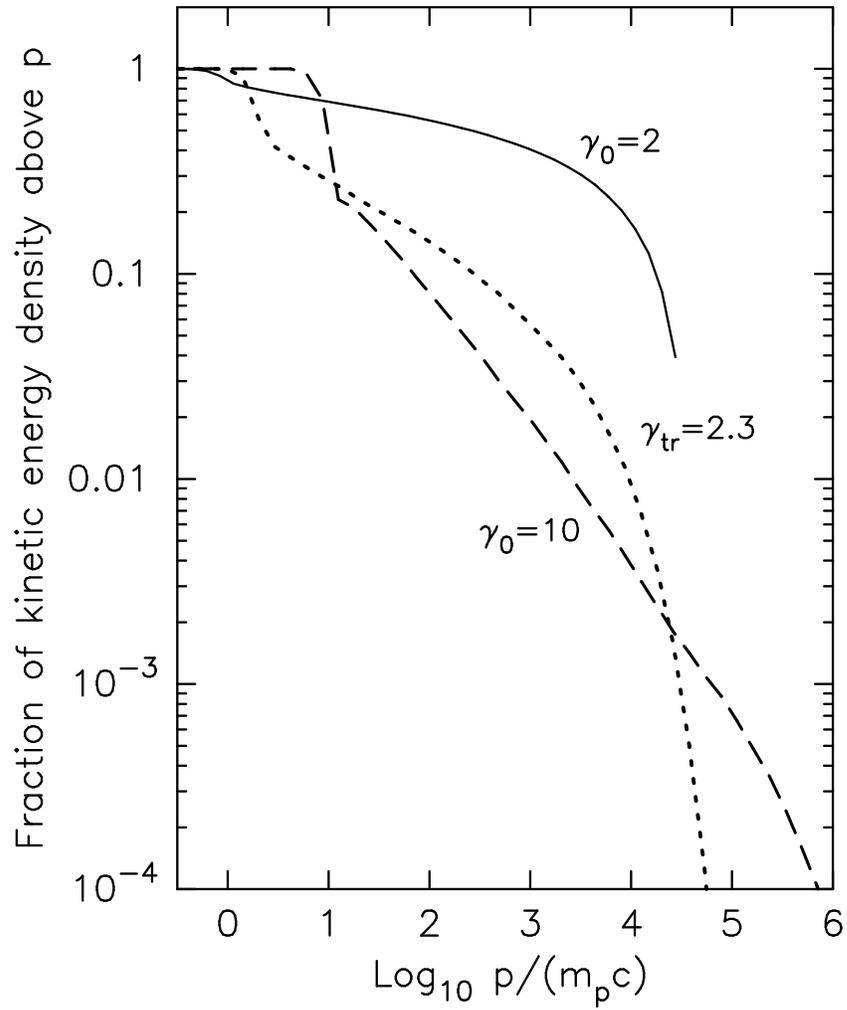}
\caption{Acceleration efficiency, $\EffP$, defined as the fraction
of total kinetic energy density above $p$ versus $p$,
is shown for two nonlinear cases.
The sharp dropoff 
at low momenta indicates the extent of the `thermal' peak.
\label{fig:eff}
}
\end{figure}

The absolute acceleration efficiency can be determined in the
self-consistent, nonlinear examples directly from the particle
distributions. In Figure~\ref{fig:eff}, $\EffP$ is plotted, i.e., the
fraction of kinetic energy density above a momentum $p$ versus $p$ for
the $\gamma_0=1.4$ and $\gamma_0=10$ examples. The fraction of kinetic
energy in the quasi-thermal part of the distribution can be determined
from the relatively sharp fall off of the distributions at low
momenta. This occurs near $m_p c$ for $\gamma_0=1.4$ and near $10\,
m_pc$ for $\gamma_0=10$. If the acceleration efficiency
is somewhat arbitrarily defined  
for these two examples to be $\EffOneFour$ and $\EffTen$, 
respectively, the result is $\EffOneFour \simeq 0.7$ and $\EffTen
\simeq 0.6$. Of course, the behavior of $\EffP$ depends on the
particle spectrum from which it is derived and thus $\EffP$ is
flatter for the $\gamma_0=1.4$ shock, with $r\simeq 4.5$, than for the
$\gamma_0=10$ shock, where $r\simeq 3.03\pm 0.01$.  Furthermore,
$\EffP$ depends strongly on $\pmax$ in the $\gamma_0 = 1.4$ shock where
particle escape plays an important role in determining
$r$. The maximum momentum has little influence for $\gamma = 10$ because
of the steep spectrum.
The fraction of kinetic energy density above $10^3 m_p c$ is about 0.3 for
$\gamma_0 = 1.4$ and about 0.1 for $\gamma_0 = 10$.
%

\section{Summary} 

Particles gain energy in collisionless shocks by scattering nearly
elastically off magnetic turbulence, back and forth between the
converging plasmas upstream and downstream from the shock.
While this basic shock acceleration physics is independent of the
speed of the shock, the mathematical modeling of the process depends
critically on whether or not the acceleration is efficient and whether
or not particle speeds, $v$, are large compared to the shock speed,
$u_0$.
The Monte Carlo techniques discussed here, which do not require $v \gg
u_0$, are well suited for the study of \rel\ shocks, and for any shock
where nonlinear effects are important and the energetic particles
originate as thermal particles in the unshocked plasma.  Except for
computational limits, these techniques allow calculations of efficient
particle acceleration in shocks of any Lorentz factor.

As a check of the \mc\ code, it was demonstrated that the well known
\TP\ power laws can be obtained in fully \nonrel\ and \ultrarel\ shocks
(Figures~\ref{fig:gyro} and \ref{fig:specTP}).  In \transrel\ shocks,
however, no such canonical results exist because the shock compression
ratio, $r$, depends on the upstream conditions in a nonlinear fashion,
even for \TP\ shocks where all effects of accelerated particles are
ignored \citep[e.g.,][]{ER91}.  it was shown how $r$, and the resulting
power law spectral index, $\sigma$, vary through the \transrel\ regime
in Figure~\ref{fig:ratio}, where $r$ has been determined by balancing
the momentum and energy fluxes across the shock in an iterative
process. 
%
These \TP\ results with self-consistently determined compression
ratios are in close agreement with the recent analytic results of 
\citet{KGGA2000} and \citet{Gallant2002}.

Despite the fact that relativistic shock theory has concentrated
almost exclusively on \TP\ acceleration, it is likely that \rel\
shocks are not \TP\ but inject and accelerate particles
efficiently. The reason is that regardless of the ambient far upstream
conditions, particles that are overtaken by an \ultrarel\ shock will
receive a large boost in energy $\sim \gamma_{\mathrm{rel}}$ 
in their first shock
crossing. Thus, virtually all of the particles in the downstream
region of an unmodified shock are strongly \rel\ with $v\sim c$ as
seen in the downstream frame.
The ability to overtake the shock from downstream and be further
accelerated depends only on the particle speed
(equation~\ref{eq:ProbRet}) and the presence of magnetic waves or
turbulence with sufficient power in wavelengths on the order of the
particle gyroradii to isotropize the downstream distributions. 
It is generally assumed that the necessary magnetic turbulence is
self-generated and if enough turbulence is generated to scatter high
momentum particles (with very low densities) that constitute a \TP\
power law, there should be enough generated to isotropize lower
momentum particles (which carry the bulk of the density). If
acceleration can occur at all, it is believed that it is likely to occur
efficiently making it necessary to calculate the shock structure and
particle acceleration self-consistently.
Furthermore, if \rel\ shock theory is to be applied to gamma-ray
bursts, where high conversion efficiencies are generally assumed,
nonlinear effects must be calculated.

When energetic particles are generated in sufficient numbers, the
conservation of momentum and energy requires that their backpressure
modify the shock structure. Two basic effects occur: a precursor
is formed when the upstream plasma is slowed by the backpressure of
the accelerated particles and the overall compression ratio changes
from the \TP\ value as a result of high energy particles escaping
and/or a change in the shocked plasma's ratio of specific heats,
$\Gamma$.
As indicated by the $\gamma_0 = 1.4$ example (Section 5.3.1), mildly \rel\
shocks act as \nonrel\ shocks, showing a dramatic weakening of the
subshock combined with a large increase in $r$
(Figure~\ref{fig:profG2}). These changes result in a particle
distribution which is both steeper than the \TP\ power law at low
momenta and flatter at high momenta (Figure~\ref{fig:specG2}).

In faster shocks (i.e., $\gamma_0 \gtrsim 3$), the initial \TP\ spectrum
is steep enough that particle escape is unimportant so only changes in
$\Gamma$ determine $r$ (equation~\ref{betasol}). In contrast to
\nonrel\ shocks where the production of \rel\ particles causes the
compression ratio to increase, it is shown that $r$ decreases smoothly to
$3$ as $\gamma_0$ increases and the fraction of fully \rel\ shocked
particles approaches one (Figure~\ref{fig:ratio}).

The most important result is that efficient, mildly \rel\ shocks do
not produce particle spectra close to the so-called `universal' power
law having $\sigma \sim 4.3$. This may be important for the
interpretation of gamma-ray bursts since the internal shocks assumed
responsible for converting the bulk kinetic energy of the fireball
into internal particle energy may be mildly \rel\ and the
external shocks, believed responsible for producing gamma-ray burst
afterglows, will always go through a mildly \rel\ phase
\citep[see][for a comprehensive review of gamma-ray bursts]{Piran99}.


\chapter{Characteristics of lepton acceleration in
relativistic parallel shocks}

The study of
lepton (primarily electron and positron) acceleration in the vicinity of
relativistic parallel shocks is part of a larger effort to investigate 
particle acceleration by relativistic oblique shocks using the equations
and techniques discussed in earlier chapters. Here, the magnetic field is
constrained to lie along the shock normal (a {\it parallel} shock),
but shock modification and
smoothing of the shock velocity profile by the
backpressure of energetic particles is allowed. 

As mentioned in Chapter 1,
it is believed that gamma-ray bursts
are dominated by leptons through pair production in the early stages
of the burst, with some unknown mix of baryons which is likely to be at 
a comparatively
low density internal to the forward shock, but possibly at a higher density
external to the shock. The gamma-ray burst spectra observed by
earth satellites are generally believed to be created by 
leptons accelerated to high energies by
the internal shocks\footnote{Refer to Figure \ref{fig:blast}}
for the initial intensity variations, and by the
forward shock to create the afterglow. 

Numerous papers \citep[e.g.,][]{Bell78b, BE87, LV96, SC02} 
have discussed the electron ``injection problem'' in which 
there may be problems in accelerating thermal electrons
(pre-acceleration) to an energy where they can be further accelerated by
the main shock.
Electrons and baryons, assuming equal temperatures,
have much different gyroradii resulting from the 
large mass difference between the two particles. The result is that
baryons can be effectively scattered by the turbulence generated by the
shock and self-generated \Alf\ waves because the  
wavelengths are of the
same order as the baryon gyroradii. 
It is not clear how electrons interact with the shock, but one assumption
that might be made is that electrons, with their much smaller gyroradii,
undergo little scattering
until they somehow increase their momentum enough to have 
gyroradii on the same order as the \Alf\ waves created by baryons
\citep{LV92}. 
Recall, from section 1.2 and also Chapter 4, that
elastic scattering is essential to first order Fermi
acceleration. 

\citet{ER91} assume that \Alf\ waves with sufficiently
short wavelengths exist and explore electron acceleration with various
injection energies, but a number of authors 
\citep[e.g.,][]{Gleev84, MDBKD97, SD00, SC02} have suggested other 
energy boosting mechanisms as a means of pre-accelerating electrons. For
example, they have suggested various types of resonances with the
magnetic field that would increase the kinetic energy of the thermal
electrons to the point where they can continue to accelerate by the
first order Fermi or diffusive process. \citet{DMCDD00} and
\citet{SC02} suggest the Buneman instability as an energy transfer mechanism. 

\citet{E81} and others have proposed that a discontinuous collisionless 
unmodified subshock exists along with the larger smoothed
shock, and the subshock continuously accelerates the lower momentum
electrons to a point where they can interact more effectively with
the smoothed shock.   
For relativistic
shocks, all particles will have roughly an average Lorentz 
factor of $\gamrel$ 
and a momentum  $p = \gamrel mv_{\mathrm{rel}}$ after the first 
shock crossing,  as discussed in section 3.2. Hence, 
downstream electrons, after the first crossing,
are more likely to have sufficiently long diffusion lengths
and may then begin to participate more effectively in the 
Fermi acceleration process, in particular, the subshock.

The diffusion length can be explained in the following way. Consider
first a {\it mean free path}, $\lambda = \eta r_g$, 
for collisionless shocks, 
defining $\lambda$ as the average distance required to deflect a charged
particle
through $90$ degrees by a magnetic field. Bohm diffusion (for the highest
magnetic turbulence and the strongest scattering) is assumed when
$\eta = 1$. $r_g = pc/(qB)$ is the gyroradius, where $p$ is the
momentum of the particle, $q$ is the unit electric charge, and
$B$ is the magnetic field strength.  
The {\it diffusion coefficient}, $\kappa = \lambda v/3$ 
\citep[a choice; for example, see][]{JE91},
is analogous to the standard diffusion coefficient used in collisional
plasmas. Given these parameters,
the {\it diffusion length}, $L = \kappa/u_0 \propto pv$, 
provides the overall length
scale and it is roughly proportional to particle energy. 

It should be emphasized that the actual 
plasma processes in the vicinity of the shock are
unknown and the resulting particle spectrum is dependent on the Monte Carlo
model of the shock. 
A current limitation to the Monte Carlo shock model and the results on
lepton characteristics presented 
here, is the absence of a cross-shock potential in combination with a
transverse magnetic field
\citep{ZR67, JE87} caused by the 
different mobilities of electrons and baryons
as the particles cross the shock. In the absence of external electric fields,
the potential difference across the shock is determined by the 
kinetic energy of the electrons, therefore the extent of the  
departure from electrical neutrality is on the order of a Debye length.
The cross-shock potential might have some  
influence on the dynamics of low-energy particles and the distribution
of energy between leptons and baryons. 
Despite this,
the Monte Carlo model, which explicitly simulates the particle
kinematics, has been very successful over
the past two decades in explaining observed spectra in 
nonrelativistic shocks. The more general
relativistic version of this model, discussed in detail in Chapter 5,
is in agreement with analytic results \citep{ED2002}
and should, at least, provide useful information
on the broad features of lepton spectral characteristics.

Electron kinematics in collisionless shocks are still poorly understood, and
the purpose here is not to address details of the electron kinematics  
directly,
but to understand the general characteristics of electron spectra, given 
shock speeds, subshock sizes and lepton densities as input parameters in
the Monte Carlo program that models the particle dynamics. 
Although the results presented here are a small part of the larger study,
it is hoped that they will contribute toward
the ultimate goal of creating computer generated
gamma-ray spectra that can be compared to actual observations. 
In this way, bounds may be placed on the gamma-ray burst model and 
increase our understanding of the physics of gamma-ray bursts.  

\citet{JE91} note that there is considerable evidence that shocks can
directly accelerate ambient thermal particles and, for this reason,
the assumptions made here are the same as those made by \citet{E81}.
It is assumed that in all
shock fronts, some discontinuous subshock can exist which has a small enough
width to allow thermal electrons to accelerate by the first-order Fermi
process. \citet{TK71} propose that very fast shocks have a very narrow
width, smaller than the
diffusion length of thermal electrons, which suggests that, even after 
the shock is modified and smoothed by the backpressure of accelerated
particles, thermal electrons
see at least some small subshock embedded in the main shock
(shown in Figure \ref{fig:subshock}), 
and that some non-negligible fraction of 
electrons should be capable of scattering downstream and return to
the (sub)shock for additional acceleration boosts until electrons have
a large enough diffusion length to interact with the entire smoothed shock. 

In this chapter,
lepton and baryon momentum flux distributions will be compared  as a function
of the subshock fraction and as a function of shock speed.
It will be shown how sensitive the lepton momentum distribution 
is to subshock size and to shock speeds over a range of
mildly relativistic shocks. Furthermore, leptons and baryons will be compared
as a function of lepton to baryon particle number density in terms of
injection efficiency and energy efficiency. The
equipartition density levels, where baryons and leptons
share energy equally, will be determined for leptons of various masses,
and it will be shown how a comparatively large 
particle density of leptons affect shock modification, as opposed to
shock modification by baryons. 

%
%
\section{Shock smoothing and the subshock}
\begin{figure}[!hbtp]              
\dopicture{0.55}{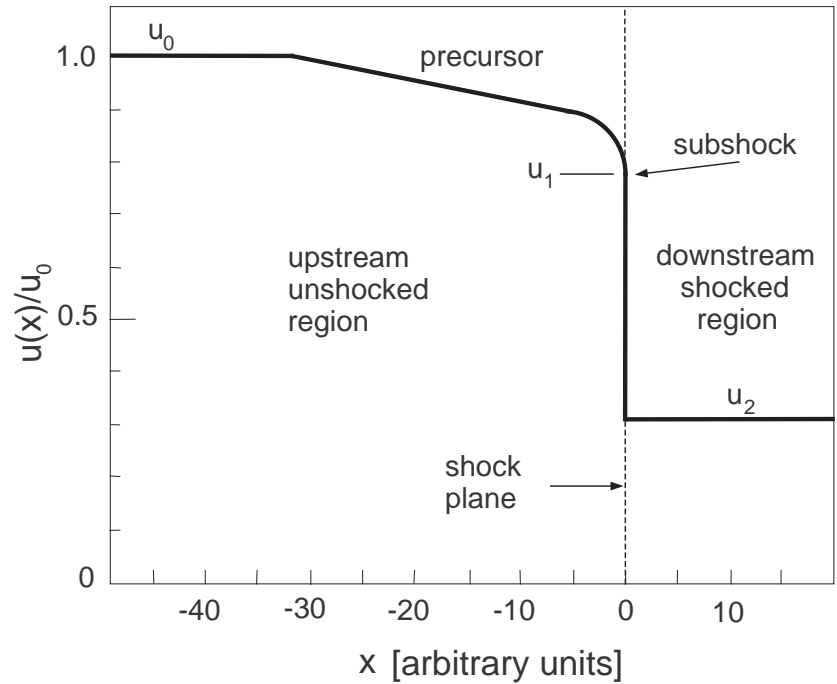}
\caption{Idealized 
modified shock velocity profile with the main shock having compression
ratio $r = u_0/u_2$,
showing the precursor
formed from the backreaction of accelerated particles, and the 
unmodified subshock
with compression ratio $r_s = u_1/u_2$. The velocities
$u_0$, $u_1$, and $u_2$ are all normalized to $u_0$, and flow from left
to right in the diagram. 
\label{fig:subshock}
}
\end{figure}

Nonlinear interaction of accelerated energetic particles on the shock 
velocity profile is well known 
\citep[e.g.,][]{Bell87, Drury83, JE91, BE99}. 
In Figure \ref{fig:subshock},
an idealized modified collisionless
shock is shown in the shock reference frame. 
Far upstream, the flow velocity is the original shock velocity, $u_0$. 
Energetic particles have diffusion lengths proportional 
to their momentum so the few
most energetic particles, carrying significant momentum, create a 
backpressure that modifies the shock on a larger scale.
The less energetic particles are more
numerous and modify the shock on a smaller scale.
In this way the precursor is formed as shown, down to $u_1$.
Below $u_1$, unaccelerated (thermal) particles have much smaller
energies and they interact with the shock on a much smaller scale.
The portion of the shock front between $u_1$ and $u_2$ is called the 
subshock; i.e., that portion of the shock that has maintained 
approximately its original narrow width.     

\begin{figure}[!hbtp]              
\dopicture{0.75}{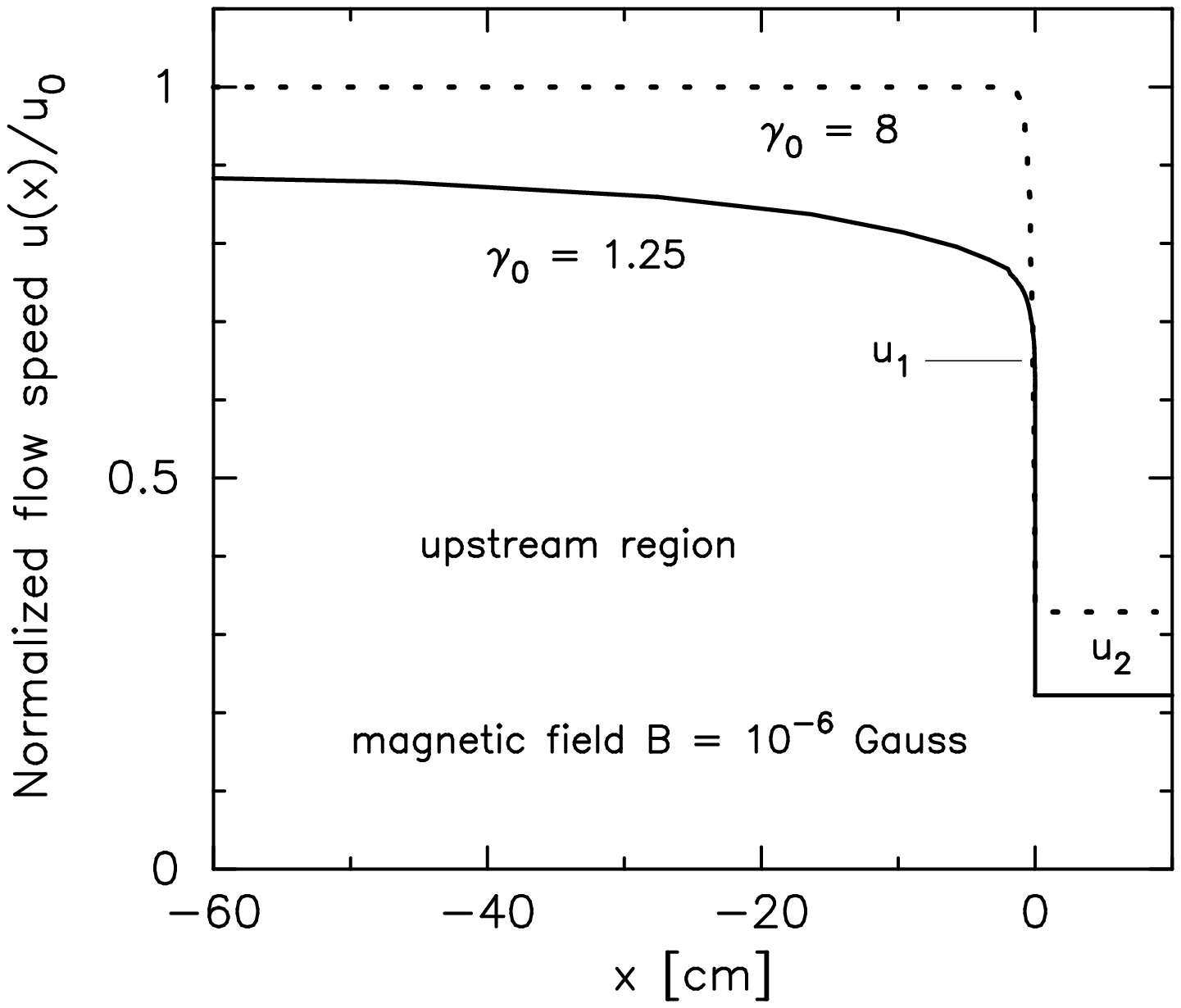}
\caption{Two modified shock velocity profiles with 
Lorentz factors of 1.25 and 8.0 as seen from the shock frame. The profile
sharpens as the shock becomes more relativistic. The subshock is at $u_1$.
The downstream flow
velocity, $u_2$, is determined by the compression ratio which varies,
depending on the speed of the shock. The shock is always modified in 
such a way as to conserve momentum and energy flux across the shock.
\label{fig:grid2}
}
\end{figure}

Slower shocks are modified on a larger scale, as shown in 
Figure \ref{fig:grid2}, because the  energetic, highly relativistic 
particles have speeds much greater than the shock speed and many particles
penetrate far into the upstream region, 
causing significant backpressure there. 
As the shock speed becomes more and more relativistic, downstream
energetic particles
scattering back towards the upstream are unlikely to cross the shock 
into the upstream region unless they lie within a critical angle 
$\theta_0 \sim 1/\gamma_0$, as discussed in detail in section 5.1. 
Particles with trajectories at larger angles are scattered back
downstream without crossing the shock 
(Refer to Figures \ref{fig:traj5000} and \ref{fig:traj09c}).
Hence, fewer energetic particles
cross the shock from downstream to upstream, the backpressure from
energetic particles is lower and the highly relativistic shocks are
modified on a smaller scale, as shown in Figure \ref{fig:grid2}
as the dotted line.   
As a result, the modified velocity profiles of highly relativistic 
shocks appear to be much sharper than the velocity profiles  
of mildly relativistic shocks when viewed with the same distance scale.

\begin{figure}[!hbtp]              
\dopicture{0.8}{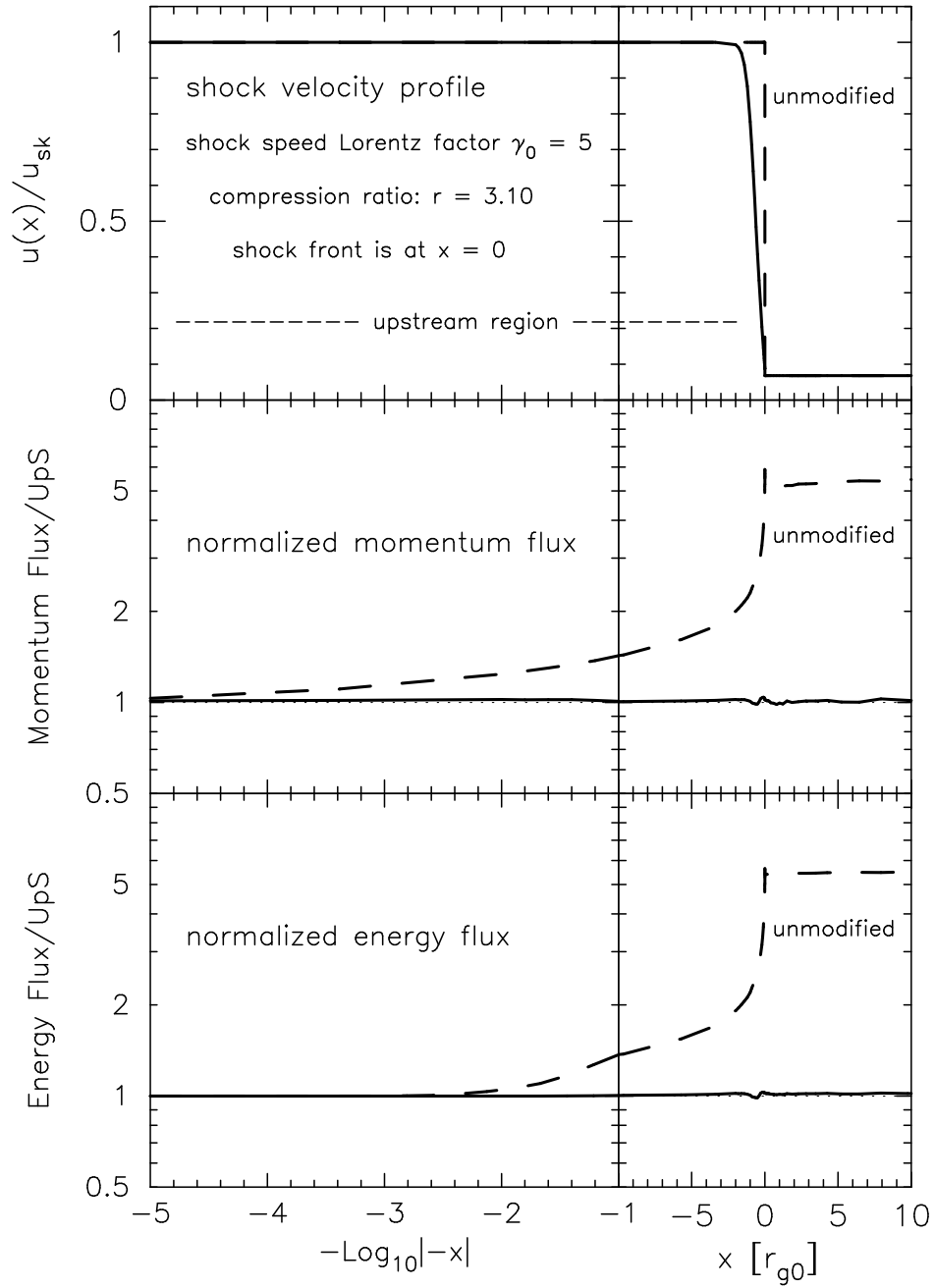}
\caption{The shock velocity profile for a modified shock (solid lines)
with no subshock, and Lorentz factor 
$\gamma_0 = 5$, plus the corresponding momentum and energy flux profiles
displayed on logarithmic scales.
These may be compared with momentum and energy flux profiles for the 
unmodified shock (dashed) where the fluxes are not conserved across the 
shock.
\label{fig:gridflx}
}
\end{figure}

The smoothing of the shock in the precursor region is such that
the momentum and energy flux are conserved across the shock as
shown in Figure \ref{fig:gridflx}. This plot was created by allowing a
large number of particles to run through the Monte Carlo model with
a shock speed Lorentz factor of 5 and compression ratio of 3.1. The 
model ran through a number of iterations, and after each
iteration the model compared the flux on both sides of the shock (actually
on both sides of every grid zone, both upstream and downstream) 
where the flux differences were 
noted and used to slightly adjust the shock velocity 
profile.
If the correct compression ratio is used, the momentum and energy flux 
differences will be negligible after the last iteration
and the fluxes will appear flat over the
entire range of $x$ values, shown as the solid lines
in the second and third frames of
Figure \ref{fig:gridflx}. The dashed lines, representing the first 
iteration and the unmodified shock, show how bad the mismatch is in
momentum and energy flux across the shock. Momentum and energy are not
conserved in an unmodified shock when real particles are accelerated.
Details of the shock modification method can be found in section 4.3.

%
 
\section{Sensitivity of lepton momentum
flux distributions to subshock size}

To determine the sensitivity of lepton injection and acceleration
characteristics to the size of the subshock in the shock velocity
profile for mildly relativistic shocks,
subshocks of various sizes were explicitly constructed 
with an input parameter as fractions of the
difference between the $u_0$ and $u_2$ flow speeds. 
After the subshock sizes were chosen, the
shocks were smoothed, either manually or with a combination of iterative
smoothing, to insure that momentum and energy fluxes were conserved,
as shown in Figure \ref{fig:gridflx}.

Manual smoothing means that a smoothing value was set as an input 
parameter. An algorithm in the Monte Carlo model uses this parameter
to create exponential smoothing of the shock velocity profile
into the upstream region.
The model 
generates momentum and energy flux values at every grid zone
based on the amount of smoothing introduced. After observing the resulting
flux profiles and their deviation from a straight line across the shock
(which means a deviation from the conservation of momentum and energy),   
the input smoothing value is adjusted and the model is rerun,
one iteration at a time,  until an
acceptable flatness of the momentum and energy flux profiles is achieved,
similar to the solid line in Figure \ref{fig:gridflx}. The manual method
is less efficient and sometimes not as accurate as the automatic iterative
method that the Monte Carlo model uses to adjust the velocity profile at
every grid zone. However, manual smoothing is necessary at high
Lorentz factors because the flux differences between grid zones become
too small and the automatic smoothing becomes less accurate. 
  
``Leptons'' 
(i.e., particles chosen  
to have mass $m_e/m_p = 0.1$, $0.01$ and $0.001$) 
were used for this study. 
Electrons
with their actual mass were not used due to computer
run time limitations. 
Referring to Figure \ref{fig:gyro_cyc01} and the related text,
the Monte Carlo model
simulates pitch angle scattering by altering the momentum direction
of a particle through
a small random angle $\delta\theta$
(which depends on the particle's gyroradius) 
in a small increment of time, $\delta t$.
The result is that lighter particles with smaller momenta
and smaller gyroradii require many more time steps to accomplish the 
same amount of scattering as heavier particles, roughly as the inverse
ratio of the masses. Hence, computer runs using very small mass 
particles become excessively long, given the present CPU speeds
that are available.     
Therefore, the strategy was to use three larger lepton masses to 
expeditiously generate data, and then 
extrapolate the results down to those for the mass of the electron. 
The results are straightforward and shown in Figure \ref{fig:mass_ss1}.   


\subsection{Lepton flux distributions} 
\begin{figure}[!hbtp]              
\dopicture{0.8}{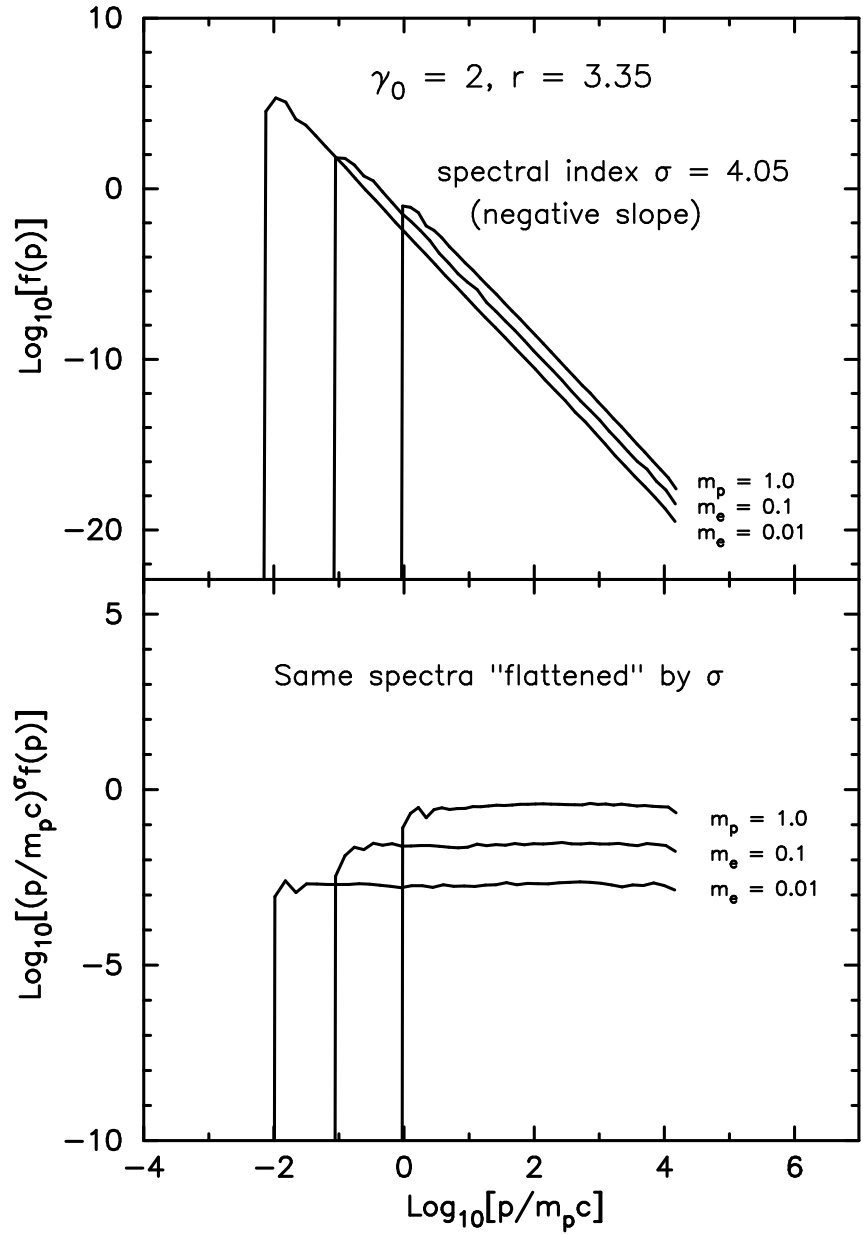}
\caption{Spectra for an unmodified relativistic ($\gamma_0 = 2$) shock.
The baryon and lepton spectra have the same slope, indicating that
all particles have a sufficient diffusion length to interact
with the shock.
The bottom frame shows the same spectra, but flattened by multiplying
the distribution function by normalized
momentum raised to the power of $\sigma$.
\label{fig:TP2f2s3}
}
\end{figure}

Momentum or energy flux distributions provide insight into how the
particles are interacting with the shock and undergoing Fermi acceleration.
The momentum flux distribution 
is measured by keeping track of the particles that cross a chosen
boundary with the momentum associated with each particle, similar to
the flux measurement at every grid zone as described in \citep{ER91}.
But, rather than simply summing
the total flux in the $x$ direction, the momentum for the flux distribution
is made to be omnidirectional with a cosine function and is
sorted into bins depending on the magnitude of the
momentum. The result is a plot of the
momentum flux distribution, $f(p)$, as a function of momentum, $p$. 
Figure \ref{fig:TP2f2s3}
shows a momentum flux distribution of particles with three different
masses accelerated by an unmodified shock. 
An unmodified shock implies that the 
subshock size is the same size as the main shock, and implies that
the entire shock width is much smaller than
the diffusion length of all particles in the system, therefore 
particles of all energies (determined by the input parameters)
can be accelerated by the shock. Also, because all particles see the same
entire shock width, the momentum fluxes all have the same slope. 

In the top frame of Figure \ref{fig:TP2f2s3}, the momentum flux
distributions $f(p) \propto p^{-\sigma}$, i.e., equation (\ref{eq:fp}),
of accelerated baryons and leptons are shown simply as the logarithms
of $f(p)$ vs $p$.  
Their slope is called the spectral index, $\sigma$. 
In the bottom frame,  
the distributions are flattened by multiplying 
them by $p^{\sigma}$, which allows more direct comparisons
of subtle variations in the momentum distributions.
The point here is to show the
concept for unmodified shocks, and the flattening will become more
useful for modified nonlinear shocks to be discussed later.   


As the particles scatter back and forth
across the shock, they pick up increments of momentum, 
depending on their pitch angle at the time of crossing. 
Recalling the discussion at the beginning of this chapter,
all particles pick up average energy 
of $\gamma_{\mathrm{rel}}mc^2$ and average
momentum of $\gamma_{\mathrm{rel}}mv_{\mathrm{rel}}$
at the first shock crossing, hence the
more massive particles initially have, on the average,
more momentum than less
massive particles in the ratio of their masses. 
For any given particle speed, the
probability of return of the particle
from downstream depends almost entirely on the
compression ratio across the 
shock\footnote{Note equation (\ref{bell25}) where $u_2$ 
is determined by the compression ratio $r$.}.
Since all particles have the same
probability of return for a given particle speed, 
the momentum distributions as a function of momentum will all 
have the same slope, as described in section 1.2. However,
the distribution flux levels for the species were set at the earliest
stages of acceleration, referring to equation (\ref{bell50}), and
they follow their corresponding flux levels for all higher momenta.    
For example, in the top frame of Figure \ref{fig:TP2f2s3}, the less massive
leptons (i.e., $m_e = 0.1$ and $0.01$) scatter and accelerate at lower
momenta with initially large momentum flux 
[again, equation (\ref{bell50})], but due to the constant spectral index,
there are significantly fewer leptons and correspondingly lower lepton
momentum flux at the initial momentum of accelerated baryons.
 

%

\begin{figure}[!hbtp]              
\dopicture{0.8}{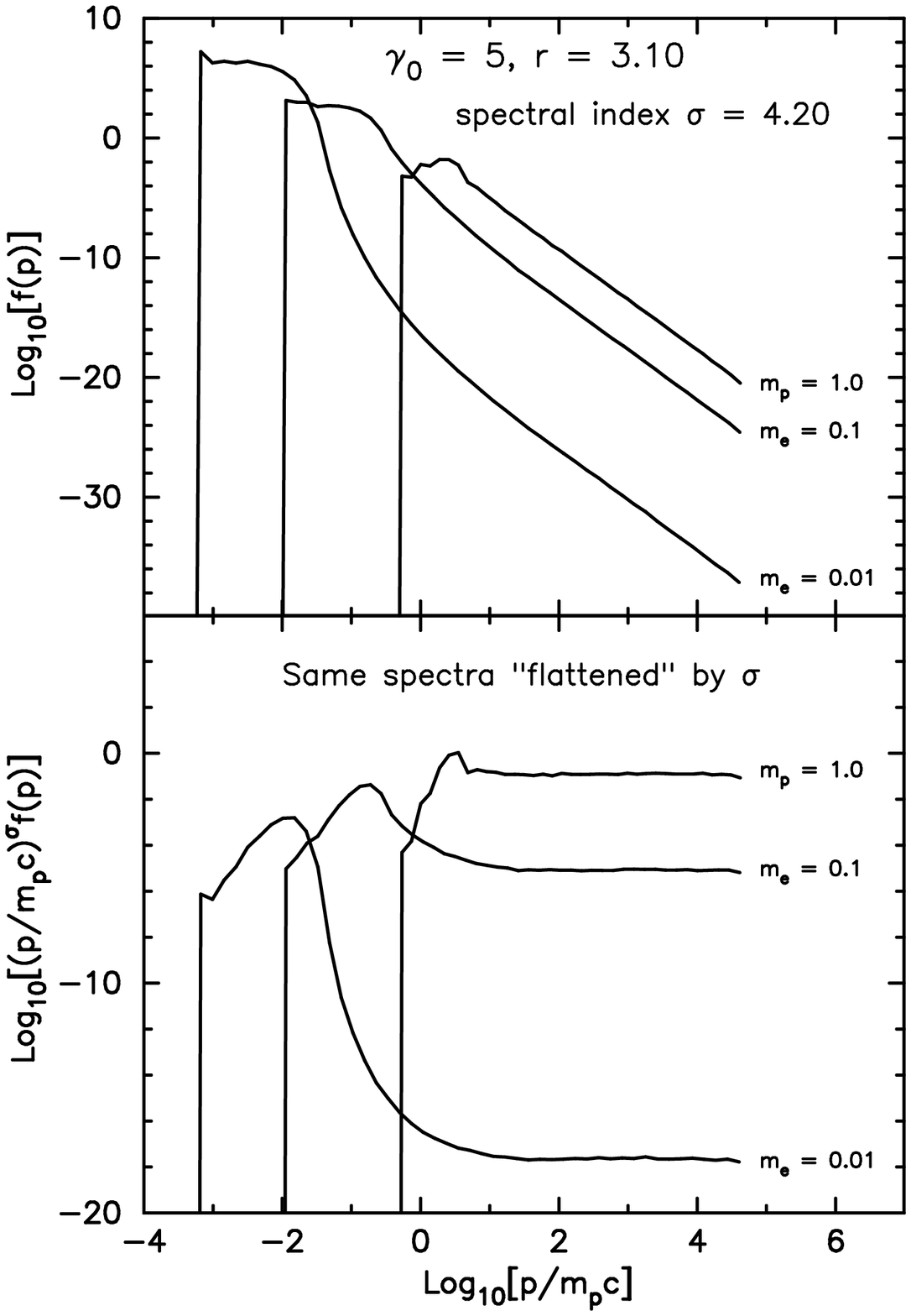}
\caption{Spectra for the modified relativistic ($\gamma_0 = 5$) shock
shown in Figure \ref{fig:gridflx}.
The baryon and lepton spectra develop the same slope, indicating that
the leptons have achieved a long enough diffusion length to interact
with the entire modified shock, but at greatly reduced numbers compared
to baryons.
The bottom frame shows the same spectra, but flattened by multiplying
the distribution function by normalized momentum 
raised to the power of $\sigma$.
\label{fig:g5f2s3}
}
\end{figure}

Referring to Figure \ref{fig:grid2},
shocks modified by the backpressure of energetic baryons are
smoothed and the size of the subshock is correspondingly reduced to $u_1$. 
As a consequence, the less massive leptons,
with initially less momentum and shorter
diffusion lengths than baryons, cannot interact effectively
with the entire shock because the shock is
smoothed on the longer scale of the baryons.    
As a consequence, far fewer leptons reach energies where their diffusion
lengths are comparable to those of baryons and the overall lepton flux 
is much lower. This effect can be observed in 
Figure \ref{fig:g5f2s3}. The leptons with mass $m_e = 0.01$ are affected
much more drastically than the leptons with mass $m_e = 0.1$ 
by the smoothed shock.

To determine the sensitivity of the lepton momentum flux to the size of
the subshock, subshocks of various sizes were explicitly constructed
with an input parameter, as described above, in
smoothed shocks with various Lorentz factors. The subshock size is
defined in the following way. 
Referring to Figure \ref{fig:subshock} again, the subshock fraction 
of the main shock is $f_{s} = (u_1 - u_2)/(u_0 - u_2)$. 
The compression ratio of the
subshock is $r_s = u_1/u_2$ and the compression ratio of the
main shock is $r = u_0/u_2$; 
hence, $f_{s} = (r_s - 1)/(r - 1)$, and $f_{s}$ is the normalized 
or relative subshock size. Subshock size can also be stated simply by just 
comparing compression ratios, $r_s/r$; however, this method gives a
value of $1$ when the subshock disappears at $u_1 = u_2$ and does not 
show the fraction of the main shock ascribed to the subshock.   

The velocity profiles containing constructed subshocks were adjusted to
conserve momentum and energy fluxes, using baryons only.
The resulting velocity profiles were then used
as a template to run the model with pairs of particle species, 
the first
being baryons with mass equal to $1.0$ and the second  being lighter
particles with mass of either $0.1$, $0.01$, or $0.001$, 
each with its own number
density of 1 particle per cubic centimeter. Little error was introduced
by this method because the baryons are primarily responsible for the
shock smoothing, and the smaller mass leptons were shown to have only a
minor effect on the overall shock velocity profile smoothing 
and the related momentum and energy flux conservation profiles
when the densities of baryons and leptons are approximately equal.
   
\begin{figure}[!hbtp]              
\dopicture{0.9}{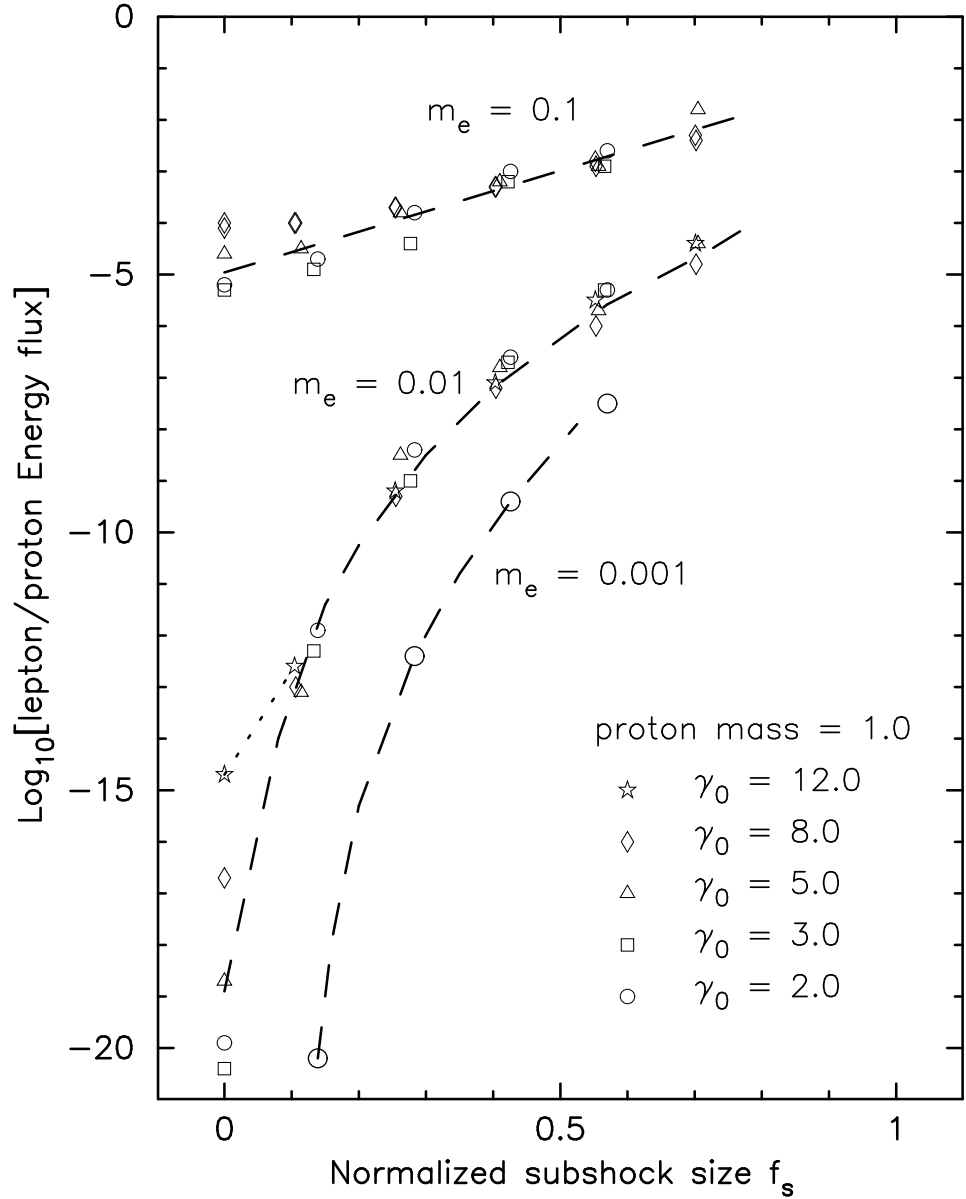}
\caption{The ratio of lepton to baryon momentum flux as a function of
subshock size is shown for particles with masses of $0.1$, $0.01$,
and $0.001$.
The lighter leptons are far more sensitive to subshock size than the heavier 
lepton, spanning 
fifteen orders of magnitude for small subshocks. 
Sensitivity to shock speed is low except for very small subshocks where
the flux is seen to increase with higher Lorentz factors.
\label{fig:sbshck1}
}
\end{figure}

\begin{figure}[!hbtp]              
\dopicture{0.9}{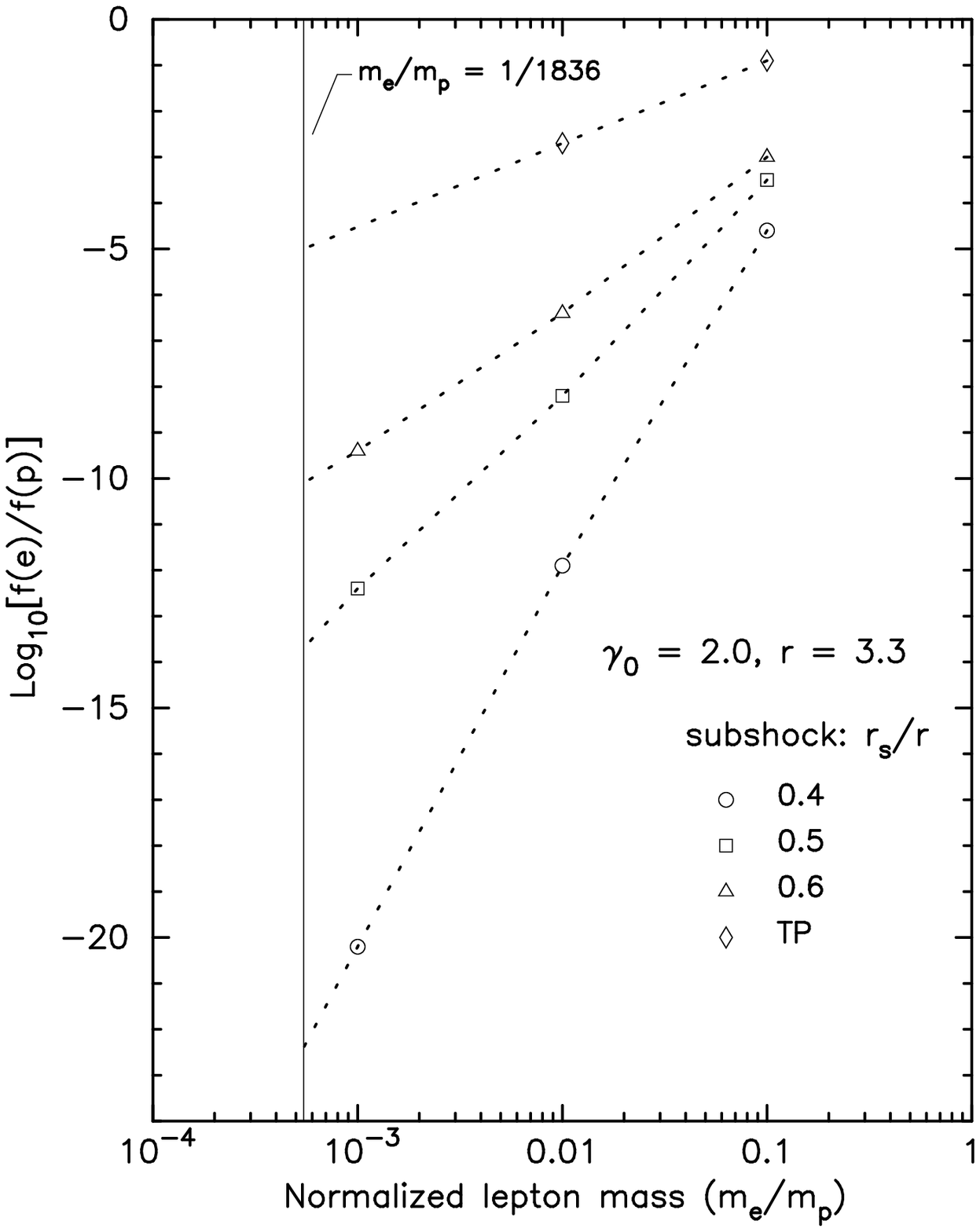}
\caption{Projections of lepton flux from larger 
lepton masses down to the mass of the electron show what flux to
expect for real electrons. Note that the subshock sizes are in terms of
the compression ratios, $r_s/r$.  
\label{fig:mass_ss1}
}
\end{figure}

After each run was completed,
the flux spectra was generated and measured from plots similar to the one
shown in the bottom frame of Figure \ref{fig:g5f2s3}. The vertical axes
of the plots show the logarithms of the fluxes, therefore the differences
of the measured values between baryons and the lighter leptons 
directly give the logarithm of the ratio of their flux distributions,     
$log_{10}[f(e)/f(p)]$. A number of subshock 
sizes were explored, from $f_s = 0$ to $f_s = 0.8$, 
for several shock speeds with Lorentz factors ranging from 
$\gamma_0 = 2$ to $\gamma_0 = 12$. The results, shown in 
Figure \ref{fig:sbshck1}, indicate a great increase in sensitivity to
subshock size as the particles become less massive, but show little 
sensitivity to shock speed for the larger subshock sizes
over the range studied here. 
For large subshock size, the flux
values approach those of an unmodified shock, such as those shown 
in Figure \ref{fig:TP2f2s3}.
For small subshock size, lepton flux is extremely small  
and, as shown in Figure \ref{fig:mass_ss1}, with 
projections to the electron rest mass, $m_e/m_p = 1836$, 
electrons would require a significant subshock size, or 
would need to be initially
pre-energized by some other means, in order to  
enter into the first-order Fermi
shock acceleration process. However, 
highly relativistic shocks tend to become very sharp on the scale of
baryon diffusion lengths, as shown in Figure \ref{fig:grid2}, 
and, for ultrarelativistic shocks, may be sharp on the scale of 
electron diffusion lengths and thereby have a large effective
subshock capable of accelerating electrons. This may be the reason that
there appears to be little dependence on shock speed 
when the constructed subshock is made large.

%

\subsection{Lepton injection efficiency}
\begin{figure}[!hbtp]              
\dopicture{0.9}{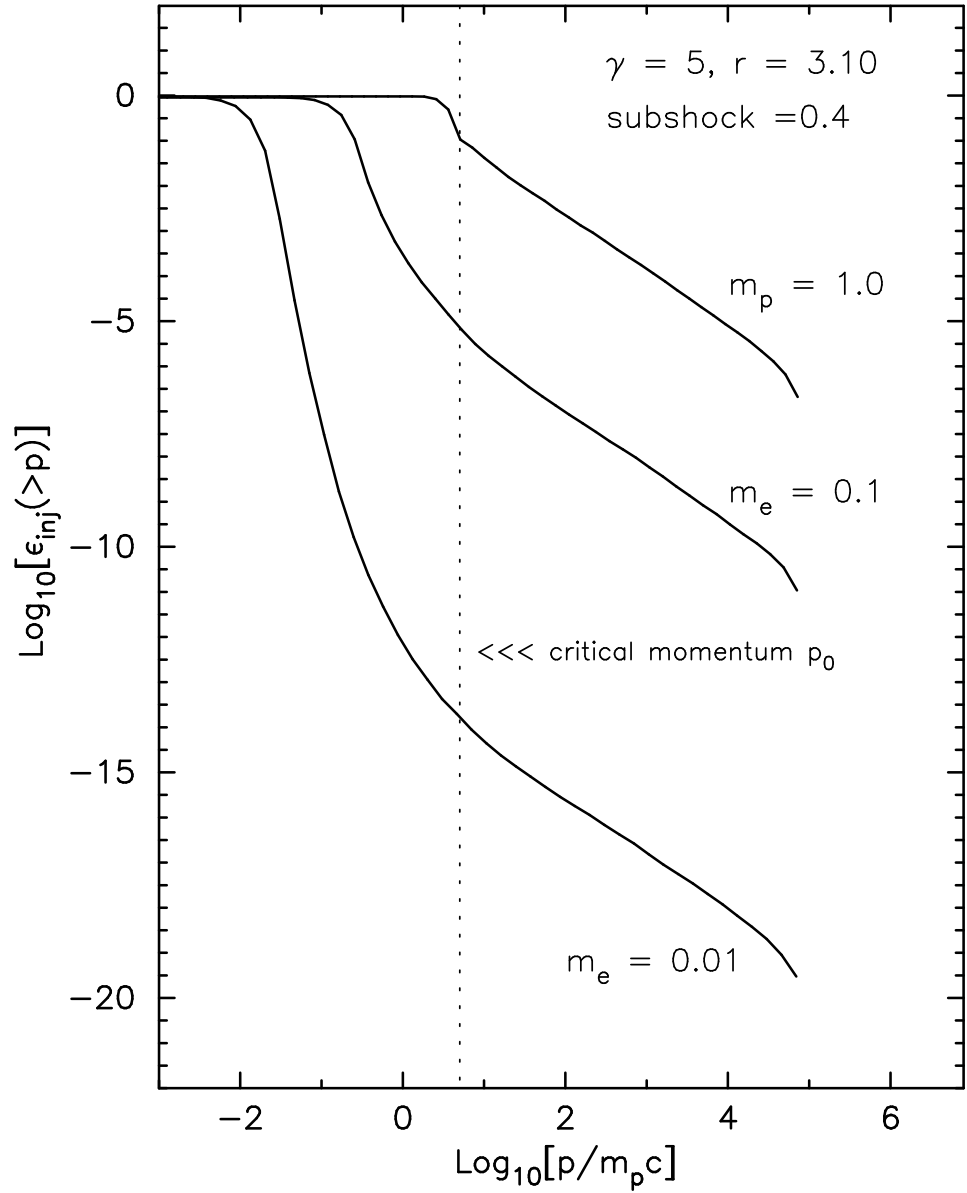}
\caption{Injection efficiency $\epsilon_{inj}(>p)$, 
i.e., the fractional
number density of particles above a given momentum, is plotted against
normalized momentum. The critical momentum, $p_0$, 
is chosen at the point where
baryons clearly obtain a power law dependence. Lepton power law 
dependence is also established around $p_0$.   
\label{fig:injeff}
}
\end{figure}

The lepton momentum flux is directly related to the injection efficiency
of leptons taking part
in the first-order Fermi acceleration process. Based on the concept 
expressed by
\citet{EBJ95}, the relative injection efficiency, 
$\epsilon_{\mathrm{inj}}(p > p_0)$,
can be defined as the number density of particles accelerated above a 
critical momentum, $p_0$,   
compared to the number density of particles $n_0$ flowing toward the shock,
or 
\begin{equation}
\label{eq:inj_eff}   
\epsilon_{inj}(> p_0) 
\equiv \bfrac{n_{\mathrm{inj}}(p > p_o)}{n_0}\ .
\end{equation}
The critical momentum, shown as the dotted line in 
Figure \ref{fig:injeff}, is approximately at 
$p/m_pc = 5$ (i.e., $\mathrm{log}_{10}[p/m_pc]\approx 0.7$). 
This is the momentum where the 
thermal distribution of baryons ends and
the power law distribution begins. 
The same momentum can be found
in the lower frame of Figure \ref{fig:g5f2s3}
at the beginning of the flat portion of the baryon flux distribution and 
where all three momentum flux distributions flatten to the 
same power law for this particular case of a modified
relativistic shock with Lorentz factor of $\gamma_0 = 5$. The efficiencies
of each type of particle was normalized to 1.0 to allow the injection
efficiency of each type to be measured against its own initial particle
density. This also allows direct comparisons between injection efficiencies
of the various particle types.  
 
The injection efficiencies for leptons and baryons may be determined as
follows. By using the critical momentum, $p_0$, as the defining point for
baryons, the absolute injection efficiency for baryons is read off the 
plot shown in Figure \ref{fig:injeff}. Then, using
the same momentum, for leptons, the relative injection efficiencies
for leptons are the differences (in logarithms) between the lepton 
efficiencies and the absolute baryon efficiency at $p_0$. 
   
\begin{figure}[!hbtp]              
\dopicture{0.9}{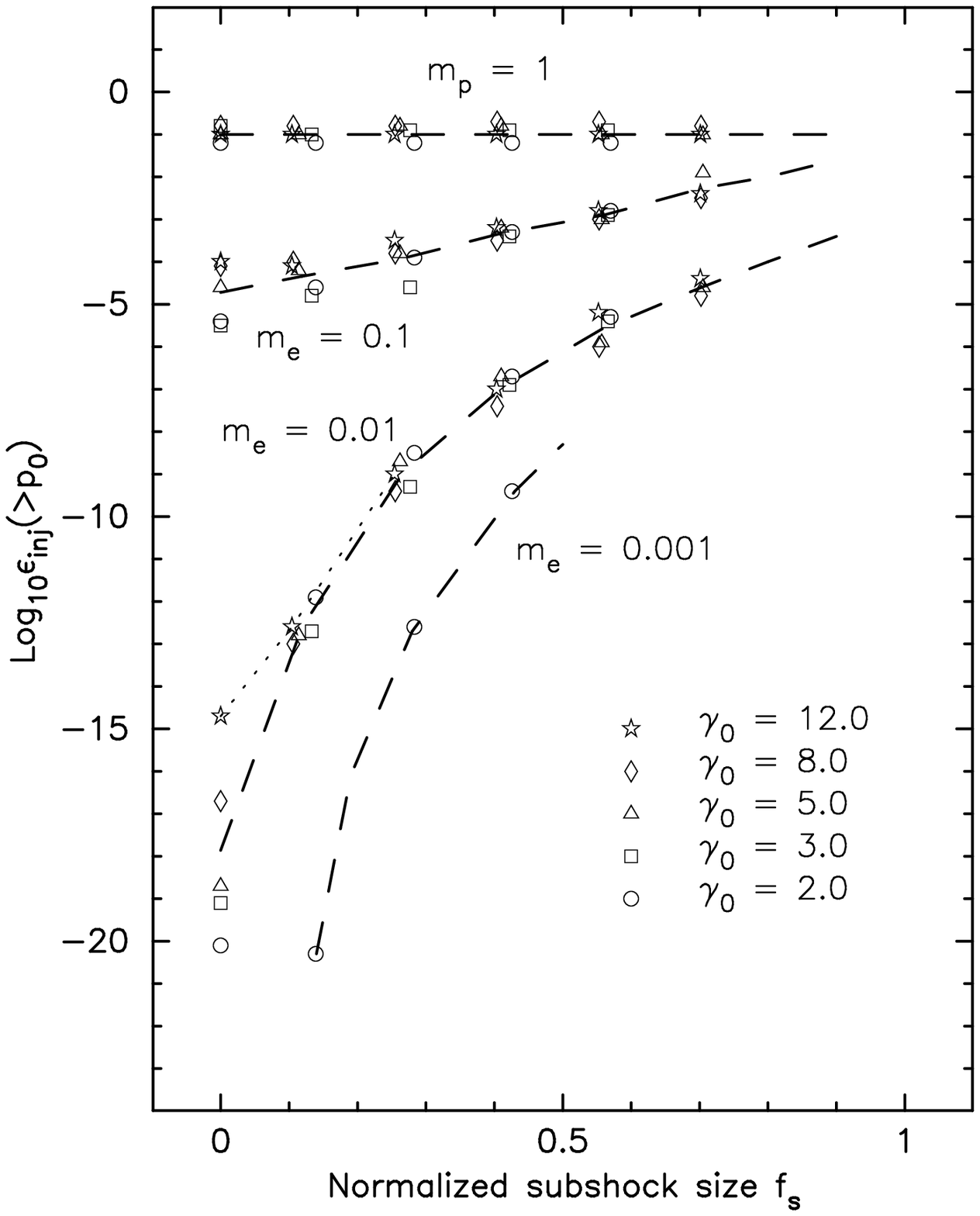}
\caption{The top curve shows the absolute injection efficiency of baryons
in modified shocks as a function of subshock size.
The curves for $m_e = 0.1$, $0.01$, and $0.001$ are the relative injection 
efficiencies normalized to the baryon injection efficiency. The 
injection efficiencies tend to be higher for higher $\gamma_0$ shocks
when the subshock is small.
\label{fig:injeff3}
}
\end{figure}

\begin{figure}[!hbtp]              
\dopicture{0.9}{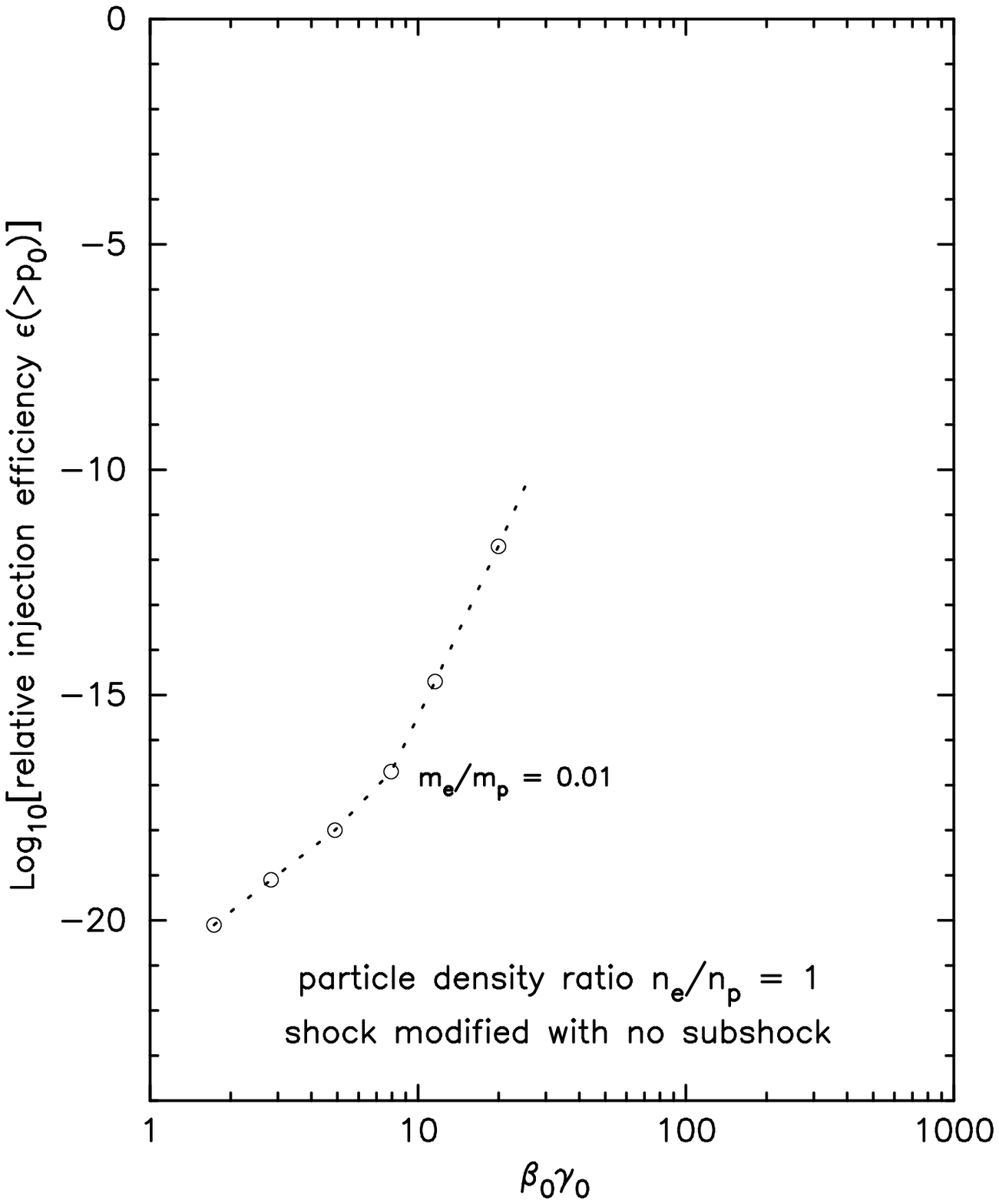}
\caption{Relative injection efficiency, 
normalized to baryon injection efficiency,
increases with shock speed if there is no subshock.
Leptons with mass  $m_e/m_p = 0.01$ 
may reach their maximum injection efficiency 
above a Lorentz factor of $\sim 100$.
\label{fig:injeff_gam}
}
\end{figure}

The critical momentum was found for
each subshock size and for each Lorentz factor to provide the absolute
injection efficiencies for baryons. Then, all of the relative injection
efficiencies were found for leptons with masses of 
$0.1$, $0.01$, and $0.001$ to create the injection efficiencies 
shown as a function of fractional subshock size, $f_s$,
in Figure \ref{fig:injeff3}. The lepton efficiencies are
normalized to the baryon injection efficiency, and the baryon efficiency 
is shown as the absolute efficiency on the same graph. The normalization
factor is the ratio of the baryon to lepton densities, taken in pairs. 
The injection efficiencies tend to follow the same pattern as did the
fluxes. This is expected because the fluxes above the critical momentum
generally follow the same power law for all particles. 
It may be noted in 
Figure \ref{fig:injeff3} that the lepton injection efficiencies tend toward 
those of test particles in unmodified shocks (i.e., where $f_s = 1$)
at the highest subshock sizes at all shock speeds used in this study.
Also note, at the
smallest subshock sizes, there is a greater dependency on shock speed.
This characteristic is shown more clearly in Figure \ref{fig:injeff_gam}.
This figure shows the relative injection efficiency
for leptons with masses $0.1$ and $0.01$ 
as a function of shock speed in shocks where
there was no explicit subshock established. For the 
shock speeds with Lorentz factors of $2$ - $20$, 
there is a dramatic improvement in lepton injection efficiency through
eight orders of magnitude in the $m_e/m_p = 0.01$ mass. 
Shock speed dependence occurs because, 
as the speed increases, the shock front 
becomes sharper on the length scale of the lepton 
and more of the lower energy leptons can enter
into the acceleration process. 
Once a particle reaches a momentum where its diffusion length is large
enough to interact with the entire smoothed shock width
(i.e., through the entire
precursor region in Figure \ref{fig:subshock}), 
the power law portion of the flux distribution
curve begins and the injection efficiency is established.
Baryons reach this 
``critical momentum'' point easily, compared to
leptons. Lepton momentum is lower than baryon momentum at the same speeds
due to their smaller masses. At shocks with low Lorentz factors,
the diffusion lengths of all leptons increase 
substantially after the first crossing, but not enough to interact with
the entire smoothed shock width. Leptons  require additional shock 
crossings to reach the required energies to interact with the entire
shock width and far fewer leptons can accomplish this, 
therefore the lepton injection efficiency is lower. 
%
The theoretical maximum relative injection efficiency
would occur when the shock front is
so narrow that the diffusion lengths of all particles, leptons included,
is greater than the shock width after the first crossing. By the same
argument presented earlier, the ratio of lepton to baryon momenta is
lower by the ratio of their masses after the first crossing. 
Then, given equal number densities
of leptons and baryons, the maximum relative 
injection efficiency of leptons depends only on the injection efficiency
power law through the relation
\begin{equation}
\label{eq:maxeff}
\mathrm{log}_{10}[\epsilon_{\mathrm{inj}}(\mathrm{max})] = 
\sigma_{\mathrm{inj}}\mathrm{log}_{10}\bfrac{m_e}{m_p}
\ .
\end{equation}
The average measured $\sigma_{\mathrm{inj}} \sim 1.3 \pm 0.1$.
Computer results for leptons
with mass $m_e/m_p = 0.01$ extend only to a Lorentz factor of $20$, 
therefore it is not clear how their injection efficiency behaves at higher
shock speeds, but based on the relationship above, the 
maximum relative
injection efficiency for these particles may be approximately
$10^{-2.6}$ at some Lorentz factor 
above $100$ in Figure \ref{fig:injeff_gam}. It appears that
the maximum relative injection efficiency (i.e., at the same momentum) 
of leptons will always be far below that of baryons when their 
number densities are equal.
 

\section{Characteristics of increased lepton densities}

Injection efficiencies, as defined in the previous section show the fraction
of particles that are incorporated into the Fermi acceleration process,
but a more important question is how much energy is carried by the 
particles.  Present gamma-ray burst models require leptons to 
carry a significant fraction of the total energy density of the burst, and
there are a number of ways 
this cound occur. One possibility is through rapid energy transfer from
baryons to leptons  by some unknown means. Another possibility is through a 
mechanism other than shocks to directly energize leptons; for example,
magnetic reconnection. The possibility that seems most likely, and the one
that will be investigated here, is that there are 
more leptons than baryons, and they are energized by diffusive shock
acceleration. One of the requirements in the
Fireball model of a gamma-ray burst is that the baryon mass fraction must be
low to allow the fireball to expand relativistically \citep[e.g.,][]{MR97}.
In addition, in the Fireball model large numbers of 
electron-positron pairs are produced. This suggests that the energy is
carried predominately by the leptons due to their large 
particle number density. In this section, the characteristics of energy
distributions are explored for various lepton to baryon particle
number density ratios.

\begin{figure}[!hbtp]              
\dopicture{0.8}{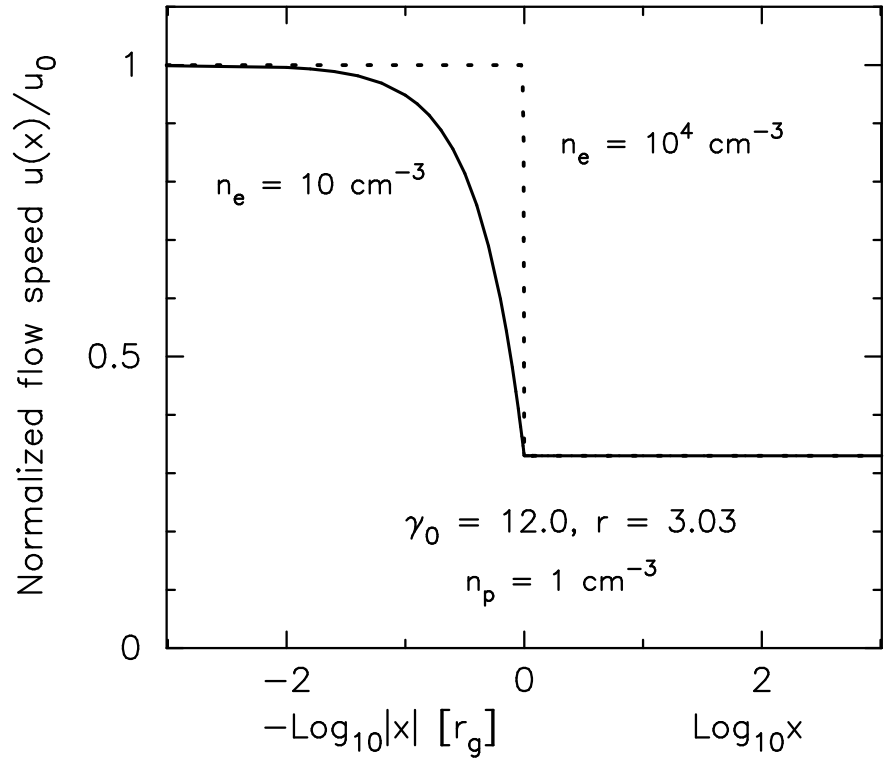}
\caption{Velocity profiles steepen with increased lepton/baryon particle
density ratios. The profile for a lepton density of 
$10^4\ \mathrm{cm}^{-3}$ with a lepton mass of $0.01m_p$ is
also smoothed, but on the scale of the lepton gyroradius and therefore
it appears very sharp to baryons, whose gyroradius is used here for units
on the horizontal axis.
\label{fig:densvprof}
}
\end{figure}

Clearly, the energy carried by particles 
is in the ratio of their masses for particles with the same velocity.
It was shown that when the leptons and baryons have equal particle
densities, the baryons,
and in particular the highly accelerated ones, carried significant
energy and exerted the predominant backpressure on the shock to modify
its velocity profile. As the lepton to baryon density ratio increases,
the situation changes. Large numbers of leptons, compared to baryons,
begin to exert their own backpressure
on the shock velocity profile. Referring to Figure \ref{fig:densvprof},
for a lepton density a little higher than the baryon density, say
for a case where $n_e/n_p \sim 10$,
the shock velocity profile is
still dominated by baryon momentum and the smoothing takes place on the
scale of the baryon's gyroradius. However, when leptons dominate, as is
the case of a lepton to baryon density ratio of $10^4$
and a lepton mass of $0.01m_p$, 
the smoothing results more from the backpressure of the 
energetic leptons and on the scale of the
lepton gyroradius, recalling that smoothing always occurs in such a way
as to conserve momentum and energy flux across the shock. The resulting
smoothed shock has a shock width very small on the scale of the baryon
diffusion length, and is on the order of the lepton
diffusion length. Leptons now interact with the shock as baryons did
when baryon momentum was dominant. Clearly, all baryons returning from
downstream see the entire shock width.  

\begin{figure}[!hbtp]              
\dopicture{0.9}{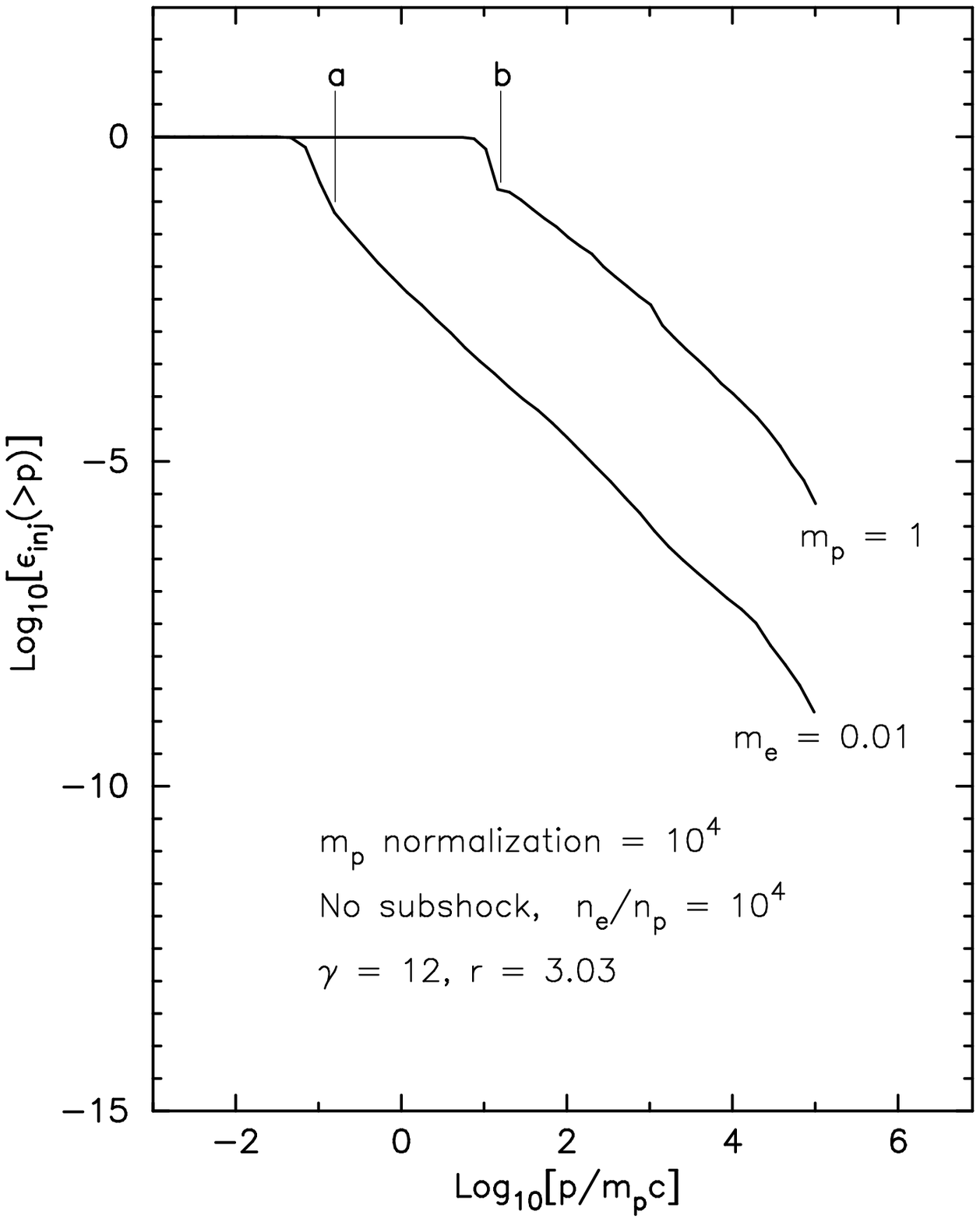}
\caption{Initial injection efficiencies are shown at $a$ for leptons 
and at $b$ for baryons after the
first shock crossing of each species in a scenario where leptons dominate 
baryons by $10^4$. Due to the narrow or sharp shock, modified primarily by
leptons, the injection efficiencies of all particles is relatively high
and leptons achieve a power law after the first crossing, similar to baryons.
Note that the baryon data were normalized by multiplying by the
$n_e/n_p$ ratio.
\label{fig:lepieff}
}
\end{figure}

In Figure \ref{fig:injeff}, the relative
injection efficiency was defined at the
onset of the power law for baryons, and a critical momentum $p_0$ was
used to define the lepton injection efficiency because the lepton power
law started at approximately that point. In other words, few leptons
received enough initial momentum after the first crossing to return for
additional acceleration and develop the power law. The leptons that
remained had to undergo repeated accelerations to develop enough momentum
with their related diffusion length to see the entire shock, hence the 
momentum flux of leptons in the power law region
was very low, due to both their small mass and
relatively few particles remaining, compared to baryons.
This definition of (relative) injection efficiency is somewhat misleading
when the shock front becomes narrow on the scale length of most particles, 
and this situation occurs 
when the lepton to baryon number density ratio increases beyond $10^3$. As
leptons begin to have more influence over the shock modification and carry
more total energy, and as the shock sharpens 
as shown in Figure \ref{fig:densvprof}, far more leptons can
enter into the shock acceleration process after the first shock crossing.
In other words, after the first shock crossing, the diffusion length
of a much larger population of leptons extends over the entire 
lepton-modified shock width
and the leptons start their power law, as shown in Figure \ref{fig:lepieff}
at point $a$. Point $a$ is determined by the ratio of the masses because
after the first crossing, as discussed earlier, the average speed for all
particles is $v_{\mathrm{rel}}$ and the momentum of a particle is 
$\gamma_{\mathrm{rel}}mv_{\mathrm{rel}}$. 

All baryons have a sufficient
diffusion length to see the entire shock and a large population of leptons
can now see the entire shock. Therefore, both baryons and leptons have the
same power law dependence after the first crossing and the relation,
$p_a/p_b = m_e/m_p$ results, as shown
at the top of Figure \ref{fig:lepieff}.
This point becomes well defined and takes the shape of the baryon injection
efficiency plot as the lepton
to baryon density ratio continues to increase. The {\it initial} particle  
injection efficiency (i.e., in those cases where the power law begins
after the first shock crossing) is defined as points $a$ and $b$. 
Defining the initial injection efficiency in this way indicates
that the lepton and baryon particle
injection efficiencies are nearly equal in 
Figure \ref{fig:lepieff}, but they occur at different momenta as mentioned
above. In the previous section
the mass difference was tied to the maximum relative injection efficiency
through equation (\ref{eq:maxeff}). Now the initial injection efficiency
is connected to the maximum relative injection efficiency through this
equation.

\begin{figure}[!hbtp]              
\dopicture{0.9}{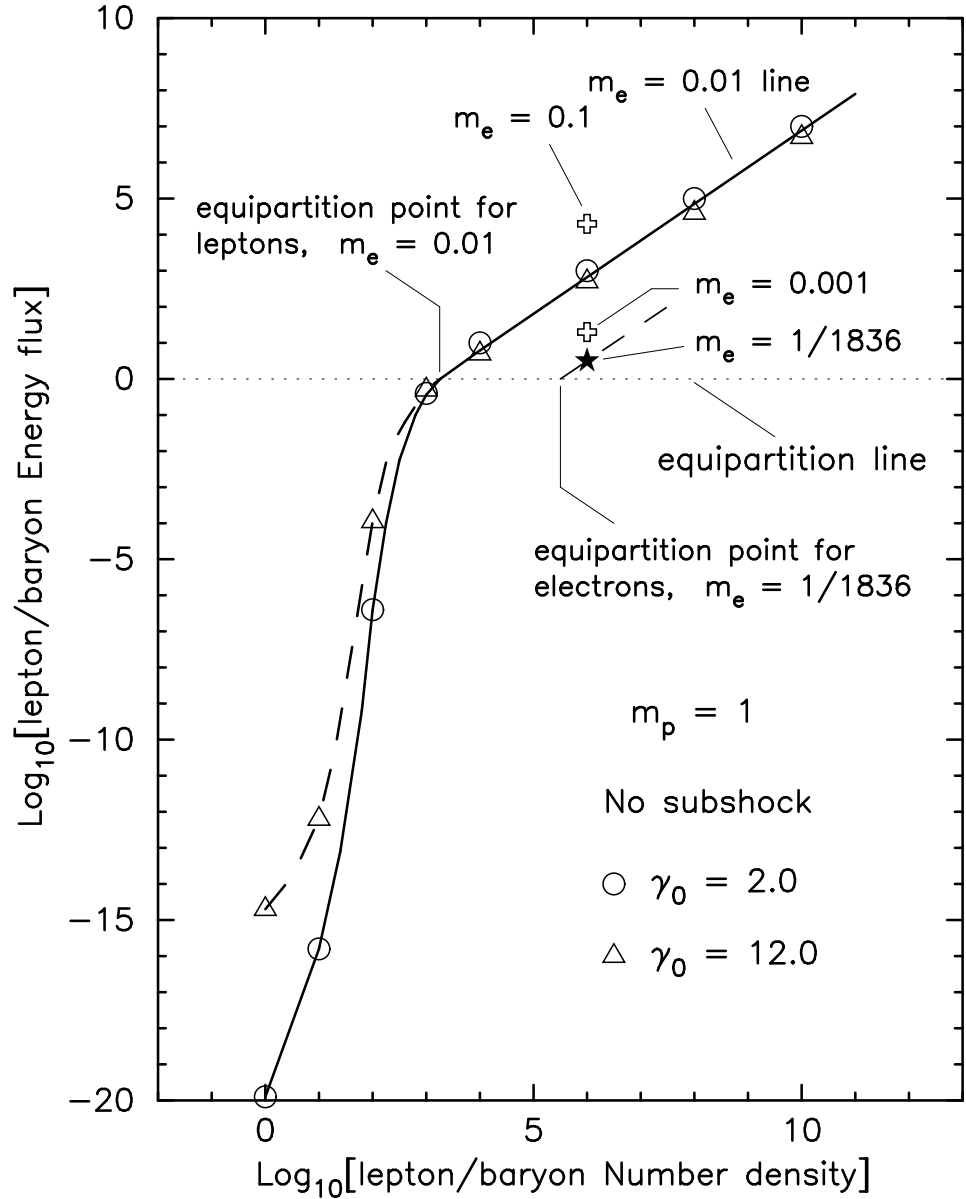}
\caption{Partitioning of energy among leptons and baryons as a function 
of the relative number density of leptons. The energy density is split
equally between leptons of mass $0.01m_p$
and baryons when lepton density exceeds baryon
density by approximately 1600. Leptons with the rest mass of the electron
reach equipartition at a density ratio of $3.2 \times 10^5$.
\label{fig:lepdens}
}
\end{figure}

In order to make comparisons between lepton and baryon flux characteristics,  
they need to be measured at the same momentum, therefore the
relative injection efficiency will be used as before,
by taking the differences in logarithms between the lepton and baryon
curves along the critical momentum line at $p_0$ (i.e., the momentum at
point $b$ in Figure \ref{fig:lepieff}). This was done for a
range of lepton to baryon particle number densities and for a range of
shock speeds, using a lepton to baryon mass ratio of $0.01$.
The results are expressed in Figure \ref{fig:lepdens} as the ratio
of lepton to baryon energy flux as a function of lepton to
baryon particle number density.
When the particle number density is low,
flux characteristics are dominated by baryons and the baryon-smoothed shock.  
As the leptons become more important and the shock smoothing begins to be
influenced by leptons as well as baryons, 
leptons begin to share more of
the energy density, and eventually there will be a point where the leptons
and baryons share energy equally. This point, on the equipartition line in 
Figure \ref{fig:lepdens}, appears to be at approximately
$n_e/n_p \sim 1600$ for a lepton mass of $0.01m_p$. In other words, 
as the lepton to baryon particle number density increases from 
$\sim 10^3$ to $\sim 10^4$, the spectral characteristics of shock
accelerated baryons and leptons with
mass $m_e = 0.01m_p$, and the shock profile itself,
makes a transition from those dominated by baryons 
to those dominated by leptons.
At higher number density ratios, both leptons and baryons behave the
same. Leptons dominate completely above a number density ratio of about
$10^4$ and, from that point on, the energy flux is simply proportional 
to the number density, hence the straight line (with a slope of 1). 

Additional energy fluxes were found
at a number density ratio of $10^6$ for
lepton masses of $0.1m_p$ and $0.001m_p$, and from these a projection was 
possible down to the electron mass of $(1/1836)m_p$. 
Straight lines, each with a slope of 1, 
can be projected back to the equipartition line and, if this is done
for the electron mass (designated as the solid star in 
Figure \ref{fig:lepdens}, it can be estimated that the equipartition point
for particles having the rest mass of an electron
is at a lepton to baryon particle number density ratio of
$3.2\times10^5$. The measurements were collected from shocks with
Lorentz factors of $2$ and $12$, the range of shock speeds in this study,
and the results appear to be independent of these shock speeds above the
equipartition line.

\begin{figure}[!hbtp]              
\dopicture{0.9}{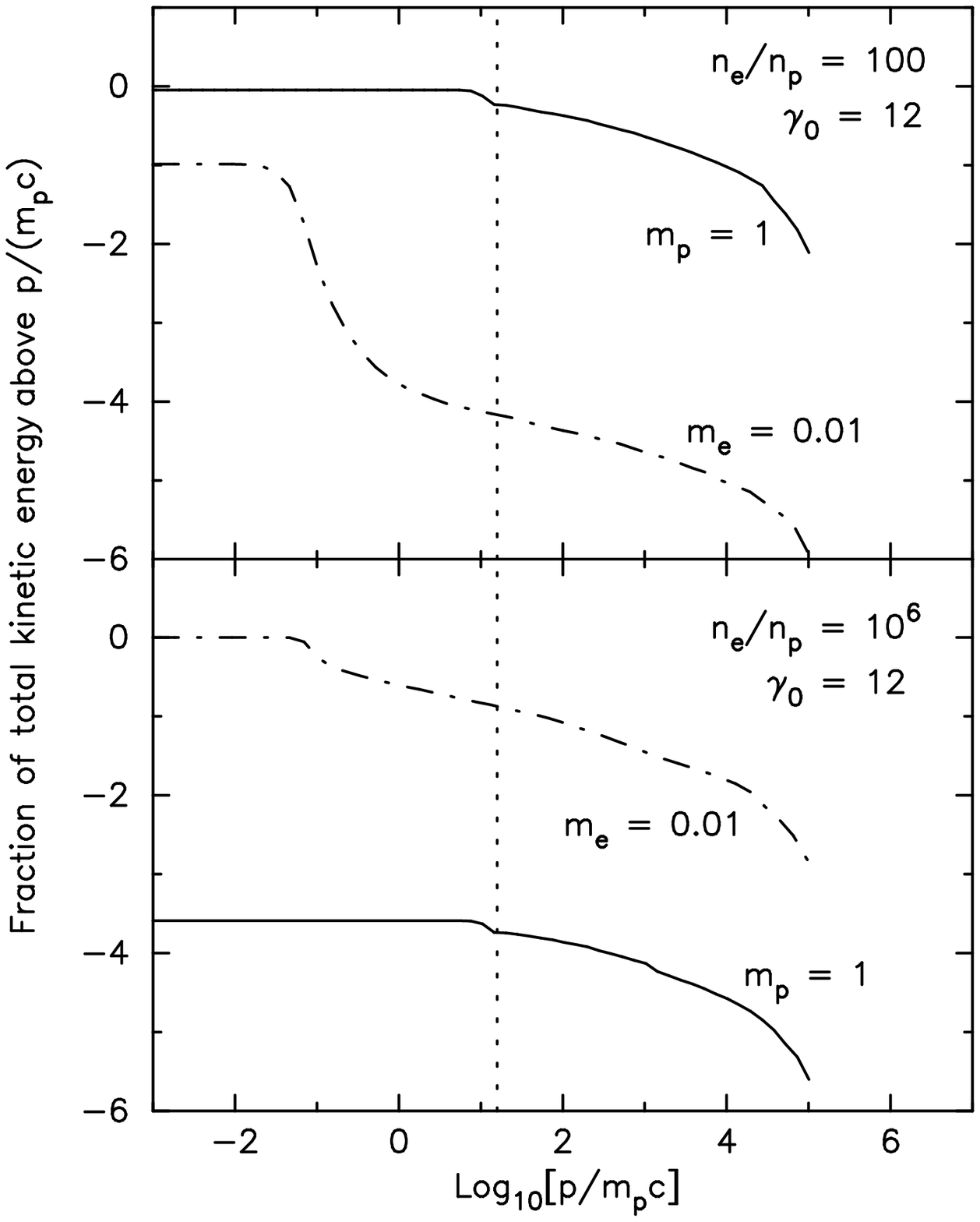}
\caption{Fraction of total kinetic energy
carried by baryons and leptons is shown for
two lepton to baryon density ratios. In the top frame where the lepton
density is 100, the kinetic energy carried by leptons is up to four orders
of magnitude below the faction carried by baryons. When the lepton density
increase to $10^6$, as shown in the lower frame,
the fraction of energy carried by leptons dominates.  
\label{fig:acceff2f}
}
\end{figure}

These concepts are expressed in a different way in Figure \ref{fig:acceff2f}.
Here, fractional total energy above momentum $p$
is shown for leptons and baryons for the case of relatively small lepton to
baryon density in the top frame, and for a large lepton to baryon density
in the lower frame. In the top frame, baryons carry the bulk of the total
energy, even though lower in particle number because they entered into the
acceleration process early and have a much higher density of accelerated
particles as a result of baryons still dominating the shock smoothing.
In the lower frame, leptons dominate the shock smoothing, also enter the 
acceleration process early, have a much higher density of accelerated 
particles (compared to leptons in the top frame) and, by their sheer
numbers, now carry the bulk of the energy. In Figure \ref{fig:acceff2f},
the curves are not normalized because the point here is to emphasized the
actual fraction of total energy that is associated with each particle type.

Noting the vertical dotted line in Figure \ref{fig:acceff2f}, another
parameter can be defined as {\it energy efficiency} (sometimes called 
acceleration efficiency as it was defined in section 5.3.3), similar to 
what was done in Figure \ref{fig:injeff_gam}. Energy efficiency is the
fraction of total energy remaining above momentum p. Since the dotted
line in Figure \ref{fig:acceff2f} is chosen, as before, at the onset of
the power law portion of the baryon curve, this momentum is also the 
critical momentum, $p_0$. Hence, energy efficiency (or acceleration 
efficiency) is
\begin{equation}
\label{eq:acceff}
\epsilon_{\mathrm{acc}}(p > p_0) \equiv 
\bfrac{\mathrm{fraction\ of\ total\ kinetic\ energy\ density\ above}\ p_0}
{\mathrm{total\ kinetic\ energy\ density}}   
\end{equation}
Using this definition, ratios of the energy efficiencies can be measured
for various lepton baryon density ratios and for various shock speeds,
similar to what was done for injection efficiencies. These results were
compiled for a lepton mass of $0.01$. When they are plotted on the same
graph in Figure \ref{fig:lepdens}, they overlay the $m_e = 0.01$ curve
exactly, and this is not surprising. Injection efficiency and
energy efficiency are closely related because injection efficiency is
concerned with the fraction of particles that get injected into the
acceleration process and energy efficiency is concerned with the fraction
of kinetic energy carried by these same particles. Since they were defined at
the point where each curve entered the (constant) power law dependence,
the most they could vary is by a constant. The constant is $1$. 
%

\section{Summary}

Leptons with various masses 
and baryons were subjected to a range of plane parallel modified shocks 
having Lorentz factors from $2$ up to $20$. 
The energy and momentum fluxes of these
particles were studied as a function of unmodified subshock size 
and particle number density to try to better understand the parameters
required for leptons to carry a significant portion of the energy in 
a gamma-ray burst. 

Assuming equal numbers of leptons and baryons, baryons dominate the shock
smoothing process and lepton injection efficiencies
are very sensitive to the size of any 
embedded subshock, the sensitivity increasing with decreasing lepton mass.

For subshock fractional sizes, $f_s$, above $0.4$, the injection efficiencies
of leptons appear to be insensitive to shock speeds over the range used in this
study. In the case of negligible
subshock, the injection efficiencies of leptons
are sensitive to the shock speed, and projections suggest that for shock
Lorentz factors on the order of $100$, the shock front may be sufficiently
narrow to increase the relative injection efficiency 
of leptons with mass $0.01m_p$ to a maximum of $10^{-2.6}$ of the baryon
injection efficiency.  
A lepton mass of $m_e \sim (1/1836)m_p$ may require a shock Lorentz factor
greater than $1000$
to reach a maximum relative injection efficiency
of about $10^{-4.2}$ that of baryons.  
The overall conclusion is that, given the assumptions in the Monte Carlo
model, leptons can never carry enough energy to
explain the spectra observed in gamma-ray bursts if the lepton to baryon
particle densities are of the same order, even for \ultrarel\ shock speeds.

The injection efficiencies of leptons and baryons were studied in the 
case where the lepton to baryon number density ratios were varied and
a lepton to baryon mass ratio of $0.01$ was used. 
It was
found that, as expected, for low number density ratios, baryons dominate
the shock smoothing and the lepton injection efficiencies and energy
efficiencies remained low. As the number density ratio increased to the
$10^3$ range, leptons began to have a strong influence on the modification
of the shock velocity profile, the shock became much sharper, and a much
larger population of leptons were injected into the shock process. 
The energy equipartition point of leptons 
with mass of $0.01m_p$ was at the particle number
density ratio of $1600$. Above a number density ratio of $10^4$, leptons
were found to dominate the shock smoothing process
and the energy flux was directly proportional to number density ratio.
Using this method, an extrapolation was made down to the mass of the electron,
and it was determined that the energy equipartition point for 
electrons (and positrons) in plane parallel shocks is at a lepton to baryon
number density ratio of $3.2\times10^5$. This result was independent of
shock speeds with Lorentz factors of $2$ to $12$ used in this study.
However, the energy equipartition point may be independent of shocks with 
any higher Lorentz factor for the following reason. 
When the lepton to baryon particle densities are
of the same order, the maximum lepton relative injection 
efficiency  occurs at \ultrarel\ shock speeds and it is far below baryon 
injection efficiency (at the same momentum). Hence, the only way the
lepton efficiency can dominate, under the same acceleration assumptions,
is by increasing the lepton to baryon
number density ratio. When the lepton density ratio is high enough that
leptons dominate the shock modification process,
shock speed no longer plays a roll in increasing injection efficiency.
The high number density ratio of leptons to baryons is required for two
reasons. The low mass of the leptons means that each lepton carries
far less energy than baryons, in proportion to their masses at the same
speeds, therefore large numbers of leptons are necessary to carry an
equal portion of the total energy. More importantly, there must be a
sufficient lepton density to dominate the shock smoothing so that larger
numbers of leptons can be injected into the shock acceleration process.

The primary finding in this study was that for plane modified shocks
with the magnetic field parallel to the shock normal, and with shock
Lorentz factors in the $2$ to $12$ range, the particle number density
ratio between leptons and baryons must be above
$3.2\times10^5$ for leptons with the rest mass of the electron
to carry sufficient kinetic energy in gamma-ray bursts and account for the
observed GRB spectra. The 
energy equipartition point appears to be independent of shock speed.
%
%
\chapter{Conclusions}

Chapter summaries can be found at the end of each chapter, 
but the accomplishments and
main conclusions of the entire dissertation are summarized here.
Despite the fact that relativistic shock theory has concentrated
almost exclusively on \TP\ acceleration, it is likely that \rel\
shocks are not \TP\ but inject and accelerate particles
efficiently. The reason is that regardless of the ambient far upstream
conditions, all particles that are overtaken by a \rel\ shock
will receive a large boost in energy 
$\Delta E \sim \gamma_{\mathrm{rel}} m c^2$ in
their first shock crossing. Thus, virtually all of the particles in
the downstream region of an unmodified shock are strongly \rel\ with
$v\sim c$.
The ability to overtake the shock from downstream and be further
accelerated depends only on the particle speed
(Eq.~\ref{eq:ProbRet}), the presence of magnetic waves or
turbulence with sufficient power in wavelengths on the order of the
particle gyroradii to isotropize the downstream distributions, and the 
pitch angle of the returning particle (Eq.~\ref{eq:ret_pitch}).
It is generally assumed that the necessary magnetic turbulence is
self-generated and if enough turbulence is generated to scatter high
momentum particles (with very low densities) that constitute a \TP\
power law, there should be enough generated to isotropize lower
momentum particles (which carry the bulk of the density). If
acceleration can occur at all, it is likely to occur
efficiently making it necessary to calculate the shock structure and
particle acceleration self-consistently.
Furthermore, if \rel\ shock theory is to be applied to gamma-ray
bursts, where high conversion efficiencies are generally assumed,
nonlinear effects must be calculated.

When energetic particles are generated in sufficient numbers, the
conservation of momentum and energy requires that their backpressure
modify the shock structure. Two basic effects occur: a precursor
is formed when the upstream plasma is slowed by the backpressure of
the accelerated particles and the overall compression ratio changes
from the \TP\ value as a result of high energy particles escaping
and/or a change in the shocked plasma's 
adiabatic index $\Gamma$.
As indicated by the $\gamma_0 = 1.4$ example (Section 5.3.1), mildly \rel\
shocks act as \nonrel\ ones showing a dramatic weakening of the
subshock combined with a large increase in $r$
(Fig.~\ref{fig:profG2}). These changes result in a particle
distribution which is both steeper than the \TP\ power law at low
momenta and flatter at high momenta (Fig.~\ref{fig:specG2}).

In faster shocks (i.e., $\gamma_0 \gtrsim 3$), the initial \TP\ spectrum
is steep enough that particle escape is unimportant so only changes in
$\Gamma$ determine $r$ (Eq.~\ref{betasol}). In contrast to
\nonrel\ shocks where the production of \rel\ particles causes the
compression ratio to increase, it is shown that $r$ decreases smoothly to
$3$ as $\gamma_0$ increases and the fraction of fully \rel\ shocked
particles approaches one (Fig.~\ref{fig:ratio}).

An important result is that efficient, mildly \rel\ shocks do
not produce particle spectra close to the so-called `universal' power
law having $\sigma \sim 4.23$. This may be important for the
interpretation of gamma-ray bursts since the internal shocks assumed
responsible for converting the bulk kinetic energy of the fireball
into internal particle energy may be mildly \rel\ and the
external shocks, believed responsible for producing gamma-ray burst
afterglows, will always go through a mildly \rel\ phase
\citep[see][for a comprehensive review of gamma-ray bursts]{Piran99}.


\section{Relativistic magnetohydrodynamic Monte Carlo model}

The basic shock acceleration physics is independent of the speed of the
shock, but the mathematical modeling of the shock process depends on
whether or not particle speeds are large compared to the shock speed.
The relativistic Monte Carlo model used for this study, 
allows particle speeds to be comparable to shock speeds
and, aside from computational limits, the
model allows shocks to have arbitrary Lorentz factors.

The fully relativistic nonlinear Monte Carlo model, under development from
a nonrelativistic model, will
simulate kinematic particle acceleration in steady-state 
modified shocks with a gyrotropic pressure tensor 
and magnetic fields at oblique angles when it is completed. 
In its present state, operating with
the magnetic field parallel to the shock normal, 
the shocks can be self-consistently modified while conserving particles, 
momentum, and energy across the shock, and while allowing charged
particles to gain energy by first-order Fermi acceleration, and
escape when they reach a high enough energy. The model can also be used with
oblique angles and anisotropic pressure for unmodified test particle shocks. 

New flux conservation laws of energy and momentum were developed
by setting the divergence of the stress-energy tensor equal to zero and
applying the 4-dimensional version of Gauss' theorem. Combining these equations
with Maxwell equations yields a completely general
set of relativistic magnetohydrodynamic
jump conditions across the shock. A new equation of state
was developed from the conservation of total energy
that incorporates oblique magnetic fields and gyrotropic
pressure. A new equation was developed from kinetic theory that gives an
excellent approximation to the adiabatic index at intermediate shock
speeds. The adiabatic index becomes the normal ratio of specific heats 
at the nonrelativistic limit ($5/3$) and at the ultrarelativistic limit
($4/3$). An anisotropy parameter was introduced for the gyrotropic pressure
tensor to allow an analytic solution to the equations.
The Monte Carlo code was further developed with new 
relativistic momentum transformation equations and a
method of handling fluxes at grid zones for relativistic shock modification. 

The Monte Carlo model was used first to duplicate well known test
particle power law results in fully nonrelativistic and ultrarelativistic
parallel shocks and was shown to move all parameters smoothly between
these two shock speed regimes. 
The model was used in the trans-relativistic regime
and can self-consistently determine compression ratios and adiabatic
indices in close agreement with recent analytic results of \citet{KGGA2000}.  
A number of solutions were presented
where the downstream pressure is anisotropic.  This is believed to be
the first presentation of oblique shock solutions
which apply smoothly for \nonrel, \transrel, and \ultrarel\ shocks.

The results presented in Chapter 5 assumed that no first-order Fermi
acceleration occurs, but they apply directly to test-particle
acceleration where the energy density in accelerated particles is
small. In that case, the differences in the compression ratio,
$r$, produced by large magnetic
fields and/or anisotropic pressures, even at \ultrarel\ speeds, become
important since changes in $r$ map directly to changes in the power
law index of the accelerated spectrum. This power law is the most
important characteristic of test-particle Fermi acceleration and is
often associated with \ultrarel\ internal shocks in gamma-ray burst
(GRB) models.
In standard GRB models, the rapidly expanding fireball cools,
converting the internal energy of the hot plasma into kinetic energy
of the relativistically moving ejecta and electron-positron pairs. At
the point where the fireball becomes optically thin and the GRB we see
is emitted, the matter is too cool to emit gamma-rays unless some
mechanism can efficiently re-convert the kinetic energy back into random
internal energy, i.e., unless some particle acceleration process takes
place. Fermi shock acceleration is widely believed to be this
mechanism \citep[e.g.,][]{RM92,Piran99}.

Internal flows in GRBs may be \ultrarel, but when a fast flow overtakes
a slower flow, the resulting shocks may be in the \transrel\ range
and may be responsible for the early transient gamma-ray intensity peaks.
The outer blast wave shock slows as it 
sweeps up and accelerates interstellar material. This
shock, believed to produce long-lasting afterglows, will eventually
always go through a \transrel\ phase, and the results apply to
particle acceleration here as well.
%

\section{Application to lepton and baryon acceleration}
The Monte Carlo model was applied to lepton and baryon
particle acceleration for the case of moderately relativistic,
smoothed shocks with the shock normal parallel to the magnetic field. 
Using leptons of various light masses, and equal particle densities,
it was found that unmodified subshocks of various fractional sizes 
within the main modified shock produced a large change in the lepton flux.
The sensitivity to the larger fractional sizes of the subshock was little
affected by the speed of the shock
in the range of Lorentz factors between
2 and 12. However, in the case of no subshock, or small subshock,
it was found that the
lepton to baryon relative injection efficiency
ratio, based on the critical momentum, $p_0$, where baryons start the
power law distribution,
increased substantially as the shock speed
increased. For leptons with mass of $0.01m_p$ that of the baryon,
leptons reach their maximum relative injection
efficiency of approximately $10^{-2.6}$ that of baryons 
around a shock Lorentz factor of
$100$. For a lepton mass of $m_e \sim (1/1836)m_p$, the shock Lorentz factor
may need to be greater than $1000$ to reach a 
maximum relative injection efficiency
of about $10^{-4.2}$ that of baryons. Such speeds may briefly 
occur in the initial outflow of a GRB, but it would require the
entrained total baryonic mass to be very small 
\citep[i.e., less than $10^{-4}M_{\odot}$;][]{Rees00}. 
The conclusion here is that,
given the assumptions of this model with particles energized by diffusive
shock acceleration, leptons cannot carry sufficient
energy to explain the observed spectra if the lepton
and baryon densities are of the same order.

Present gamma-ray burst models require leptons to 
carry a significant fraction of the total energy density of the burst.
The most likely possibility is that
there are far more leptons than baryons. Arguments connected with opacity
and variability timescales, etc \citep[e.g.,][]{Piran97} require highly
relativistic outflow in GRBs and this is only possible if 
the total baryonic mass is low. If the 
total baryonic mass is low and if 
electron-positron pairs are created in large
numbers by the initial gamma-ray photon burst while the burst is still
opaque \citep{RML02} and by neutrinos if the baryonic density is low
\citep{Rees00}, then lepton to baryon particle density ratio should
be high. If it is high enough, then leptons can carry the bulk of the 
available burst energy.
 
The question then is what ratio of leptons to baryons would be
required for equal sharing of the energy between the two species. 
The characteristics of leptons and baryons with
various particle number densities were studied for plane parallel,
relativistic, modified shocks in the same speed range. 
Initial energy and momentum of the particles, after the first shock crossing, 
was found to be in the ratio of their masses as expected.
For moderately high lepton to baryon number density ratios 
(e.g., $10$ - $100$),
the smoothed shock velocity profile modification was completely 
dominated by baryons. Using a lepton mass of $0.01m_p$, it was
found that the shock and spectral characteristics make a transition from
baryon dominated to lepton dominated in the 
lepton to baryon number density ratio
range of $10^3$ to $10^4$. Leptons with mass $0.01m_p$ shared
energy density equally with baryons when the particle number density ratio
was approximately 1600. Above this point, leptons begin to completely 
dominate the energy and energy flux, and was found to be directly proportional
to number density ratio.

By using leptons of various masses, a projection was possible down to
the mass of the electron ($m_e/m_p \sim 1/1836$). An important and new 
result was that the energy density equipartition point between 
electron-positron pairs and baryons was at a lepton to baryon 
particle number density ratio of $3.2\times10^5$, and it appears to be
independent of the shock Lorentz factor. This suggests the possibility 
that, over a wide range of shock conditions, 
the number density ratio of leptons to baryons required for 
equipartition of kinetic energy in a GRB may stay roughly constant,
similar to the constant value found above for parallel shocks.  

After the Monte Carlo model is completed, the
model will be used to further investigate lepton and baryon acceleration
using oblique nonlinear shocks. This may better establish the 
lepton to baryon number density ratio required for equipartition. 

%

\appendix

\chapter{Definitions and Fundamental Relationships}

\begin{description}
\item[\Alf\ velocity] $v_A \propto \bfrac{B}{\sqrt{\gamma\rho}}$ 
where $B$ is the magnetic field strength, $\gamma$ is the pertinent
Lorentz factor and $\rho$ is the rest mass density. The \Alf\ velocity
refers to the speed of propagation of a transverse magnetic disturbance
along the direction of the magnetic field.
 
\item[Gyroradius]
 $r_g = \bfrac{p}{ZqB}$ for charged particles in the presence
of a magnetic field,
where $p$ is the particle's momentum, $q$ is the unit electric charge, 
$Zq$ is the particle's total charge ($Z$ is the total charge number), and 
$B$ is the magnetic field strength.

\item[Mean free path]
 $\lambda = \eta r_g$, is the average distance between
collisions, if the particles undergo collisions. For collision{\it less} 
interactions (e.g., through magnetic field tubulence), a similar concept
can be defined as the average distance required for the particle to
change direction by $90$ degrees. Bohm diffusion is assumed when the 
gyrofactor $\eta = 1$.
Higher values of $\eta$ imply smoother magnetic fields, less
scattering, and longer mean free paths.

\item[Diffusion coefficient] 
$\kappa = \bfrac{1}{3}\lambda v = 
\bfrac{1}{3}\eta r_gv = \bfrac{1}{3}\bfrac{\eta pv}{ZqB} = 
\bfrac{1}{3}\bfrac{\eta\gamma\beta^2mc^2}{ZqB}$, where $\beta$ is the 
speed of the particle.

\item[Diffusion length] 
$L = \bfrac{\kappa}{u_0} = \bfrac{1}{3}\bfrac{\lambda v}{u_0} = 
\bfrac{1}{3}\bfrac{\eta r_gv}{u_0} \propto pv$, 
where $u_0$ is the shock speed.

\item[Collision time] 
$t_c = \bfrac{\lambda}{v} = \bfrac{\eta r_g}{v}$ is a 
means of estimating the time for a particle to scatter through an angle
of $90$ degrees.

\item[Debye length]
$\lambda_D \propto \sqrt{\bfrac{E}{nq^2}}$ is a measure of the
electrostatic shielding distance around a charged particle in a plasma 
\citep{shu92}.
$E$ refers to the kinetic energy of electrons, $n$ is the total
particle number density, and $q$ is the unit electric charge.

\item[Rigidity] 
$R = \bfrac{pc}{Zq}$ provides a measure of the sensitivity of the momentum
of a particle to a magnetic field. The higher the rigidity, 
the less influence a magnetic field has on the path of the particle.
 
\item[Energy and momentum flux distributions]
distributions $f(p)$ or $g(E)$ are defined here as the total number of 
passages of particles through a grid plane from any direction.
\begin{figure}[!hbtp]             
\dopicture{.5}{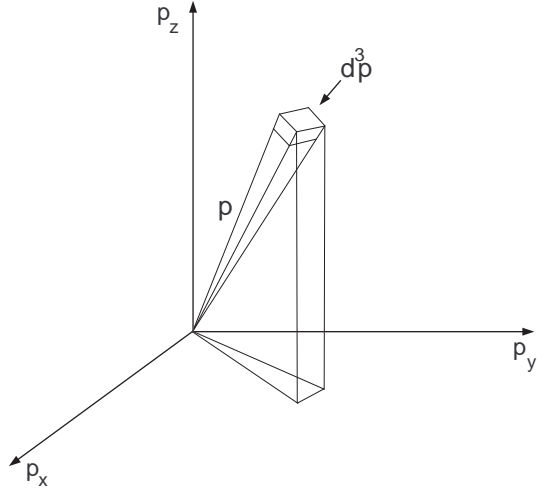}         
\caption{Momentum space diagram showing  an element of momentum space 
$d^3p$
\label{fig:f_of_p}
}
\end{figure}
Referring to Figure \ref{fig:f_of_p}, the total number of particles in
momentum space is
\begin{equation}
\label{eq:Intf1}
\int_{-\infty}^{\infty}f(\vec{p})d^3p = 4\pi\int_0^{\infty}p^2f(p)dp
\end{equation}
provided the integration takes place where the distribution is isotropic;
then
\begin{equation}
\label{eq:densf1}
4\pi p^2 f(p) dp\ \mathrm{gives\ the}\  
\left[\bfrac{\mathrm{\#\ \ of\ \ particles}}{\mathrm{cm^3}}\right] \ 
\mathrm{in\ \ dp}
\end{equation}
In the same way, if 
\begin{equation}
\label{eq:Intf2}
\int_0^{\infty}g(E)dE =
\bfrac{4\pi}{v}\int_0^{\infty}\bfrac{dJ}{dE}dE 
\end{equation}
represents the total number of particles in energy space, where
$\bfrac{dJ}{dE}$ represents the differential flux in
particles/($\mathrm{cm}^2\cdot 
\mathrm{second}\cdot\mathrm{ster}\cdot\mathrm{E}$),
then
\begin{equation}
\label{eq:dJdE}
\bfrac{4\pi}{v}\bfrac{dJ}{dE}dE\ \mathrm{gives\ the}\ 
\left[\bfrac{\mathrm{\#\ \ of\ \ particles}}{\mathrm{cm^3}}\right] \ 
\mathrm{in\ \ dE}
\end{equation}
and the densities can be equated, yielding:
\begin{equation}
\label{eq:de_dp}
\bfrac{dJ}{dE} = vp^2f(p)\bfrac{dp}{dE}
\end{equation}
Finally, using the relation $E^2 = p^2c^2 + (mc^2)^2$, 
equation (\ref{eq:de_dp}) simplifies to
\begin{equation}
\label{eq:J_p}
\bfrac{dJ}{dE} = p^2f(p)
\end{equation}
\end{description}
%

\chapter{Determining the ratio of specific heats}

\section{Kinetic pressure and the equation of state}  

The total distribution of particle momenta in a plasma
gives the plasma  a certain internal energy which has associated
with it an adiabatic exponent or index. The discussion that
follows develops a relationship between this adiabatic index
and the calculated average momentum, or kinetic energy of the particles,
based on simple statistical physics. The relationship can be used to
estimate the adiabatic index in the downstream plasma frame of a
relativistic shock. The concepts are well known and versions of the
derivation below are found in most basic texts that discuss kinetic
theory \citep[e.g.,][]{Feyn63}. 
This overview is for the convenience of the reader.
  
The adiabatic equation of state for an ideal gas of point particles,
relating the pressure and volume of a closed system, may be written as 
\begin{equation}
\label{ad_eos1}
PV^{\Gamma} = C 
\end{equation} 
where $P$ is the kinetic pressure, $V$is the volume under consideration,
$\Gamma$ is the adiabatic index, and $C$ is a proportionality constant.
The significance of point particles is that the particles
have no rotational or vibrational contributions and all of the 
kinetic energy of the particles is due to translation.

Taking logarithms and then the derivatives in the equation above, one has
\begin{equation}
\label{ad_eos2}
\frac{dP}{P} + \Gamma\frac{dV}{V} = 0 \qquad {\mathrm or} 
\qquad VdP + \Gamma PdV = 0
\end{equation} 
Adding $PdV$ to both sides and rearranging,
\begin{equation}
\label{ad_eos3}
\frac{PdV + VdP}{\Gamma - 1} = -PdV = dU.
\end{equation} 

Since $-PdV$ represents an adiabatic incremental compression of a volume at
constant pressure and $dU$ is the corresponding increase in total internal
energy, and since the left hand side has the total derivative of $PV$,
one can integrate, with the appropriate limits, and write:
\begin{equation}
\label{adiabat_eos6}
\frac{PV}{\Gamma -1} = U \qquad \mathrm{or} \qquad \frac{U}{V} = u = 
\frac{P}{\Gamma - 1}.
\end{equation} 
where $u$ is the total internal energy density of the
particle ensemble.

The exponent, $\Gamma$, is sometimes called the 
{\it ratio of specific heats}, but this is only accurate in the 
nonrelativistic ($\Gamma = 5/3$) and ultrarelativistic ($\Gamma = 4/3$)
limits. Between these two limits, $\Gamma$ is simply the exponent
for the adiabatic relation or the adiabatic index.

The total energy density $e$ and isotropic pressure $P$ are related through
a combination of the adiabatic equation of state 
(i.e., equation (\ref{adiabat_eos6}) and the conservation of energy,
which forms another equation of state:
\begin{equation}
\label{eq:tot_eos}
e = \bfrac{P}{\Gamma - 1} +\rho c^2
\end{equation}
%

\section{Nonrelativistic case}
\begin{figure}[!hbtp]             
\dopicture{.45}{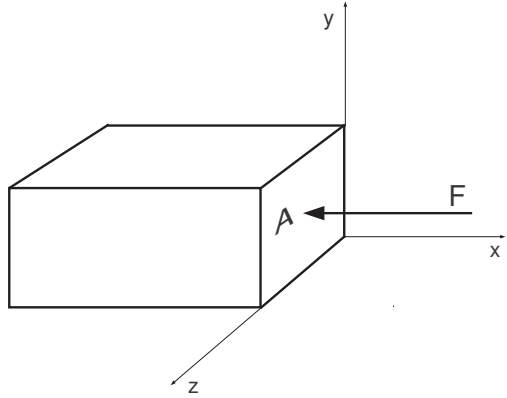}         
\caption{A rectangular volume containing ideal particles under adiabatic
compression from a force $F$.
\label{fig:pbox}
}
\end{figure}

Consider a rigid rectangular chamber with a movable wall on the right end
as shown in Figure \ref{fig:pbox}. The chamber  
contains ideal point particles of 
particle density $n$. The volume of the chamber has 
dimensions $x,y,z$  and the movable wall with area $A = yz$
on the right end is pushed into the volume a distance $-dx$ by a 
normal force $F$. Then the reversible 
work done on the volume is $dW = F(-dx) = PA(-dx) = -PdV$, 
where $-dV$ is the corresponding decrease in volume and 
P is the pressure inside the box. 

The total momentum change by a particle moving in the $x$ direction 
inside the volume and elastically colliding with the movable wall is 
$\delta p = 2p_x$. The total number of particles that collide with
the movable wall in time $\delta t$ is $nv_x\delta tA$. 
Hence, the force on the movable wall inside the volume is
$F = \bfrac{\delta p_{tot}}{\delta t} = (nv_xA )(2p_x)$ and the pressure is 
$P = 2np_xv_x$. The particle velocities have a wide range of speeds and
directions and, in particular, only half the particles in the $x$ direction
move toward the movable wall. Taking these considerations into account while
averaging over the velocities, the pressure becomes 
$P = n<p_xv_x>$.
If there are no magnetic fields and the 
$x,y \mathrm{\ and\ } z$ velocity distributions are equivalent, then 
$<p_xv_x> = \bfrac{1}{3}<\vec{p}\cdot\vec{v}>$, and the {\it isotropic}
pressure can be written as
\begin{equation}
\label{eq:adiabat_P1}
P = \bfrac{1}{3}n<\vec{p}\cdot\vec{v}> 
\end{equation}
It might be noted here, that if magnetic fields were present, one might
take averages of the momenta parallel and perpendicular to the magnetic
field, assuming rotational symmetry about the field vector. 
In that case one could have a {\it gyrotropic} pressure (tensor)
with the following space components along the diagonal:
\begin{equation}
\label{eq:P_para}
P_{\parallel} = n<\vec{p}\cdot\vec{v}>_{\parallel}
\end{equation}
and
\begin{equation}
\label{eq:P_perp} 
P_{\perp} = \bfrac{n}{2}<\vec{p}\cdot\vec{v}>_{\perp}
\end{equation}

In general, the momentum for a particle with mass $m$ and speed $v$ is
\begin{equation}
\label{adiabat_mom1}
p = \bfrac{\sqrt{E_{tot}^2 - (mc^2})^2}{c} = mc\sqrt{\gamma^2 - 1} = 
mc\sqrt{\gamma^2\beta^2} = \gamma mv
\end{equation}
where the Lorentz factor $\gamma = (1 - \bfrac{v^2}{c^2})^{-\frac{1}{2}}$.

\noindent
Using this relationship for momentum one can write, for isotropic pressure,
\begin{equation}
\label{adiabat_P21}
P = \bfrac{1}{3}n\left<\gamma mv^2\right>\ .
\end{equation}
For speeds where $\gamma \sim 1$, equation (\ref{P21}) can be approximated
to first order as
\begin{equation}
\label{adiabat_P22}
P = \bfrac{2}{3}n\left<\bfrac{mv^2}{2}\right> = 
\bfrac{2}{3}\bfrac{U}{V}\ ,
\end{equation}
where $U$ is the total internal energy and is the total kinetic energy
of ideal point particles in the volume under consideration. 
 
Comparing equation (\ref{adiabat_P22}) to equation (\ref{adiabat_eos6}) 
above, it
can be seen that $\Gamma - 1$ corresponds to $\bfrac{2}{3}$, which yields
$\Gamma = \bfrac{5}{3}$. This is the value for the adiabatic ratio of
specific heats for ideal point
particles having nonrelativistic speeds where $\gamma \sim 1$.

An error is introduced if the expression $\gamma mv^2/2$ is interpreted to
be kinetic energy at mildly  
relativistic speeds where the appropriate expression
for kinetic energy should be $mc^2(\gamma - 1)$. For example,
a 0.1 per cent error occurs, and will continue to increase for 
$v \gtrsim 0.1c$. At $v\sim 0.6c$, the error increases to about five per cent.
%

\section{Ultrarelativistic case}

The same ideas can be used to derive an expression for isotropic pressure
for ultrarelativistic particles in 
the volume described above, and the resulting relationship
will be the same as in equation (\ref{eq:adiabat_P1}), i.e. 
$P = \bfrac{1}{3}n<\vec{p}\cdot\vec{v}>$. However, for true ultrarelativistic
particles, the momentum of the particles is 
$p = \bfrac{E_{tot}}{c}$, kinetic energy $E_k = E_{tot} = \gamma mc^2$
and $v \sim c$. 
The equation for isotropic pressure becomes:
\begin{equation}
\label{eq:urP}
P = \bfrac{1}{3}n\left<\gamma mc^2\right> = \bfrac{1}{3}\bfrac{U}{V}
\end{equation}
and by the same comparison, this time equation (\ref{eq:urP})
 to equation (\ref{adiabat_eos6}), 
$\Gamma - 1 = \bfrac{1}{3}$ and
$\Gamma = \bfrac{4}{3}$. 

An error is introduced if the ultrarelativistic case is assumed at less
relativistic speeds where the rest mass cannot quite be neglected.
For example, a 0.1 per cent error occurs, 
and will continue to slowly increase for Lorentz factors 
$\gamma \lesssim 500$, but at $\gamma \sim 10$, the error has only increased
to approximately five per cent.

%

\newcommand{\aaDE}[3]{ 19#1, A\&A, #2, #3}
\newcommand{\aatwoDE}[3]{ 20#1, A\&A, #2, #3}
\newcommand{\aasupDE}[3]{ 19#1, {\itt A\&AS,} {\bf #2}, #3}
\newcommand{\ajDE}[3]{ 19#1, {\itt AJ,} {\bf #2}, #3}
\newcommand{\anngeophysDE}[3]{ 19#1, {\itt Ann. Geophys.,} {\bf #2}, #3}
\newcommand{\anngeophysicDE}[3]{ 19#1, {\itt Ann. Geophysicae,} {\bf #2}, #3}
\newcommand{\annrevDE}[3]{ 19#1, {\itt Ann. Rev. Astr. Ap.,} {\bf #2}, #3}
\newcommand{\apjDE}[3]{ 19#1, {\itt ApJ,} {\bf #2}, #3}
\newcommand{\apjtwoDE}[3]{ 20#1, {\itt ApJ,} {\bf #2}, #3}
\newcommand{\apjletDE}[3]{ 19#1, {\itt ApJ,} {\bf  #2}, #3}
\newcommand{\apjlettwoDE}[3]{ 20#1, {\itt ApJ,} {\bf  #2}, #3}
\newcommand{\apjpress}{{\itt ApJ,} in press}
\newcommand{\apjletpress}{{\itt ApJ(Letts),} in press}
\newcommand{\apjsDE}[3]{ 19#1, {\itt ApJS,} {\bf #2}, #3}
\newcommand{\apjsubDE}[1]{ 19#1, {\itt ApJ}, submitted.}
\newcommand{\apjsubtwoDE}[1]{ 20#1, {\itt ApJ}, submitted.}
\newcommand{\appDE}[3]{ 19#1, {\itt Astroparticle Phys.,} {\bf #2}, #3}
\newcommand{\apptwoDE}[3]{ 20#1, {\itt Astroparticle Phys.,} {\bf #2}, #3}
\newcommand{\araaDE}[3]{ 19#1, {\itt ARA\&A,} {\bf #2}, #3}
\newcommand{\asrDE}[3]{ 19#1, {\it Adv. Space Res.,} {\bf #2}, #3}
\newcommand{\assDE}[3]{ 19#1, {\itt Astr. Sp. Sci.,} {\bf #2}, #3}
\newcommand{\icrcplovdiv}[2]{ 1977, in {\itt Proc. 15th ICRC(Plovdiv)},
   {\bff #1}, #2}
\newcommand{\icrcsaltlake}[2]{ 1999, {\itt Proc. 26th Int. Cosmic Ray Conf.
    (Salt Lake City),} {\bf #1}, #2}
\newcommand{\icrcsaltlakepress}[2]{ 19#1, {\itt Proc. 26th Int. Cosmic Ray Conf.
    (Salt Lake City),} paper #2}
\newcommand{\jgrDE}[3]{ 19#1, {\itt J.G.R., } {\bf #2}, #3}
\newcommand{\jppDE}[3]{ 19#1, {\it J. Plasma Phys., } {\bf #2}, #3}
\newcommand{\mnrasDE}[3]{ 19#1, {\itt M.N.R.A.S.,} {\bf #2}, #3}
\newcommand{\mnrastwoDE}[3]{ 20#1, {\itt M.N.R.A.S.,} {\bf #2}, #3}
\newcommand{\mnraspress}[1]{ 20#1, {\itt M.N.R.A.S.,} in press}
\newcommand{\natureDE}[3]{ 19#1, {\itt Nature,} {\bf #2}, #3}
\newcommand{\nucphysA}[3]{#1, {\itt Nuc. Phys. A,} {\bf #2}, #3}
\newcommand{\pfDE}[3]{ 19#1, {\itt Phys. Fluids,} {\bf #2}, #3}
\newcommand{\phyreptsDE}[3]{ 19#1, {\itt Phys. Repts.,} {\bf #2}, #3}
\newcommand\prDE[3]{ 19#1, {\itt Phys. Rev.,} {\bf #2}, #3}
\newcommand{\physrevEDE}[3]{ 19#1, {\itt Phys. Rev. E,} {\bf #2}, #3}
\newcommand{\prlDE}[3]{ 19#1, {\itt Phys. Rev. Letts,} {\bf #2}, #3}
\newcommand{\revgeospphyDE}[3]{ 19#1, {\itt Rev. Geophys and Sp. Phys.,}
   {\bff #2}, #3}
\newcommand{\rppDE}[3]{ 19#1, {\itt Rep. Prog. Phys.,} {\bf #2}, #3}
\newcommand{\ssrDE}[3]{ 19#1, {\itt Space Sci. Rev.,} {\bf #2}, #3}
%
%
\setlength{\baselineskip}{12pt}


\begin{thebibliography}{List of References}    
\addcontentsline{toc}{chapter}{\bf 8\hspace{.18cm}  Bibliography}

\bibitem[Achterberg \etal(2001)]{AGKG2001}
Achterberg, A., Gallant, Y. A., Kirk, J. G., \& Guthmann, A. W.,
{\it Particle acceleration by ultra-relativistic shocks: theory and
simulations}, 
\mnrastwoDE{01}{328}{393}

\bibitem[Appl \& Camenzind(1988)]{AC88}
Appl, S. \& Camenzind, M.,
{\it Shock conditions for relativistic MHD jets},
\aaDE{88}{206}{258}

\bibitem[Axford, Leer, \& Skadron(1977)]{ALS77}
Axford, W. I., Leer, E., \& Skadron, G.,
{\it The Acceleration of Cosmic Rays by Shock Waves},
\icrcplovdiv{11}{132}

\bibitem[Ballard \& Heavens(1991)]{BH91}
Ballard, K.R. \& Heavens, A.F.,
{\it First-order Fermi Acceleration at Oblique Relativistic 
Magnetohydromagnetic Shocks},
\mnrasDE{91}{251}{438}

\bibitem[Bednarz \& Ostrowski(1996)]{BedOstrow96}
Bednarz, J., \& Ostrowski, M.,
{\it The acceleration time-scale for first-order Fermi acceleration in
relativistic shock waves},
\mnrasDE{96}{283}{447}

\bibitem[Bednarz \& Ostrowski(1998)]{BedOstrow98}
Bednarz, J. \& Ostrowski, M.,
{\it Energy Spectra of Cosmic Rays Accelerated at 
Ultrarelativistic Shock Waves},
\prlDE{98}{80}{3911}

\bibitem[Bell(1978a)]{Bell78a}
Bell, A. R.,
{\it The Acceleration of Cosmic Rays in Shock Fronts - I},
\mnrasDE{78a}{182}{147-156}

\bibitem[Bell(1978b)]{Bell78b}
Bell, A. R.,
{\it The Acceleration of Cosmic Rays in Shock Fronts - II},
\mnrasDE{78b}{182}{443-445}

\bibitem[Bell(1987)]{Bell87}
Bell, A.R., 
{\it The non-linear self-regulation of cosmic ray acceleration at shocks},
\mnrasDE{87}{225}{615-626}

\bibitem[Berezhko \& Ellison(1999)]{BE99}
Berezhko, E.~G., \& Ellison, D.~C., 
{\it A Simple Model of Nonlinear Diffusive Shock Acceleration},
\apjDE{99}{526}{385}

\bibitem[Berezhko \& V\"olk(1997)]{BerezV97}
Berezhko, E. G., \& V\"olk, H. J., 
{\it Kinetic Theory of Cosmic Rays and Gamma
Rays in Supernova Remnants. I. Uniform Interstellar Medium}, 
\appDE{97}{7}{183}

\bibitem[Blandford \& Eichler(1987)]{BE87}
Blandford, R. D., \& Eichler, D,
{\it Cosmic Ray Acceleration in Astrophysics},
\phyreptsDE{87}{154}{1-75}

\bibitem[Blandford \& McKee(1976)]{BM76}
Blandford, R. D. \& McKee, C. F., 
{\it Fluid Dynamics of Relativistic Blast Waves},
\pfDE{76}{19, No.8}{1130}

\bibitem[Blandford \& Ostriker(1978)]{BO78}
Blandford, R.D., \& Ostriker, J.P.,
{\it Particle Acceleration by Astrophysical Shocks},
\apjletDE{78}{221}{L29}
 
\bibitem[Cabannes(1970)]{Cab70}
Cabannes, H.,
{\it Theoretical Magnetofluid Dynamics} Academic Press (1970)

\bibitem[Cheng \& Lu(2001)]{CL2001}
Cheng, K.S. \& Lu, T., 
{\it Gamma-Ray Bursts: Afterglows and Central Engines},
2001, {\itt Chin. J. Astron. Astrophys., }{\bf 1, No. 1}, 1-20

\bibitem[de Hoffmann \& Teller(1950)]{dHT50}
de Hoffman, F. \& Teller, E.,
{\it Magneto-Hydrodynamic Shocks},
\prDE{50}{80}{692}

\bibitem[Dieckmann, et.al.(2000)]{DMCDD00}
Dieckmann, M.E., et.al.,
{\it Electron Acceleration due to High Frequency Instabilities at 
Subpernova Remnant Shocks},
\aatwoDE{00}{356}{377-388}

\bibitem[Drury(1983)]{Drury83}
Drury, L.~O'C., 
{\it An Introduction to the Theory of Diffusive Shock Acceleration
of Energetic Particles in Tenuous Plasmas},
\rppDE{83}{46}{973-1027}

\bibitem[Dwarkadas \& Chevalier(1998)]{DC98}
Dwarkadas, V.V. \& Chevalier, R.A.,
{\it Interaction of Type Ia Supernovae with their Surroundings},
\apjDE{98}{497}{807-823}

\bibitem[Ellison(1981)]{E81} 
Ellison, D.C.,
{\it Monte Carlo Simulation of Collisionless Shock Acceleration},
PhD. Thesis, 1981, Catholic University

\bibitem[Ellison(1985)]{Ellison85}
Ellison, D.C.,
{\it Shock Acceleration of Diffuse Ions at the Earth's Bow Shock:
Acceleration Efficiency andA/Z Enhancement},
\jgrDE{85}{90-A1}{29-38}

\bibitem[Ellison(1991a)]{EllisonJapan91}
Ellison, D. C. 1991a,
{\it Fermi Particle Acceleration in Relativistic Shocks:
Spectra and Efficiencies from Modified Shocks},
Proc. ICRR International Symposium
on ``Astrophysical Aspects of the Most Energetic Cosmic Rays,''  
p. 281, Eds., M. Nagano and F. Takahara, Wold Scientific, Bangalore.

\bibitem[Ellison(1991b)]{EllisonPoland91}
Ellison, D. C. 1991b,
{\it Fermi Particle acceleration in relativistic shocks: Preliminary
nonlinear results},
in ``Relativistic Hadrons in Cosmic Compact Objects,'' p. 101, Eds.,
A. A. Zdziarski and M. Sikora, Springer-Verlag, Berlin.

\bibitem[Ellison, Baring, \& Jones(1995)]{EBJ95}
Ellison, D. C., Baring, M. G., \& Jones, F. C.,
{\it Acceleration Rates and Injection Efficiencies in Oblique Shocks},
\apjDE{95}{453}{873-882}

\bibitem[Ellison, Baring, \& Jones(1996)]{EBJ96}
Ellison, D. C., Baring, M. G., \& Jones, F. C.,
{\it Nonlinear Particle Acceleration in Oblique Shocks},
\apjDE{96}{473}{1029}

\bibitem[Ellison \& Double(2002)]{ED2002}
Ellison, D.D. \& Double, G.P.,
{\it Nonlinear Particle Acceleration in Relativistic Shocks},
\apptwoDE{02}{18-3}{213-228}

\bibitem[Ellison \& Eichler(1984)]{EE84}
Ellison, D. C., \& Eichler, D.
{\it Monte Carlo Shock-Like Solutions to the Boltzmann Equation 
With Collective Scattering},
\apjDE{84}{286}{691}

\bibitem[Ellison et al.(1993)]{EGBS93}
Ellison, D. C., Giacalone, J., Burgess, D., and Schwartz, S. J., 
{\it Simulations of Particle Acceleration in Parallel Shocks:
Direct Comparison Between Monte Carlo and One-Dimensional Hybrid
Codes}, 
\jgrDE{93}{98}{21,085}

\bibitem[Ellison, Jones, \& Baring(1999)]{EJB99} Ellison, D. C.,
Jones, F. C., \& Baring, M. G.,  
{\it Direct Acceleration of Pickup Ions at the Solar Wind
Termination Shock: The Production of Anomalous Cosmic Rays},
\apjDE{99}{512}{403}

\bibitem[Ellison, Jones, \& Reynolds(1990)]{EJR90}
Ellison, D.C., Jones, F.C., and Reynolds, S.P.,
{\it First-Order Fermi Particle Acceleration at Relativistic Shocks},
\apjDE{90}{360}{702}

\bibitem[Ellison, M\"obius, \& Paschmann(1990)] {EMP90} 
Ellison, D.C., M\"obius, E., \& Paschmann, G., 
{\it Particle Injection and Acceleration at the Earth's Bow Shock: 
Comparison of Upstream and Downstream Events},
\apjDE{90}{352}{376} 

\bibitem[Ellison \& Reynolds(1991)]{ER91}
Ellison, D. C., \& Reynolds, S. P.,
{\it A determination of relativistic shock jump conditions 
using Monte Carlo techniques},
\apjDE{91}{378}{214}

\bibitem[Fermi(1949)]{Fermi49}
Fermi, E., 1949, {\itt Phys. Rev.}, {\bf 75}, 1169-74 
(Collected Papers, vol II, pp.655-55) 

\bibitem[Feynman, Leighton \& Sands(1963)]{Feyn63}
Feynman, R.P., Leighton, R.B. \& Sands, M.,
{\it Feynman Lectures on Physics},
1963, Addison-Wesley

\bibitem[Gallant(2002)]{Gallant2002}
Gallant, Y. A.,
{\it Particle Acceleration at Relativistic Shocks},
preprint, 2002, {\tt astro-ph/0201243}

\bibitem[Gallant \& Achterberg(1999)]{GA99}
Gallant, Y. A. \& Achterberg, A., 
{\it Ultra-high-energy cosmic ray acceleration by relativistic blast waves},
\mnrasDE{99}{305}{L6}

\bibitem[Gleeve(1984)]{Gleev84}
Gleeve, A.A., 
{\it The Heating and Acceleration of Electrons by Shocks},
\asrDE{84}{4, No.2-3}{255-263}

\bibitem[Heavens \& Drury(1985)]{HD88}
Heavens, A. F. \& Drury, L. O'C., 
{\it Relativistic shocks and particle acceleration},
\mnrasDE{88}{235}{997}

\bibitem[Hollweg(1982)]{hollweg}
Hollweg, J.,  
{\it On the Origin of Solar Spicules},
\apjDE{82}{257}{345-353}

\bibitem[Jackson(1975)]{Jksn75}
Jackson, J.D., 
{\it Classical Electrodynamics, 2nd ed.},
1975, John Wiley \& Sons, NY

\bibitem[Jones \& Ellison(1987)]{JE87}
Jones,F.C. \& Ellison, D.C., 
{\it Noncoplanar Magnetic Fields, Shock Potentials, and Ion Deflection},
\jgrDE{87}{92 No. A10}{11205}

\bibitem[Jones \& Ellison(1991)] {JE91} 
Jones, F.C., \& Ellison, D.C., 
{\it The Plasma Physics of Shock Acceleration},
\ssrDE{91}{58}{259} 

\bibitem[Kirk(1988)]{Kirk88}
Kirk, J. G., 1988, 
{\it Particle Acceleration at Relativistic Shock Fronts},
Thesis, Dr. rer. nat. habil.,
Ludwig-Maximillians-Universit\"at, M\"unchen, Germany.

\bibitem[Kirk \& Duffy(1999)]{KD99}
Kirk, J.G. \& Duffy, P.,
{\it Particle Acceleration and Relativistic Shocks},
J. Phys. G: Nucl. Part. Phys., 1999, {\bf{25}}, R163 

\bibitem[Kirk et al.(2000)]{KGGA2000}
Kirk, J. G., Guthmann, A. W., Gallant, Y. A., Achterberg, A.,
{\it Particle Acceleration at Ultrarelativistic Shocks: An Eigenfunction
Method}, 
\apjtwoDE{00}{542}{235}

\bibitem[Kirk \& Schneider(1987a)]{KS87a}
Kirk, J. G. \& Schneider, P., 
{\it On the acceleration of charged particles at relativistic shock fronts},
\apjDE{87a}{315}{425}

\bibitem[Kirk \& Schneider(1987b)]{KS87b}
Kirk, J. G., \& Schneider, P.,
{\it Particle Acceleration at Shocks: A Monte Carlo Method},
\apjDE{87b}{322}{256}

\bibitem[Kirk \& Webb(1988)]{KW88}
Kirk, J.G. \& Webb, G.M.,
{\it Cosmic-ray Hydrodynamics at Relativistic Shocks},
\apjDE{88}{331}{336}

\bibitem[Krymsky(1977)]{Krym77}
Krymsky,G.F., 1977, {\itt Dokl. Akad. Nauk SSSR}, {\bf 234}, 1306
(Engl. transl. {\itt Sov. Phys.-Dokl.}, {\bf 23}, 327), 1981,
{\itt Izv. Akad. Nauk SSSR Ser. Fiz.}, {\bf 45}, 461

\bibitem[Landau \& Lifshitz(1959)]{LL59}
Landau, L.D. \& Lifshitz, E.M.,
{\it Fluid Mechanics}, 
1959, Adison-Wesley Pub.

\bibitem[Landau \& Lifshitz(1962)]{LL62}
Landau, L.D. \& Lifshitz, E.M.,
{\it The Classical Theory of Fields}, 
1962, Adison-Wesley Pub.

\bibitem[Levinson(1992)]{LV92}
Levinson, A.,
{\it Electron Injection in Collisionless Shocks},
\apjDE{92}{401}{73-80}

\bibitem[Levinson(1996)]{LV96}
Levinson, A.,
{\it On the Injection of Electrons in Oblique Shocks},
\mnrasDE{96}{278}{1018-1024}

\bibitem[Lichnerowicz(1967, 1970)]{Lich67}
Lichnerowicz, A., 1967,
{\it Relativistic Hydrodynamics and Magnetohydrodynamics}, Benjamin

\bibitem[Malkov(1998)]{Malkov98}
Malkov, M.,
\physrevEDE{98}{58}{4911}

\bibitem[McClements, et.al(1997)]{MDBKD97}
McClements, K.G., et.al.,
{\it Acceleration of cosmic ray electrons by ion-excited waves at
quasi-perpendicular shocks},
\mnrasDE{97}{291}{241-249}

\bibitem[M\'{e}sz\'{a}ros \& Rees(1997)]{MR97}
M\'{e}sz\'{a}ros, P. \& Rees, M.J.,
{\it Poynting Jets from Black Holes and Cosmological Gamma-ray Bursts},
\apjDE{97}{482}{L29-L32}

\bibitem[Michel(1981)]{Michl81}
Michel, F.C.,
\apjDE{81}{247}{664}

\bibitem[Ostrowski(1991)]{Ostrow91}
Ostrowski, M., 
{\it Monte Carlo simulations of energetic particle transport in weakly
inhomogeneous magnetic fields - I. Particle acceleration in 
relativistic shock waves with oblique magnetic fields},
\mnrasDE{91}{249}{551}

\bibitem[Peacock(1981)]{Peacock81}
Peacock, J. A.,
{\it Fermi acceleration by relativistic shock waves},
\mnrasDE{81}{196}{135}

\bibitem[Pelletier(1999)]{Pelletier99}
Pelletier, G.,
{\it Cosmic ray acceleration and nonlinear relativistic wavefronts},
\aaDE{99}{350}{705}

\bibitem[Pelletier \& Marcowith(1998)]{PellMar98}
Pelletier, G., \& Marcowith, A.,
{\it Nonlinear dynamics in the relativistic plasma of astrophysical
high-energy sources},
\apjDE{98}{502}{598}

\bibitem[Piran(1999)]{Piran97}
Piran, T., 1997
{\it Unsolved Problems in Astrophsics},
ed. J.N. Bahcall \& J.P. Ostriker, Princeton University Press

\bibitem[Piran(1999)]{Piran99}
Piran, T., 
{\it Gamma-ray bursts and the fireball model},
\phyreptsDE{99}{314}{575}

\bibitem[Ramirez-Ruiz, MacFadyen, \& Lazzati(2002)]{RML02}
Ramirez-Ruiz, E., MacFadyen, A.I., \& Lazzati,D.,
{\it Precursors and $e^{\pm}$ pair loading from erupting fireballs}
\mnrastwoDE{02}{331}{197-202}


\bibitem[Rees(2000)]{Rees00}
Rees, M.J.,
{\it A Review of Gamma Ray Bursts},
\nucphysA{2000}{A663\&664}{42c-55c}

\bibitem[Rees \& M\'esz\'aros(1992)]{RM92}
Rees, M.J., \& , M\'esz\'aros, P. 
{\it Relativistic fireballs - Energy conversion and time-scales}
\mnrasDE{92}{258}{41}

\bibitem[Rybicki \& Lightman(1979)]{RL79}
Rybicki, G.B. \& Lightman, A.P., 
{\it Radiative Processes in Astrophysics},
1979, John Wiley \& Sons (ISBN 0-471-82759-2)

\bibitem[Scheuer(1987)]{Sch87}
Scheuer, P.A.G. (1987) {\it Astrophysical Jets and Their Engines},
ed. W. Kundt, NATO ASI SEries, Reidel Dordrecht, {\bff 208}, 129

\bibitem[Schlickeiser \& Dermer(2000)]{SD00}
Schlickeiser, R. \& Dermer, C.D.,
{\it Proton and electron acceleration through magnetic turbulence in
relativistic outflows},
\aatwoDE{00}{551}{946-972}

\bibitem[Schmitz \& Chapman(2002)]{SC02}
Schmitz, H \& Chapman, S.C., 
{\it Electron Preacceleration Mechanisms in the Foot Region of
High Alfv\'{e}nic Mach Number Shocks},
\apjtwoDE{02}{579}{327-336}

\bibitem[Schneider \& Kirk(1987)]{SK87}
Schneider, P. \& Kirk, J. G., 
{\it Fermi acceleration at shocks with arbitrary velocity profiles},
\apjDE{87}{323}{L87}

\bibitem[Shu(1992)]{shu92}
Shu, F.H., 1992,
{\it The Physics of Astrophysics, Vol. II: Gas Dynamics},
University Science Books 

\bibitem[Tan, Matzner \& McKee(2001)]{TMM01}
Tan, J.C., Matzner, C.D. \& McKee, C.F.,
{\it Trans-relativistic blast waves in supernovae as gamma-ray burst
progenitors},
\apjtwoDE{01}{551}{946-972} 

\bibitem[Taub(1948)]{Taub48}
Taub, A.H., 
{\it Relativistic Rankine-Hugoniot Equations},
\prDE{48}{74}{328-334}

\bibitem[Tidman \& Krall(1971)]{TK71}
Tidman, D.A. \& Krall, N.A. (1971),
{\it Shock waves in collisionless plasmas}, Wiley Science

\bibitem[Tolman(1934)]{T34}
Tolman, R.C., 
{\it Relativity, Thermodynamics and Cosmology},
University Press, Oxford (1934); unabridged Dover paperback reprint (1987)

\bibitem[Vietri(1995)]{Vietri95}
Vietri, M.,
{\it The acceleration of ultra-high-energy cosmic rays in gamma-ray bursts},
\apjDE{95}{453}{883}

\bibitem[Webb, Zank \& McKenzie(1987)]{WZM87}
Webb, G.M., Zank, G.P., \& MccKenzie, J.F., 
{\it Relativistic Oblique Magnetohydrodynamic Shocks},
\jppDE{87}{37, part 1}{117}

\bibitem[Zel'dovich \& Raizer(1967)]{ZR67}
Zel'dovich, Y.B. \& Raiser, Y.P.,
{\it Physics of Shock Waves and High-Temperature Hydrodynamic Phenomena},
Academic Press, 1967



\end{thebibliography}
\end{document}